\documentclass[12pt]{caltech_thesis}
\usepackage{CJKutf8}
\usepackage{graphicx}
\usepackage{dcolumn}
\usepackage{bm}
\usepackage{multirow}
\usepackage[hyphens]{url}
\usepackage{graphicx}
\usepackage{dcolumn}
\usepackage{multirow}
\usepackage{braket}
\usepackage{color}
\usepackage{ulem}

\DeclareUnicodeCharacter{2212}{-}
\DeclareUnicodeCharacter{0301}{ }
\usepackage{todonotes}
\usepackage{hyperref}
\usepackage{lipsum}
\usepackage{graphicx}

\newcommand{\calH}{\mathcal{H}}

\newcommand{\Tr}{\mathrm{Tr}}
\usepackage{braket}
\usepackage{color}
\usepackage{ulem}
\usepackage{algorithm,algorithmic}
\usepackage{indentfirst}

\usepackage[utf8]{inputenc}
\usepackage[T1]{fontenc}
\usepackage{newtxtext,newtxmath}

\usepackage[
    backend=biber,natbib,
    style=phys
]{biblatex}

\begin{document}

\title{Renormalization group flow and fixed-point in tensor network representations\\\quad\\
\begin{CJK*}{UTF8}{min}{\CJKfamily{ipxm} (テンソルネットワークによる繰り込み群フローと固定点の研究)}\end{CJK*}}
\author{Atsushi Ueda\\
\begin{CJK*}{UTF8}{min}{\CJKfamily{ipxm} 上田 篤}\end{CJK*}}
\degreeaward{Doctor of philosophy}                 
\university{The University of Tokyo}    

\orcid{0000-0003-1884-6808}

\maketitle

\begin{acknowledgements} 	 
I would like to express my profound gratitude to my supervisor, Prof. Oshikawa, for his invaluable mentorship throughout my PhD journey. His elegant and profound approach to studying physics has profoundly shaped my research philosophy and kindled my enthusiasm for discovery. His generosity in providing opportunities for international exposure and networking has been instrumental in my professional development. I particularly value his advice on mastering both numerical and analytical techniques, a strategy that marked a turning point in my research trajectory. His amiable nature and charm have not only made him a beloved figure in the physics community but have also fostered a collaborative and inspiring environment for me. The conferences and seminars I attended under his guidance were fertile grounds for the research ideas I am zealously pursuing today.

I would also like to extend my heartfelt thanks to Prof. Tada, a former assistant professor in our lab. His support and regular check-ins during the challenging COVID-19 pandemic were a great source of comfort and encouragement. His kindness, along with the support of his family, played a crucial role in helping me navigate through this difficult period. Additionally, I am deeply thankful to my friends who provided unwavering support and companionship during these trying times, making the journey less daunting.

I am grateful for the financial support provided by the MERIT-WINGS program and the JSPS fellowship (DC1), which were pivotal in facilitating my studies.

On a personal note, I owe a tremendous debt of gratitude to my family. To my parents, my sister, and her husband, whose concern and regular check-ins were a constant source of reassurance. The delicious meals and moments of joy they shared with me have been both a comfort and an inspiration. I extend my deepest appreciation to my wife, whose unwavering support, cheer, and love have been my pillars of strength. Her presence in my life is a blessing I cherish immensely.

In summary, my journey through my PhD has been enriched and made possible by each of these individuals and their unique contributions to my life, for which I am eternally grateful.
\end{acknowledgements}

\begin{abstract}
We propose the integration of an energy-based finite-size scaling methodology with tensor network renormalization (TNR) techniques. TNR, serving as a numerical implementation of real-space renormalization group (RG) methods, provides a pathway to access the low-lying energy spectrum of various systems. By melding TNR with conformal perturbation theory, we can effectively calculate running coupling constants. This combined methodology is particularly valuable in practical calculations, as it adeptly navigates around the numerical errors commonly encountered in TNR applications.

A primary objective of numerical simulations in the study of lattice models is often the precise determination of their phase diagrams. The demarcation of phase transition points within these diagrams often necessitates simulations of systems with very large sizes. For example, accurately identifying the phase boundary of the Ising model through spontaneous magnetization typically requires simulating thousands of lattice sites. While TNR is capable of handling large system sizes, it is also known to suffer from amplified numerical errors as the size of the system increases.

In contrast, our proposed methodology requires only a few steps of RG, thereby inducing fewer numerical errors and reducing computational costs. The energy-based finite-size scaling approach does not rely on large system sizes, unlike conventional methods that use observables such as magnetization and heat capacity to determine phase transitions. This approach is not only more efficient but also more resilient to the challenges posed by the scale of the simulations, offering a significant advantage in the study of critical phenomena in lattice models.

Additionally, we will delve into the origins of numerical errors in TNR simulations from a field-theoretical perspective. This analysis will shed light on how these errors scale with the approximation parameter, denoted as $D$. Understanding this scaling is critical for accurately estimating and managing errors in simulation results.

Through the application of this methodology, we aim to provide a more accurate and computationally efficient means of exploring phase transitions and the critical properties of lattice models, enhancing our understanding of these complex systems.

In the subsequent discussion, we delve into the tensor structure of fixed points in the context of lattice models. A significant challenge in this area arises from the effects of finite bond dimensions, which make the true fixed-point tensor practically unattainable through direct numerical methods. To circumvent this issue, we adopt an analytical approach, employing conformal mappings to study the fixed-point tensors.

This analytical exploration leads to a revealing insight: the tensor elements of the fixed-point tensor correspond to the four-point functions of primary operators within the framework of conformal field theory (CFT). This correspondence is not just a theoretical conjecture; it is corroborated by empirical observations showing that tensors renormalized for finite sizes tend to align with our theoretical predictions. 

The significance of this finding cannot be overstated. It suggests that the tensor representations of fixed points in lattice models embody the universality of non-trivial infrared (IR) physics at the lattice level. Our approach thus is not only a new solution to a decade-old problem, but also bridges the gap between the abstract theoretical constructs of CFT and the practical, computable structures in lattice models. By demonstrating this universal behavior, we provide robust support for the concept of universality in critical phenomena, particularly as it manifests in the intricate world of lattice models.

Through this investigation, we aim to offer a deeper understanding of the fundamental principles underlying critical phenomena, specifically highlighting how the universal aspects of CFT are reflected in the practical, numerical realm of lattice model simulations.
\end{abstract}

\begin{publishedcontent}[iknowwhattodo]
\nocite{PhysRevB.104.165132,PhysRevE.106.014104,PhysRevB.108.024413,ueda2023fixedpoint}
\end{publishedcontent}
The main content of this thesis is the first paper on this list(Phys. Rev. B 108, 024413 (2023)).\\
\underline{Contributions}\\
${}^1$ A.U participated in the conception of the project, coding of the numerical implementations, and participated in the writing of the manuscript.\\
${}^2$ A.U participated in the conception of the project, analytical calculations, coding of the numerical implementations, and participated in the writing of the manuscript.
${}^3$ A.U participated in the coding of the numerical implementations, analysis of the results, and participated in the writing of the manuscript.\\
${}^4$ A.U participated in the conception of extending level-spectroscopy to visualizing RG flows using TNR, coding of the numerical implementations, upgrading level-spectroscopy, performing third-order perturbations,  and participated in the writing of the manuscript.\\

\tableofcontents
\listoffigures
\listoftables
\printnomenclature

\mainmatter
\chapter{Introduction}
Phase transitions hold a particular fascination in the field of statistical mechanics due to their display of universality. This universality is most notably observed in the behavior of systems undergoing continuous phase transitions, which are characterized by the divergence of derivatives of the partition function. These divergences are quantified by critical exponents, which are key indicators of the system's behavior near the critical point.

What makes these critical exponents particularly intriguing is their universality across different physical systems. Despite the diverse nature of these systems, the critical exponents of certain systems tend to exhibit the same values. For instance, a striking example of this universality is seen when comparing the liquid-vapor transition in water with the ferromagnetic to paramagnetic transition in three-dimensional magnets. Remarkably, these vastly different systems share the same critical exponents. 
However, this concept of universality in phase transitions presents a somewhat counter-intuitive picture at first glance. Consider the stark differences between substances like water and magnets: water is composed of hydrogen and oxygen, while magnets can be made from materials like neodymium. Moreover, the temperatures at which these substances undergo phase transitions are vastly different. Imagine measuring the critical exponents of water in a national laboratory in the U.S. and then measuring those for magnets in a makeshift lab in your basement. Despite the differences in substances, environments, and methodologies, the results would be surprisingly consistent.

This phenomenon almost suggests that nature possesses an innate understanding of the essence of phase transitions. It is as though the underlying principles governing these phenomena inherently “\textit{know}” to discard irrelevant details, focusing instead on fundamental aspects that are common across diverse systems. This remarkable aspect of universality in phase transitions not only challenges our intuitive understanding but also highlights the profound simplicity and elegance with which nature operates at a fundamental level. 

Nature's method of simplifying complex phenomena can be likened to how we perceive images in our daily lives. Consider, for example, the iconic Windows XP wallpaper “Bliss\footnote{It was a wallpaper of my first computer.\url{https://en.wikipedia.org/wiki/Bliss_(image)}. They now have a 4K version of it.~\url{https://msdesign.blob.core.windows.net/wallpapers/Microsoft_Nostalgic_Windows_Wallpaper_4k.jpg}}.,” which depicts a serene landscape of a green hill and blue sky. At first glance, we see just these broad elements. However, upon closer inspection, one might notice finer details like yellow flowers dotting the hill. With an even more focused view, like through a microscope, one could observe bees around these flowers or even delve into the atomic structures of these elements.

Yet, from a normal viewing distance, these minute details are effectively invisible. Our perception simplifies the scene, focusing on the most significant elements while “ignoring” the smaller, less impactful ones. This is akin to how we approach phase transitions in statistical mechanics. In the study of phase transitions, we often consider the thermodynamic limit, which implies observing the system as if it were infinitely large. This perspective requires us to “step back” and view the system from a great distance, thereby making any finite-sized clusters or features with limited correlation lengths appear increasingly smaller and less significant.

This “stepping back” in observing physical systems is analogous to observing the “Bliss” wallpaper from a distance. Just as we see only the broad strokes of the hill and sky rather than the minute details, in phase transitions, the focus is on macroscopic properties that emerge when viewing the system in its entirety, from afar. Small-scale variations and details become irrelevant at this scale, allowing us to discern the universal aspects that dominate the behavior of the system as a whole. 
To quantitatively explore how systems appear to transform when we “step back" and observe them from a larger scale, the renormalization group (RG) theory becomes indispensable. This chapter is dedicated to reviewing RG theory, with a specific emphasis on its application in statistical mechanics.

RG theory provides a framework to understand how the behavior of physical systems changes across different scales. It allows us to systematically “zoom out” from the microscopic details and observe how the collective properties of a system evolve. This perspective is crucial for grasping the essence of phase transitions, as it reveals the underlying universalities that manifest when microscopic details become less relevant at macroscopic scales.

Finally, as a prelude to delving into the detailed theoretical aspects of RG theory and its applications in statistical mechanics, it can be beneficial for readers to visualize how the concept of “seeing from a distance" manifests in physical systems. To aid in this visualization, we recommend viewing a short YouTube video~\footnote{\url{https://www.youtube.com/watch?v=MxRddFrEnPc&t=1s}} that illustrates how it happens in the Ising model.

\section{Renormalization group theory in statistical mechanics}
In this section, we delve into critical aspects of phase transition and RG theory within the realm of statistical mechanics, using the Ising model as a foundational example. A key element in this context is the partition function, particularly when expressed in the transfer matrix formalism, and its intricate relationship with the corresponding action or Hamiltonian. This conceptual framework, central to our discussion, paves the way for a natural extension to and incorporation within the tensor-network language, which will be explored in the following section.

We begin our exploration with the classical one-dimensional Ising model, a fundamental and illustrative example in the study of magnetism. In this model, local spin states are characterized by a binary variable, $\sigma_i = \pm1$, where each spin can be thought of as a miniature magnet. The values of $\pm1$ can be metaphorically equated to the north (N) and south (S) poles of a magnet. This analogy is useful in visualizing how each local spin, akin to a tiny magnet, contributes to the overall magnetic properties of the system.
In the one-dimensional Ising model, each spin aligns along a single direction in an array. By considering these local spins collectively, we gain insight into the emergent magnetic behavior of the entire system, where the alignment or randomness of these “mini magnets" underpins the macroscopic properties observed. Specifically, the system is classified as \textit{ferromagnetic} if the average spin, calculated as $\frac{1}{N}\sum_{i=1}^N \sigma_i$, is non-zero. This non-zero average indicates a net alignment in one direction, characteristic of ferromagnetic ordering. Conversely, if this average equals zero, $\frac{1}{N}\sum_{i=1}^N \sigma_i =0$, the system is said to be in a \textit{paramagnetic} state. In this state, the spins are oriented randomly, resulting in no net magnetization. This binary representation not only simplifies the analysis but also provides profound insights into the underlying mechanisms of magnetic interactions and phase transitions. 

In the realm of statistical mechanics, the behavior of spins in the Ising model is governed by local Boltzmann weights. The probability of a specific spin configuration occurring is quantified by these Boltzmann weights. These weights are mathematically defined as $\exp[-\beta \epsilon\{\sigma_i\}]$, where $\beta=1/T$ represents the inverse temperature, and $\epsilon\{\sigma_i\}$ denotes the energy associated with a particular spin configuration. Consequently, the probability of observing a certain configuration,$p\{\sigma_i\}$ , is calculated using the formula $p\{\sigma_i\} = \frac{\exp[-\beta \epsilon\{\sigma_i\}]}{Z(\beta)}$. Here, $Z(\beta)=\sum_{\{\sigma_i\}}\exp[-\beta \epsilon\{\sigma_i\}]$ is the partition function, which serves as a normalization factor ensuring that the sum of probabilities over all possible configurations equals one.

The energy of the Ising model in periodic boundary condition (PBC) is defined as 
\begin{align}
    \epsilon\{\sigma_i\} = -J\sum_{i=1}^{N-1} \sigma_{i}\sigma_{i+1} - J\sigma_N\sigma_1.\label{Energy_1dIsing}
\end{align}
This definition ensures that the energy is minimized when all spins are aligned in the same direction, making this configuration highly favored at low temperatures. In contrast, at higher temperature regimes, where $\beta\sim 0$, the probability becomes \textit{ignorant} of the energy configuration. As a result, under these conditions, the preferential status of spin alignment dictated by lower energy considerations diminishes. Instead, the system tends to favor states that are more 'typical' or statistically common, reflecting a shift from energy-driven order to entropy-driven disorder.
These two distinct phases – the ordered, low-temperature phase and the disordered, high-temperature phase – are typically delineated by a critical point known as the phase transition. Around this point, various singularities emerge in the derivatives of the free energy, which is defined as $f(\beta)=-\frac{1}{\beta}\ln{Z}(\beta)$. These singularities are indicative of drastic changes in the system's behavior and are a hallmark of phase transitions in statistical mechanics. 

Let us compute the partition function for the one-dimensional Ising model as defined in Eq.~\eqref{Energy_1dIsing}. We denote the partition function for a system of $N$ spins with fixed spins at both ends as $Z_N(\sigma_1,\sigma_N)$. Calculating this function is straightforward for a system with only two spins. Let us set $J=1$ and then, we have $Z_2(+,+) =Z_2(-,-)= e^{\beta}$, $Z_2(+,-) =Z_2(-,+) = e^{-\beta}$, where $+$ and $-$ denotes $\sigma = 1$ and $-1$, respectively. Now, consider adding an additional spin at site $i=3$. The resulting partition function for three spins can be determined by considering cases where the second and third spins are either aligned or opposite. This yields the following equations:
\begin{align*}
    Z_{3}(+,+) &= e^{-\beta}Z_2(+,-) + e^{\beta}Z_2(+,+),\\
    Z_{3}(+,-) &= e^{\beta}Z_2(+,-) + e^{-\beta}Z_2(+,+).
\end{align*}
The above equations are rewritten using matrix multiplications as
\begin{align*}
    \left(
    \begin{array}{c}
      Z_{3}(+,+)\\
      Z_{3}(+,-)
    \end{array}
    \right)=
    \left(
    \begin{array}{cc}
      e^\beta&e^{-\beta}\\
      e^{-\beta}&e^\beta\\
    \end{array}
    \right)
    \left(
    \begin{array}{c}
      Z_{2}(+,+)\\
      Z_{2}(+,-)
    \end{array}
    \right).\nonumber
\end{align*}
We can repeat this procedure to obtain the partition function of $N$ spin systems as
\begin{align}
\left(
    \begin{array}{c}
      Z_{N}(+,+)\\
      Z_{N}(+,-)
    \end{array}
    \right)&=
    \left(
    \begin{array}{cc}
      e^\beta&e^{-\beta}\\
      e^{-\beta}&e^\beta\\
    \end{array}
    \right)
    \left(
    \begin{array}{c}
      Z_{N-1}(+,+)\\
      Z_{N-1}(+,-)
    \end{array}
    \right),\nonumber\\
    &=
        \left(
    \begin{array}{cc}
      e^\beta&e^{-\beta}\\
      e^{-\beta}&e^\beta\\
    \end{array}
    \right)^{N-2}
    \left(
    \begin{array}{c}
      Z_{2}(+,+)\\
      Z_{2}(+,-)
    \end{array}
    \right),\nonumber\\
    &=\left(
    \begin{array}{c}
      \frac{1}{2}\left[(2\cosh\beta)^{N-1}+(2\sinh\beta)^{N-1}\right]\\
      \frac{1}{2}\left[(2\cosh\beta)^{N-1}-(2\sinh\beta)^{N-1}\right]
    \end{array}
    \right),
\end{align}
where the remaining two boundary conditions, $Z_N(-,-)$ and $Z_N(-,+)$ are respectively equal to  $Z_N(+,+)$ and $Z_N(+,-)$. Thus, the partition function for PBC is 
\begin{align}
    Z_N(\beta) &= Z_{N+1}(-,-)+Z_{N+1}(+,+)\nonumber\\
    &= (2\cosh\beta)^{N}+(2\sinh\beta)^{N}
\end{align}
This result can be further elucidated by employing the concept of a \textit{transfer matrix}.

The transfer matrix, denoted as 
$\mathcal{T}$, is defined by the matrix:
\begin{equation}
\mathcal{T} = \left(
\begin{array}{cc}
e^\beta & e^{-\beta} \\
e^{-\beta} & e^\beta 
\end{array}
\right).
\end{equation}
This matrix effectively \textit{transfers} the partition function from 
$Z_N$ to $Z_{N+1}$, serving as a tool to calculate the partition function for a larger system based on the known results of a smaller system. The eigenvalues of the transfer matrix, $\mathcal{T}$, are particularly significant as they dictate the thermodynamic properties of the system. In fact, it can be demonstrated that the partition function can be succinctly expressed using $\mathcal{T}$ in the following manner:
\begin{align}
    Z_N(\beta) &= \Tr\ \mathcal{T}^N\\
    &= \sum_{n=0}^1\lambda_n^N,
\end{align}
where $\lambda_0=2\cosh\beta$ and $\lambda_1=2\sinh\beta$ are the eigenvalues of $\mathcal{T}$. This elucidates an important concept: the partition function is essentially the sum of the $N$-th powers of the eigenvalues of the transfer matrix. This concept is applicable to any spatial dimension.

\subsection{Real-space renormalization: Migdal-Kadanoff transformaiton}
\begin{figure}[tb]
    \centering
    \includegraphics[width=80mm]{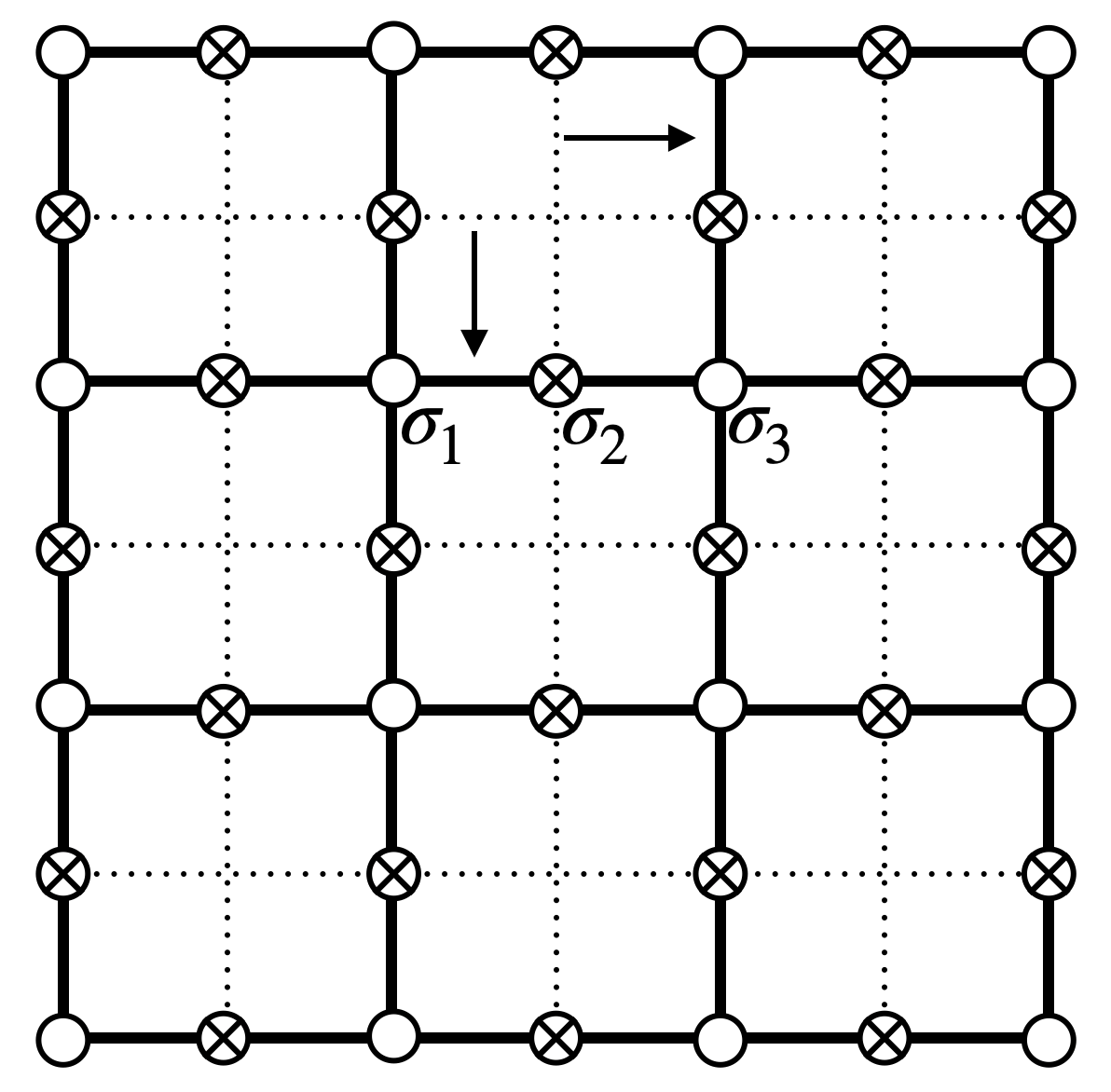}
    \caption{The sketch of Migdal-Kadanoff transformation. It aims to analytically see how the coupling $K=\beta J$ changes when the system is coarse-grained by a factor of two.}
    \label{migdal_kadanoff}
\end{figure}
The two-dimensional Ising model presents a higher level of complexity compared to its one-dimensional counterpart. To address this, Migdal and Kadanoff developed a methodology to simplify the problem~\cite{migdal1996recursion,kadanoff1976notes}. Kadanoff's approach focused on understanding how the effective coupling constant 
$K=\beta J$ evolves with changes in scale, rather than attempting to directly calculate the partition function. This is schematically illustrated in Figure \ref{migdal_kadanoff}, where the method involves tracing out the degrees of freedom at the $\otimes$ sites while retaining the spins at the $\circ$ sites. Notably, the spin at the center of each new square lattice is ignored, enabling exact treatment of the system. Kadanoff's key assumption is that during the coarse-graining of the system, the effective coupling effectively doubles. This doubling occurs as a result of bundling together two interaction bonds from the original lattice. Moreover, there are spins situated at the center of the new bonds, denoted as $\otimes$. These central spins need to be traced out to determine the new coupling constant. Then, the new coupling constant $K^{'}$is derived as follows:
\begin{align}
A\exp[K^{'} \sigma_1\sigma_3] &= \Tr_{\sigma_2}\exp[2K\sigma_1\sigma_2 + 2K\sigma_2\sigma_3] \nonumber \\
&= 2\cosh(2K(\sigma_1 + \sigma_3)) \nonumber \\
&= 2\sigma_1\sigma_3\sinh^2(2K) + 2\cosh^2(2K)
\end{align}

\begin{figure}[tb]
\centering
\includegraphics[width=80mm]{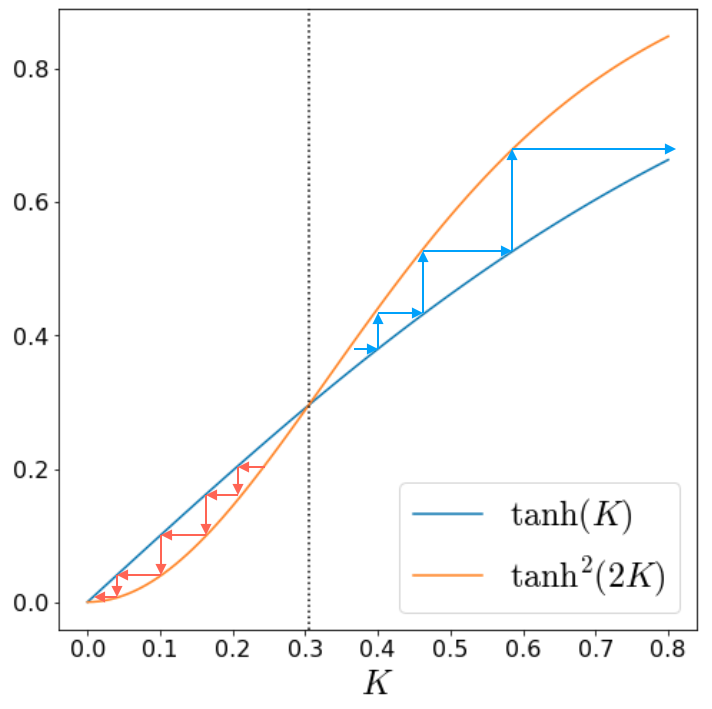}
\caption{The RG flow of the two-dimensional Ising model following Eq.~\eqref{eq:Kadanoff}.}
\label{kadanoff_rgflow}
\end{figure}

Given that $\sigma^2=1$, we can express the left-hand side of the equation as:
\begin{align}
\exp[K^{'} \sigma_1\sigma_3] = \cosh(K^{'}) + \sigma_1\sigma_3\sinh(K^{'})
\end{align}
This leads us to the relation between the old and new couplings:
\begin{align}
\tanh(K^{'}) = \tanh^2(2K)\label{eq:Kadanoff}
\end{align}
Equation \eqref{eq:Kadanoff} is fundamental to understanding how coupling constants transform under scale transformations in the two-dimensional Ising model. It indicates that with each coarse-graining of the lattice by two sites, the coupling constants evolve according to the relation $K^{'} =\text{arctanh}(\tanh^2(2K))$. This process of evolution is depicted in Fig.~\ref{kadanoff_rgflow}.

At the critical point $K_c$, marked by a black dotted line, there is no evolution of $K$ since $\tanh(K_c) = \tanh^2(2K_c)$ holds true. This point is identified as a \textit{fixed-point}, corresponding to the critical temperature. In the high-temperature regime where $K<K_c$, as shown by the red arrows, the effective coupling decreases towards zero with increasing scale. Since $K=0$ corresponds to the $J\rightarrow0$ or $T\rightarrow\infty$ limit, this indicates each spin is decoupled, being a phase of complete disorder. Conversely, starting from a low-temperature regime where $K>K_c$, $K$ increases upon coarse-graining, as indicated by the blue arrows. This behavior, aligning with the $J\rightarrow\infty$ or $T\rightarrow0$ limits, corresponds to a phase characterized by spontaneous symmetry breaking.

Figure~\ref{kadanoff_rgflow} effectively illustrates the concept of renormalization group flow (RG flow) in the context of the coupling constant $K$ in the Ising model. As depicted, when the scale increases, the couplings $K$ that start from values below the critical point $(K<K_c)$ and above it $(K>K_c)$ appear to “flow" towards $K=0$ and $K=\infty$, respectively. This dynamic behavior of the coupling constant under scaling transformations is a fundamental aspect of RG flow.

The points $K=0$, $K_c$, and $\infty$ are identified as fixed-points within this flow. These points are unique in that they remain invariant under scale transformations as formulated in Eq.~\eqref{eq:Kadanoff}. They can be conceptualized as “the terminal stations” of the scale transformation process. When reaching these points, the system has effectively discarded all irrelevant information through an infinite series of scale transformations. Drawing a parallel to the 'Bliss' wallpaper analogy, each step we take back from the image can be likened to each step of scale transformation in physical systems. As we step back, finite-sized clusters within the wallpaper appear smaller and smaller. Similarly, in physical systems undergoing scale transformations, the correlation length, denoted as $\xi$, reduces by half with each coarse-graining step. This process leads the system towards a “simpler theory”, where $\xi=0$, represents a state devoid of significant correlations at large scales.

A notable exception arises at the point of criticality, where $\xi$ reaches infinity. This infinite correlation length is a defining feature of continuous phase transitions, marking a state where correlations extend across all scales. The fixed-point associated with this criticality is of particular interest, as it embodies the key principles of universality and scale invariance fundamental to understanding these phase transitions. Despite the microscopic differences between systems like water and magnets, the macroscopic behaviors of these systems converge to the same fixed-points. This convergence to common fixed-points is what gives rise to the universal properties observed in critical phenomena across different physical systems. 

To further understand this concept, consider linearizing Eq.~\eqref{eq:Kadanoff} near the critical point $K=K_c$. This linearization yields:
\begin{align*}
(K' - K_c) \approx 1.68 (K - K_c).
\end{align*}
This relationship indicates that, after $n$-steps of scale transformation, the effective coupling constant increases exponentially as:
\begin{align*}
(K' - K_c) \approx 1.68^n (K - K_c).
\end{align*}
Extending this to a continuous scale transformation $n$, we can describe how the deviation $\delta K = K-K_c$ grows with scale:
\begin{align}
\frac{d(\delta K(n))}{dn} \approx \ln(1.68) \delta K(n). \label{eq:rg2n}
\end{align}
Since the system size scales as $L=2^n$, Eq.~\eqref{eq:rg2n} can be reformulated in terms of the logarithmic scale $l=\ln(L)$:
\begin{align}
\frac{d(\delta K(l))}{dl} &\approx \frac{\ln(1.68)}{\ln(2)}\ \delta K(l), \nonumber \\
&\approx 0.75\ \delta K(l). \label{eq:rgl}
\end{align}

This formulation represents the RG equation. Phase transitions that belong to the same universality class are characterized by the same RG equation, even though they may have different specific scale transformations as in Eq.~\eqref{eq:Kadanoff}. The coefficient $\sim0.75$ in this equation is referred to as the RG dimension, which plays a crucial role in determining the critical exponents of the system. It is noteworthy that while the exact RG dimension for the Ising universality class is one as 
\begin{align}
\frac{d(\delta K(l))}{dl} = \delta K(l) \label{eq:rgl_exact}.
\end{align}
Namely, Kadanoff's approximation yields a close value, demonstrating the effectiveness of this simplified approach. 

In the following sections, we will delve deeper into this concept, interpreting it through the lens of field theory. This approach will provide a more nuanced and comprehensive understanding of the dynamics at play in phase transitions and critical phenomena.

\subsection{Field theory describing two-dimensional fixed-point}
Universality at criticality can be exemplified by examining the behavior of the correlation function in critical systems. In the specific case of the critical two-dimensional Ising model, the spin-spin correlation function exhibits a polynomial decay characterized by:
\begin{align}
\langle\sigma(r_i)\sigma(r_j)\rangle \propto \frac{1}{|r_i - r_j|^{1/4}}\label{eq:critical},
\end{align}
where $r_i$ and $r_j$ denote the positions of spins. This equation illustrates how, at criticality, the correlation between spins decays in a manner inversely proportional to the distance raised to the power of $1/4$.

In contrast, for systems that are not at criticality, which possess a finite correlation length $\xi$, the correlation function typically exhibits an exponential decay:
\begin{align*}
\langle\sigma(r_i)\sigma(r_j)\rangle \propto e^{-|r_i - r_j|/\xi}.
\end{align*}
In this scenario, the correlation diminishes exponentially with increasing distance between spins, governed by the correlation length $\xi$.

Therefore, the correlation function in Eq.~\eqref{eq:critical} for the critical Ising model can be understood as the limit where $\xi\rightarrow\infty$. This infinite correlation length at criticality is what leads to the power-law decay of the correlation function, distinguishing it from the exponential decay observed in systems with finite correlation lengths. This behavior exemplifies the concept of universality at criticality, where different systems exhibit similar long-range correlations as they approach their critical points.

In field theory, correlation functions are expressed in terms of operators. For instance, Eq.~\eqref{eq:critical} in the field theory framework is represented as:
\begin{align}
\langle\hat{\sigma}(r_i)\hat{\sigma}(r_j)\rangle = \frac{1}{|r_i - r_j|^{1/4}}\label{eq:sigma_cft},
\end{align}
where $\hat{\sigma}$ denotes the spin operator and the brackets indicate the expectation value of inserting two such spin operators into the vacuum state. This formulation reflects how the correlation between spins decays with distance in the critical Ising model.

Another crucial operator in the critical Ising model is the energy operator, denoted as $\hat{\epsilon}$. The correlation function for this operator is given by:
\begin{align}
\langle\hat{\epsilon}(r_i)\hat{\epsilon}(r_j)\rangle = \frac{1}{|r_i - r_j|^{2}}\label{eq:epsilon_cft}.
\end{align}
This equation demonstrates that the correlation of the energy operator also decays with distance, but at a different rate compared to the spin operator.

On the lattice, the correlation function corresponding to the energy operator takes the form:
\begin{align}
\langle(\sigma(r_i)\sigma(r'_i))(\sigma(r_j)\sigma(r'_j))\rangle \propto \frac{1}{|r_i - r_j|^{2}}\label{eq:critical_epsilon},
\end{align}
where $r'_i$ is a neighboring site of $r_i$. The product $\sigma(r_i)\sigma(r'_i)$ represents the local energy in the Ising model, which explains why $\hat{\epsilon}$ is referred to as the energy operator. In addition to the above operators, there is also the identity operator denoted as $I$ that represents “inserting nothing."

The exact values of the exponents in Eqs.~\eqref{eq:sigma_cft} and \eqref{eq:epsilon_cft}, specifically $1/4$ and $2$, naturally lead to speculation about an underlying theoretical framework that explains these precise figures. Indeed, for two-dimensional critical systems, there is a profound theoretical basis behind these exponents. The concept of scale invariance, which is inherent to fixed points, extends to a broader principle known as conformal invariance in two dimensions. Conformal invariance imposes strict constraints on operators and their correlation functions, often enabling the precise determination of critical exponents.

The field theory that incorporates this invariance is known as conformal field theory (CFT). The key to CFT's effectiveness lies in its exploitation of conformal invariance, which significantly enhances the symmetry of the system and, consequently, its analytical traceability.

In CFT, each universality class corresponds to a specific CFT. For example, the Ising universality class is described by the Ising CFT, which includes three primary operators~\footnote{Primary operators are a specific class of operators that play an important role in elementary excitations. While we will only discuss this class of operators in this section, there is another class named descendants that govern higher excited states.}: $I$, $\sigma$, and 
$\epsilon$ (hereafter, we will refer to operators without the hat notation). More generally, primary operators in CFT are denoted as $\Phi_i$. 
These operators exhibit the following characteristics:
The two-point correlation function is defined as
\begin{align}
\langle\Phi_i(r_i)\Phi_j(r_j)\rangle = \frac{\delta_{i,j}}{|r_i-r_j|^{2x_i}},\label{eq:two_point}
\end{align}
where $x_i$ represents the scaling dimension of the operator $\Phi_i$. In the case of the Ising CFT, the scaling dimensions are $x_I=0$, $x_\sigma=1/8$, and $x_\epsilon=1$.
Similarly, the three-point correlation function is expressed in a universal form:
\begin{align}
\langle\Phi_i(r_i)\Phi_j(r_j)\Phi_k(r_k)\rangle = \frac{C_{ijk}}{|r_i-r_j|^{\Delta_{ij}^k}|r_j-r_k|^{\Delta_{jk}^i}|r_k-r_i|^{\Delta_{ki}^j}},\label{eq:three_point}
\end{align}
where $C_{ijk}$ is an operator product expansion(OPE) coefficient, and  $\Delta_{ij}^k = x_i+x_j-x_k$. Notable OPE coefficients for the Ising CFT include:
\begin{align}
    C_{III}&=C_{I\sigma\sigma} = C_{I\epsilon\epsilon} =1,\\
    C_{\sigma\sigma\epsilon}&=\frac{1}{2}.
\end{align}
It is important to note that the permutation of indices in these coefficients does not alter their values, preserving the $\mathbb{Z}_2$ symmetry~\footnote{For a detailed discussion on CFT, readers are referred to Ref.~\cite{francesco2012conformal}. In this chapter, we aim to give a practical introduction to CFT.}.

Emphasizing the physical significance of OPE coefficients in RG theory is crucial, as these coefficients play a pivotal role in understanding how local operators interact and combine, or “fuse," within the field theory framework. To grasp this concept, consider an analogy involving a canvas with blue and red dots placed close to each other. When viewed from a close distance, these dots are seen as distinct entities. However, as one steps back, the dots may appear to merge into a single purple dot. This phenomenon of blending or “fusion" of the dots mirrors how operators in field theory can be combined.

In field theory, the fusion of operators is mathematically represented by the mixing of two local operators situated in close proximity. Let $\Phi_i$ and $\Phi_j$ be two such operators. The fusion of these operators can be expanded in terms of the local operator basis, as illustrated below:
\begin{align}
\Phi_i(r_i)\Phi_j(r_j) \approx \sum_k \frac{C_{ijk}}{|r_i-r_j|^{x_i + x_j - x_k}} \Phi_k\left(\frac{r_i + r_j}{2}\right).
\end{align}
In this equation, the exponents with respect to $|r_i-r_j|$ are chosen to ensure consistency with the two-point correlation function, as described in Eq.~\eqref{eq:two_point}. This expansion signifies how two local operators when in close proximity, can effectively combine to form a different operator, 
$\Phi_k$, with the OPE coefficients $C_{ijk}$ dictating the nature and strength of this fusion. 

In the Ising CFT, the fusion of two spin operators $\sigma$ results in the formation of the energy operator $\epsilon$. This concept aligns intuitively with the corresponding lattice model of the Ising system. Recall that in the lattice model, the individual spin operator $\sigma_i$ and the product of adjacent spin operators $\sigma_i\sigma_{i+1}$ correspond to the CFT operators $\sigma$ and $\epsilon$, respectively. Consequently, when two spin operators positioned close to each other on the lattice are multiplied, the resulting interaction closely resembles $\sigma_i\sigma_{i+1}$, which is the lattice analog of the energy operator $\epsilon$. This interaction is mirrored in the CFT framework, as evidenced by the non-zero OPE coefficient $C_{\sigma\sigma\epsilon}$, indicating a significant fusion between two $\sigma$ operators into $\epsilon$. 

Additionally, in CFT, the identity operator $I$ represents the concept of inserting no operator into the system. As such, fusing any operator with $I$ does not result in any change to the original operator. This is reflected in the OPE coefficients, where fusing an operator with the identity operator maintains the operator unchanged, leading to $C_{iiI}=1$. This property highlights the foundational role of the identity operator in maintaining the integrity of the system's operators during the fusion process.

\begin{figure}
    \centering
    \includegraphics[width=80mm]{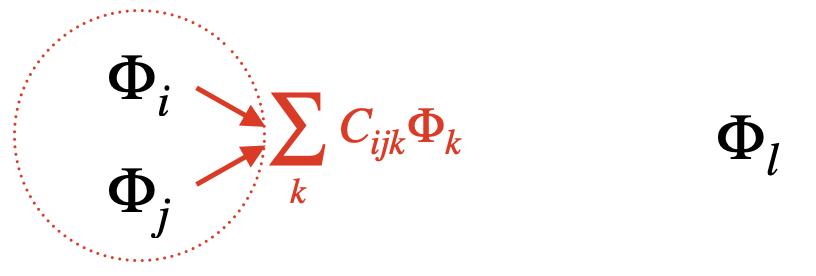}
    \caption{A schematic figure of fusing of operators. }
    \label{OPE_figure}
\end{figure}

Building upon our understanding of OPE coefficients and their role in operator fusion, we can discern the rationale behind their appearance in the three-point correlation function, as shown in Eq.~\eqref{eq:three_point}. Consider the situation where the points $r_i$ and $r_j$ are in close proximity to each other. Under this condition, Eq.~\eqref{eq:three_point} can be approximated as:
\begin{align}
\langle\Phi_i(r_i)\Phi_j(r_j)\Phi_l(r_l)\rangle \approx \frac{C_{ijl}}{|r_i-r_j|^{x_i + x_j - x_l}} \frac{1}{|r_i-r_l|^{2x_l}}.
\end{align}
This approximation effectively combines the principles outlined in Eqs.~\eqref{eq:two_point} and \eqref{eq:three_point}. Specifically, it demonstrates that a three-point correlation function can be interpreted as a two-point function following the fusion of the first two operators, $\Phi_i$ and $\Phi_j$.

This fusion process results in a single operator, which then interacts with the third operator, $\Phi_l$. The corresponding correlation function thus encapsulates this interaction, with the OPE coefficient $C_{ijk}$ playing a crucial role in quantifying the strength and nature of the fusion between $\Phi_i$ and $\Phi_j$. This concept is graphically represented in Fig.~\ref{OPE_figure}, where the fusion of the first two operators before interacting with the third is illustrated. Through this lens, the three-point function can be understood as a manifestation of the underlying fusion dynamics among the operators in the field theory framework.

In CFT, conformal mapping plays a crucial role, especially in two-dimensional contexts. In such systems, the physical $x$-$y$ plane is effectively represented using a complex plane with coordinates $z = x + iy$ and its complex conjugate $\bar{z} = x - iy$. In this framework, primary operators transform when we perform a conformal mapping $w=f(z)$ as:

\begin{align}
\tilde{\Phi}_i(w,\bar{w}) = \left(\frac{\partial w}{\partial z}\right)^{-h_i}\left(\frac{\partial \bar{w}}{\partial \bar{z}}\right)^{-\bar{h}_i}\Phi_i(z,\bar{z}),
\end{align}
In this formula, $h_i$ and $\bar{h}_i$ are known as the conformal weights of the operator $\Phi_i$, determining its scaling dimension $x_i = h_i + \bar{h}_i$ and conformal spin $s_i = h_i - \bar{h}_i$.
Consider a scale transformation described by $w = b^{-1}z$~\footnote{$b$ is the factor of scale transformation, where Kadanoff's RG corresponds to $b=2$. This is because $2cm$ in $z$-plane becomes $1cm$ in $w$-coordinate. }. Under this transformation, the primary operator $\Phi_i$ behaves as:
\begin{align}
\tilde{\Phi}_i(w,\bar{w}) = b^{x_i}\Phi_i(z,\bar{z}).\label{eq:conformal_mapping_primary_operator}
\end{align}
This relationship illustrates how the operator scales with the transformation factor \( b \). Setting $b = |z|$, we can derive the power-law decay of the two-point correlation function, as observed in Eq.~\eqref{eq:two_point}:
\begin{align}
\langle\Phi_i(z,\bar{z})\Phi_i(0)\rangle = |z|^{-2x_i}\langle\Phi_i(1)\Phi_i(0)\rangle.
\end{align}
This outcome highlights how the decay rate of the two-point function is directly linked to the scaling dimension of the operator, demonstrating the profound influence of conformal mapping in CFT. Such insights are pivotal for understanding correlation functions in critical phenomena and the symmetry principles that underpin them.

The collection of scaling dimensions $x_i$ and OPE coefficients $C_{ijk}$, collectively referred to as CFT data, is crucial for a comprehensive understanding of critical phenomena. Therefore, the determination of this CFT data, particularly from a numerical perspective, is a fundamental objective in the study of critical systems.

\subsection{Renormalization group and CFT}
The concept of CFT is intrinsically linked to the RG theory. Specifically, CFT provides a framework to calculate how deviations from critical values, such as $\delta K(l)$ in Eq.~\eqref{eq:rgl_exact}, evolve through scale transformations. When considering Kadanoff's real-space RG approach, one might question why the analysis is focused solely on the coupling constant $K$. In an exact coarse-graining of the model, other coupling constants, like the next nearest-neighbor coupling, could emerge. However, Eq.~\eqref{eq:rgl} still qualitatively represents the correct RG flow for Ising criticality. So, why is it valid to concentrate only on $K$? Field theory answers this by identifying $K$ as a relevant parameter, whereas others are not. To understand this in detail, Wilson first introduced the concept of considering \textit{all} possible perturbations that might arise from the microscopic details of the Hamiltonian~\cite{RevModPhys.55.583,Zamolodchikov:1987ti}. Deviations such as $\delta K$ from the scale-invariant action $S^{*}$ are expressed by $g_j$ with corresponding operators $\Phi_j$. Hence, the action can be formulated as:
\begin{align}
S = S^{*} + \int d^d r \sum_j g_j \Phi_j,
\end{align}
where $\Phi_j$ are normalized operators, and the sum of $j$ goes through all possible perturbations. We are interested in how only a few of these become relevant during the scale transformation. To do this, we expand the Euclidean action around the fixed point as follows:
\begin{align*}
\Tr e^{-S} = \Tr e^{-S^{*}}\left\langle1-\sum_i\int d^d r g_i \Phi_i+\frac{1}{2}\sum_{i,j}\int d^d r_id^d r_j g_ig_j \Phi_i\Phi_j+\cdots\right\rangle_{S^{*}}
\end{align*}
Under scale transformations $r\rightarrow br$, the second and third terms transform as:
\begin{align*}
\sum_i\int d^d r g_i \Phi_i&\rightarrow \sum_i\int d^d r b^{d-x_i}g_i\Phi_i,\\
\frac{1}{2}\sum_{i,j}\int d^d r_id^d r_j g_ig_j \Phi_i\Phi_j
&\rightarrow \frac{S_d}{2}(b-1)\sum_{i,j}\int d^d r_i g_ig_j C_{ijk}\Phi_k,
\end{align*}
where $S_d$ is the surface area of a $(d-1)$-dimensional sphere, with $S_2=2\pi$ in two dimensions.
In the analysis of the third term, we integrate over the region $a < |r_i - r_j| < ba$, thereby accounting for the short-range physics. Here, $a$ represents the lattice spacing at the current scale. We define $g_k$ such that $a$ is normalized to unity. Upon applying an infinitesimal scale transformation $b\approx1$, the effective coupling $g_k$ evolves to $g_k(b)$, given by 
$$
g_k(b) = b^{d-x_k}g_k - \frac{S_d}{2}(b-1)\sum_{i,j} C_{ijk}g_ig_j,
$$
Consequently, the RG equation up to 1-loop expansion becomes:
\begin{align}
\frac{d g_k}{dl} &= b\frac{d g_k}{db}\nonumber\\
&=(d-x_k) g_k - \frac{S_d}{2} \sum_{i,j} C_{ijk} g_i g_j,
\label{eq.RGeq}
\end{align}
 To this order, the RG equation (beta function) is universally determined by the scaling dimension $x_k$ and the OPE coefficients $C_{ijk}$. 

Revisiting the justification for concentrating solely on $\delta K$ in Eq.~\eqref{eq:rgl_exact}, it is essential to understand the dynamics of operators near criticality, as defined by the fixed-point CFT. The Ising CFT, although potentially allowing for infinite kinds of perturbations, has a specific criterion for which perturbations are significant during scale transformations. According to Eq.~\eqref{eq.RGeq}, the running coupling constants scale as
$$g_k(L)\propto L^{d-x_k},$$
and thereby only operators with scaling dimensions $x_k < d$ remain influential through these transformations. Such operators are termed “relevant operators.”

The rationale behind focusing on a limited number of parameters in RG analysis is rooted in the fact that typically, only a few relevant operators exist in the fixed-point CFT. In the case of the Ising CFT, for instance, the operator $\epsilon$ is the primary relevant operator that persists~\footnote{$\epsilon$ and $\sigma$ respectively corresponds to the shift of temperature and applying a uniform magnetic field. Under the $\mathbb{Z}_2$ spin flip symmetry, only $\epsilon$ is allowed. }, which corroborates the validity of Kadanoff's approach. This focus on a few relevant parameters simplifies the RG analysis while still capturing the critical behavior of the system.

Conversely, operators with scaling dimensions $x_k > d$, known as “irrelevant operators," tend to diminish and become negligible throughout scale transformations. These operators do not significantly influence the macroscopic properties of the system and therefore do not need to be considered in the RG analysis. This selective approach, focusing only on relevant operators like $\epsilon$ in the Ising CFT, is what makes critical phenomena universal across diverse systems.

To summarize this section, the universality observed in critical phenomena can be traced back to the shared RG fixed points among different systems. These RG fixed points, conceptualized as 'terminal stations' in the scale transformation process, are characterized by a limited set of parameters corresponding to relevant operators. In contrast, other microscopic details, which are associated with irrelevant operators, diminish and lose significance through the coarse-graining process.

This paradigm highlights the fundamental principle that, at the macroscopic level, the critical behavior of a system is governed not by the myriad of its microscopic details, but by a select few relevant parameters. These parameters, represented by the relevant operators at the RG fixed point, dictate the universal aspects of the system's critical behavior. As a result, systems with different microscopic structures can exhibit the same macroscopic critical phenomena, provided they converge to the same RG fixed point. This convergence is what underlies the universality of critical phenomena, emphasizing the profound impact of scale and relevant operators in determining the nature of phase transitions and critical behavior.

Moving to the next section, our focus shifts to the methodologies for extracting universal information from critical lattice models. A particularly promising approach is the tensor network formalism. This method can be seen as a generalization of the transfer matrix formalism, which we previously discussed in the context of the one-dimensional Ising model. When combined with recent advances in computational techniques, tensor network formalism enables a more sophisticated form of real-space RG analysis.

This enhanced RG approach transcends Kadanoff's methodology by its ability to retain multiple parameters throughout the scale transformation process. This characteristic aligns more closely with Wilson's original vision for RG theory, providing a more comprehensive and faithful representation of scale transformations in physical systems.

In the upcoming section, we will review the fundamental concepts underlying tensor network-based RG. This will set the stage for the main chapter, where we propose a novel approach to calculate the RG flow using tensor networks. This approach not only leverages the strengths of tensor network formalism but also addresses some of the limitations of previous methodologies, offering a more robust and accurate tool for analyzing critical behavior in lattice models.

\section{Review on tensor network renormalization}
\subsection{What is a tensor?}
Before delving into the intricacies of our study, it is essential to establish a fundamental understanding of what a tensor is. A tensor can be seen as a generalized form of vectors and matrices. To illustrate, consider the following examples, which are referred to as one-leg and two-leg tensors, respectively:
\begin{align*}
V&=\begin{pmatrix}
1 \\
2 \\
3
\end{pmatrix},\\
M&=\begin{pmatrix}
a_{00} & a_{01} \\
a_{10} & a_{11}.
\end{pmatrix}
\end{align*}
For clarity in this thesis, these tensors are denoted as $V_i$ and $M_{ij}$, indicating their respective number of legs. The indices run as $i=0,1,2$ for $V_i$ and $(i,j) = (0,0)$, $(0,1)$, $(1,0)$, and $(1,1)$ for $M_{ij}$. This notation allows us to interpret the values as $V_i = i+1$ and $M_{ij} = a_{ij}$. The dimension associated with each index is known as the bond dimension $D$. Accordingly, we denote $D=3$ for $V_i$ and $D=2$ for $M_{ij}$.

Another example is the anti-symmetric tensor $\epsilon_{ijk}$, defined as:
\begin{align*}
\epsilon_{ijk}=\begin{cases}
1 & \text{for }(i,j,k) = (0,1,2),(1,2,0),(2,0,1), \\
-1 & \text{for }(i,j,k) = (2,1,0),(1,0,2),(0,2,1),\\
0 & \text{otherwise}.
\end{cases}
\end{align*}
This tensor represents a 3-leg tensor with a bond dimension of $D=3$. The concept of tensors, as previously introduced, extends to structures with any number of legs, known as $n$-leg tensors. These multi-dimensional tensors play a pivotal role in the discussions that follow in this thesis. To aid in the comprehension of tensors, especially in more complex scenarios, we often employ a graphical representation. In this visual depiction, tensors are illustrated as circles with a corresponding number of legs emanating from them.

For instance, the tensors $V$, $M$, and $\epsilon$, previously defined, would be represented graphically as follows:
\begin{align*}
    \includegraphics[width=60mm]{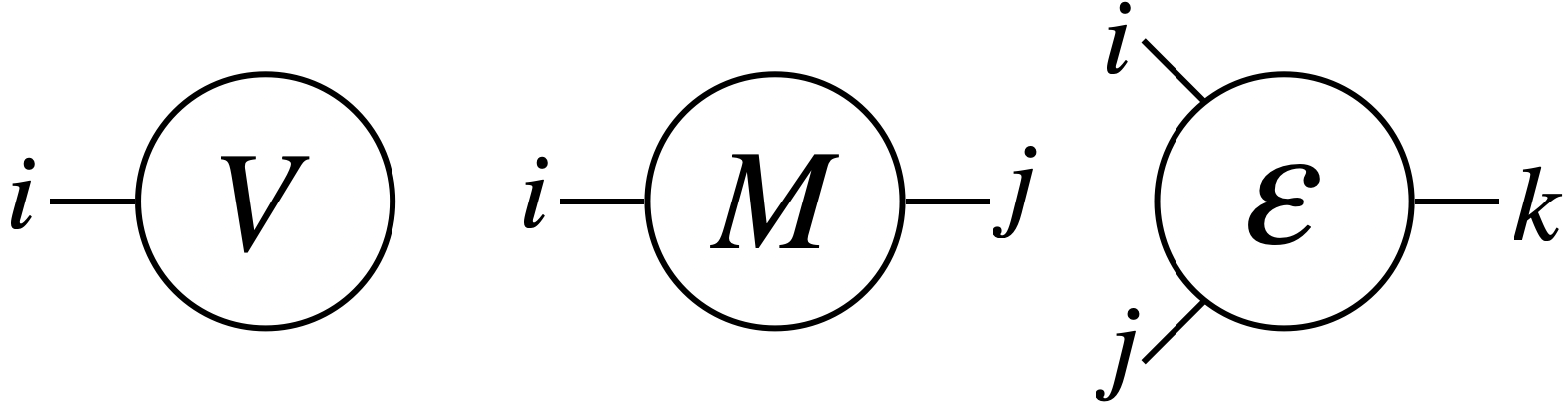}.
\end{align*}
Similarly, the multiplication of tensors is represented as a contraction of legs.
\begin{align*}
    \includegraphics[width=30mm]{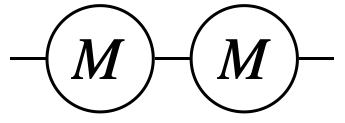}.
\end{align*}
These graphical representations provide an intuitive way to visualize the connections between tensors, particularly when dealing with complex tensor networks or operations involving multiple tensors. By utilizing these diagrams, we can more easily conceptualize the multi-dimensional relationships and transformations that tensors undergo in our analyses.
\subsection{Tensor network renormalization}
\begin{figure}[tb]
    \centering
    \includegraphics[width=86mm]{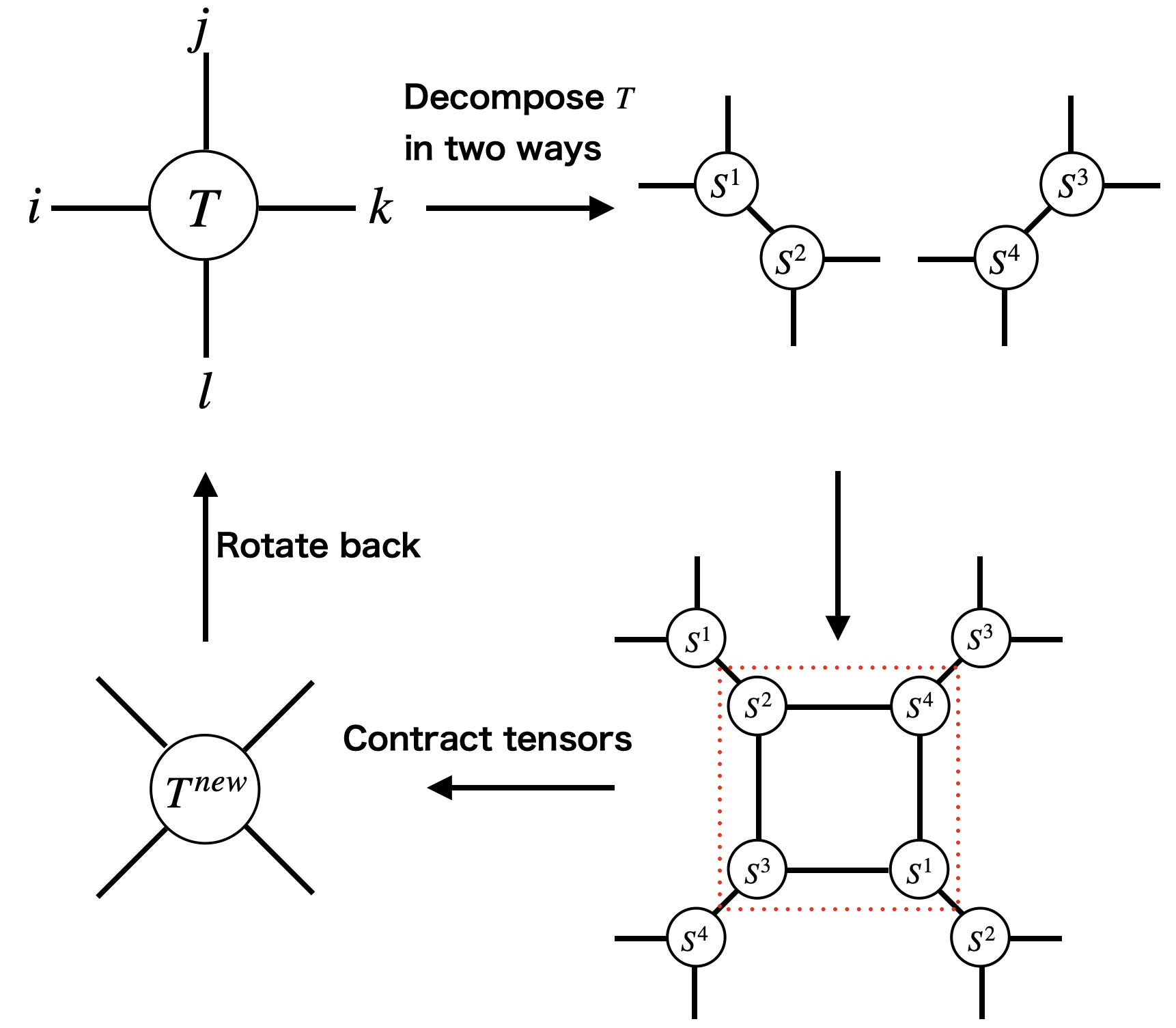}
    \caption{A schematic picture of the tensor network renormalization. The effective local Boltzmann weight at $n$-th RG step $T^{(n)}$ is decomposed into the two three-leg tensors and recombined as $T^{(n+1)}$. The effective system size enlarges by $\sqrt{2}$ each RG step. }
    \label{tnr}
\end{figure}

The tensor network is a numerical technique used to represent the partition function of statistical models. The partition functions of two-dimensional statistical models with a system size of $L$ can be expressed through the contraction of $L^2$ tensors~\footnote{Tensors can be understood as a generalization of vectors and matrices, extending into higher dimensions and complexities. A tensor characterized by $n$ indices is referred to as an $n$-leg tensor. In this framework, vectors, and matrices are special cases of tensors: a vector is a 1-leg tensor, possessing a single index, while a matrix is a 2-leg tensor, defined by two indices.}. Each tensor represents a local Boltzmann weight, and its dimensions correspond to physical degrees of freedom. For instance, the local tensor of the Ising model on the square lattice is a $D=2$ four-leg tensor $T^{(1)}_{ijkl}=e^{\beta(s_is_j+s_js_k+s_ks_l+s_ls_i)}$, where $s_\alpha = 2\alpha -1$. The tensor network representation often provides an efficient method for simulating complex systems.\\
However, the exact contraction of $L^2$ tensors is generally impracticable for larger system sizes due to the constraints imposed by the high-dimensional Hilbert space~\footnote{The one-dimensional Ising model was the simplest case, where we could perform exact contractions}. TNR aims to circumvent this issue by utilizing the principles of renormalization group theory. During each step of the RG process, $T^{(n)}$ is coarse-grained to $T^{(n+1)}$ via a series of decompositions and recombinations, as illustrated in Fig.~\ref{tnr}.
Starting from the local tensor $T^{(1)}$, we can simulate a system size of $L=\sqrt{2}^n$ after $n$ RG steps. The coarse-graining process in TNR involves numerical truncation, reducing the number of degrees of freedom while preserving essential physics. Consequently, TNR facilitates efficient numerical simulation of complex systems.

\subsection{Tensor Renormalization Group}
Levin and Nave were the pioneers in applying the technique of singular value decomposition (SVD) to the decomposition of tensor networks~\cite{PhysRevLett.99.120601}.
SVD offers a straightforward yet effective method for the tensor decomposition of a four-leg tensor $T_{ijkl}$. 
\begin{center}
    \includegraphics[width=50mm]{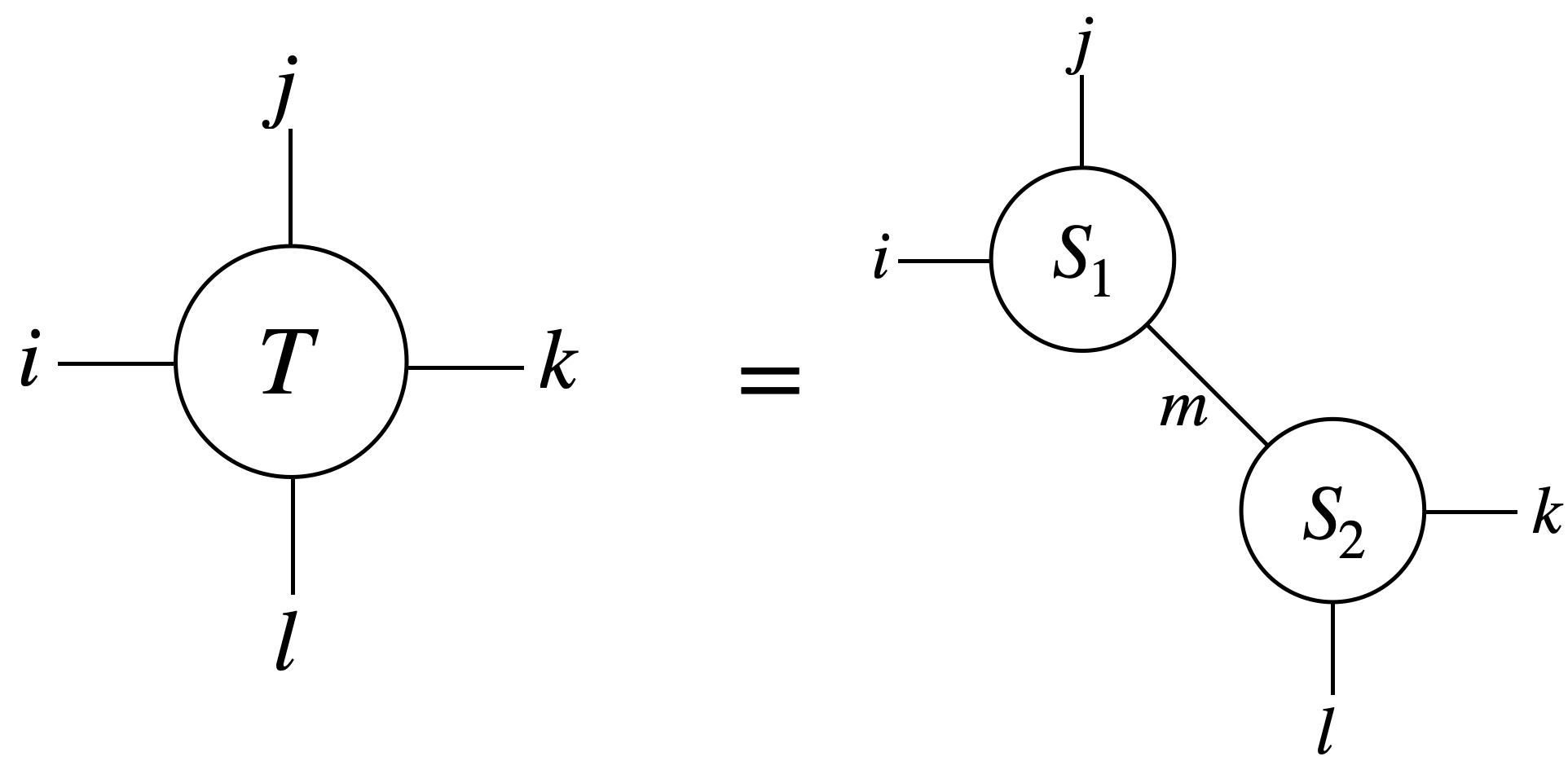}
\end{center}
The decomposition of $T_{ijkl}$ using SVD can be represented as follows:
\begin{align}
    T_{ijkl} = \sum_{m,n=1}^{d^2} U_{ijm}\Sigma_{mn}V^{\dagger}_{nkl},\label{svd}
\end{align}
where $U_{ijm}$ and $V_{nkl}$ are unitary matrices, and $\Sigma_{mn} = \delta_{m,n}s_m$ is a diagonal matrix. The diagonal elements of $\Sigma$ are the singular values that is non-negative real numbers, and SVD effectively generalizes the concept of matrix diagonalization to rectangular matrices.
In the context of tensor networks, when the bond dimension of 
$T_{ijkl}$ is $d$, the summation in Eq.~\eqref{svd} runs over $d^2$ terms. This new index $m$ represents a new bond dimension, which necessitates a truncation process to prevent the bond dimension from becoming excessively large during successive coarse-graining steps. Truncation is thus a crucial aspect of maintaining computational efficiency and feasibility in TRG.

Now, we want to truncate the index $m$ to minimize the Hilbert-Schmidt norm of the difference between the original and truncated tensors as follows:
\begin{align}
    \includegraphics[width=50mm]{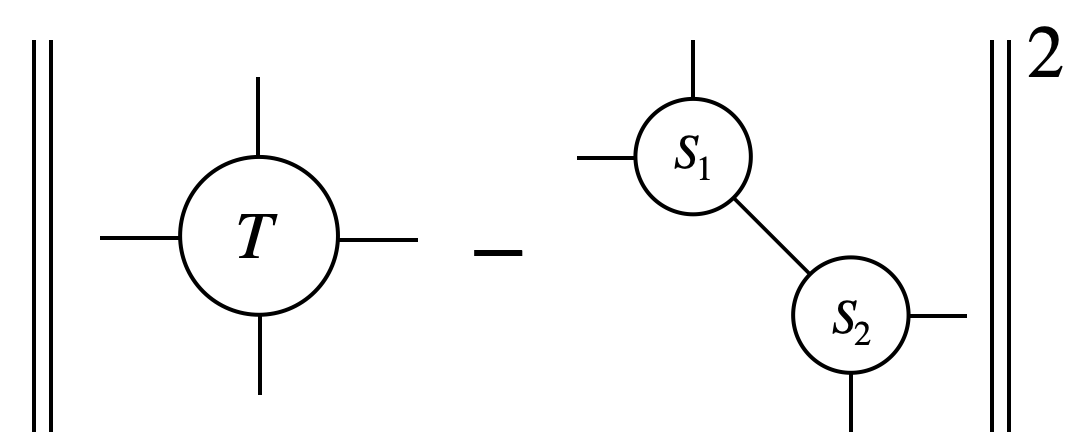}\label{eq:svd_norm}
\end{align}
We truncate the singular values when $d^2$ is larger than the desired bond dimension $D$. Let $\Sigma_{mn}^D$ be the truncated matrix that keeps only $D$ leading singular values. Then, Eq.~\eqref{eq:svd_norm} is
\begin{align}
     &\includegraphics[width=50mm]{figures/svd_norm.png} \\
     &=\Tr \left[U^{\dag}U(\Sigma^2+(\Sigma^D)^2-2\Sigma\Sigma^D)V^{\dag}V\right]\nonumber\\
     &=\sum_{m=1}^{d^2} s_m^2 + \sum_{m=1}^{D} s_m^2 - 2\sum_{m=1}^{D} s_m^2\nonumber\\
     &= \sum_{m=D+1}^{d^2} s_m^2\label{eq:svd_error}
\end{align}
In statistical mechanics systems, the singular values $s_m$ typically exhibit an exponential decay as a function of $m$. This characteristic decay pattern implies that, as $m$ increases, the singular values become progressively smaller. Consequently, when evaluating the sum in Eq.~\eqref{eq:svd_error}, this exponential decay of 
$s_m$ ensures that the cumulative sum remains extremely small, particularly when a sufficiently large bond dimension $D$ is considered. Therefore, for practical computations, a large 
$D$ effectively captures the significant contributions of the sum, while the contributions from higher values of 
$m$ become negligibly small due to this exponential decay.
\begin{figure}[tb]
    \centering
    \includegraphics[width=80mm]{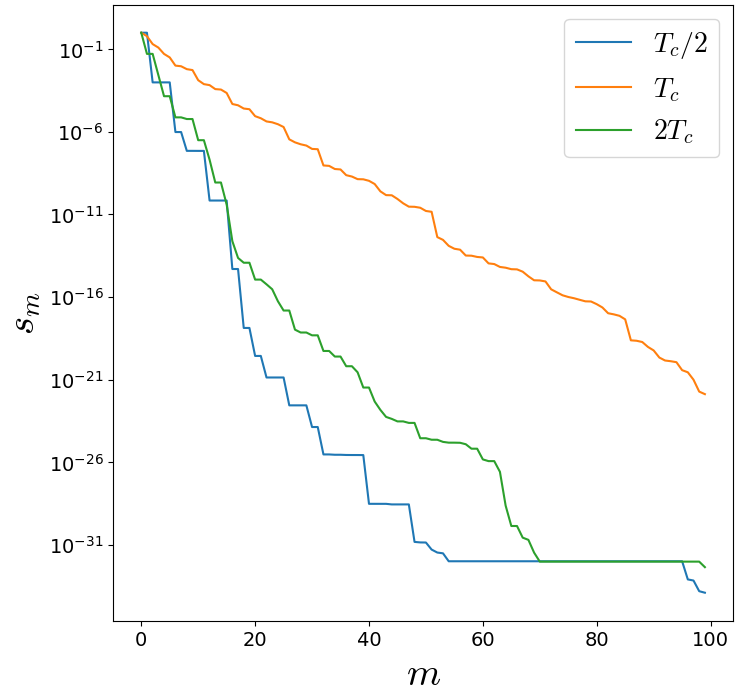}
    \caption{An example of singular values of the four-leg tensor $T_{ijkl}$. The blue, orange, and green lines show the decay of $s_m$ for the classical Ising model at the low, critical, and high-temperature regimes, respectively at $d=D=10$ after six RG steps. }
    \label{svd_decay}
\end{figure}
Figure~\ref{svd_decay} presents a typical example of singular value distributions in TRG for the two-dimensional classical Ising model. The blue and green lines represent the singular values $s_m$ for the low and high-temperature phases, respectively. In this case, the bond dimension is set to ten, resulting in a total of one hundred singular values. However, it is observed that for $m>10$, the values of $s_m$ drop below 
$10^{-6}$, affirming the efficacy and validity of this truncation method in these phases.

Conversely, the critical phase, depicted by the orange line, exhibits a slower decay in 
$s_m$. At the same bond dimension of $D=10$, the singular values are still around 
$10^{-2}$, highlighting a marked difference from the non-critical phases. This slower decay at criticality underscores a crucial challenge in TRG calculations: the difficulty in effectively capturing critical phenomena. Truncation of tensors at criticality can lead to systems with finite correlations, potentially introducing significant errors.

Given the importance of accurately estimating errors, especially in the context of critical systems, we will delve into a more detailed discussion on this topic in a later section. This analysis is crucial for both understanding the limitations of TRG at criticality and for developing strategies to mitigate error propagation in practical calculations.

Given the validity of the SVD truncations in statistical mechanics, Levin and Nave proposed the following algorithm:
 \begin{algorithm}[H]
 \caption{Algorithm for TRG}
 \begin{algorithmic}
 \renewcommand{\algorithmicrequire}{\textbf{Input:}}
 \renewcommand{\algorithmicensure}{\textbf{Output:}}
 \REQUIRE Initial tensor $T^{(0)}$ representing local Boltzmann weights
 \ENSURE Renormalized tensor $T^{(n)}$
  \STATE TRG 1-step:
First, decompose the four-leg tensor $T$ with bond dimension $d$ as follows:
  \begin{align*}
    T_{ijkl} = \sum_{m=1}^{d^2} U_{ijm}\Sigma_{mn}V^{\dagger}_{nkl},\\
    S^1_{ijn} = \sum_{m=1}^{min(d^2,D)} U_{ijm}\sqrt{\Sigma_{mn}}\\
    S^2_{nkl} = \sum_{m=1}^{min(d^2,D)} \sqrt{\Sigma_{mn}}V^{\dagger}_{nkl}\\
    \sqrt{\Sigma_{mn}} = \delta_{m,n}\sqrt{s_{m}}
\end{align*}
  Repeat the same procedure to make $S^3$ and $S^4$ by doing SVD with a pair of indices $\{jk\}$ and $\{li\}$. Then, contract $S^1\sim S^4$ inside the red dotted square and rotate by $45$ degrees as below:
  \begin{center}
  \includegraphics[width = 100mm]{figures/trg_algorithm.png}
  \end{center}
  \FOR {$i = 1$ to $n$}
  \STATE Repeat TRG 1-step
  \ENDFOR
 \RETURN $T^{(n)}$ 
 \label{trg_algorithm}
 \end{algorithmic} 
 \end{algorithm}
The practical implementation in Python is presented in my GitHub repository (\url{https://github.com/dartsushi/Loop-TNR_RGflow/tree/main/TRG_tutorial}). 
Applications to a honeycomb lattice are also discussed in the original paper~\cite{PhysRevLett.99.120601}. 

\subsection{Limitation of TRG}
Despite the successes of TRG in various applications, it is not without its limitations. A significant shortcoming of TRG, as noted in references~\cite{PhysRevLett.99.120601,PhysRevB.80.155131}, is its inability to accurately reproduce the fixed-points of disorder phases. Levin attributed this limitation to the inherent nature of SVD-based TRG, which tends to produce unphysical fixed-point tensors, referred to as corner double line (CDL) tensors. The structure of CDL tensors is illustrated as follows:
\begin{align}
\includegraphics[width=60mm]{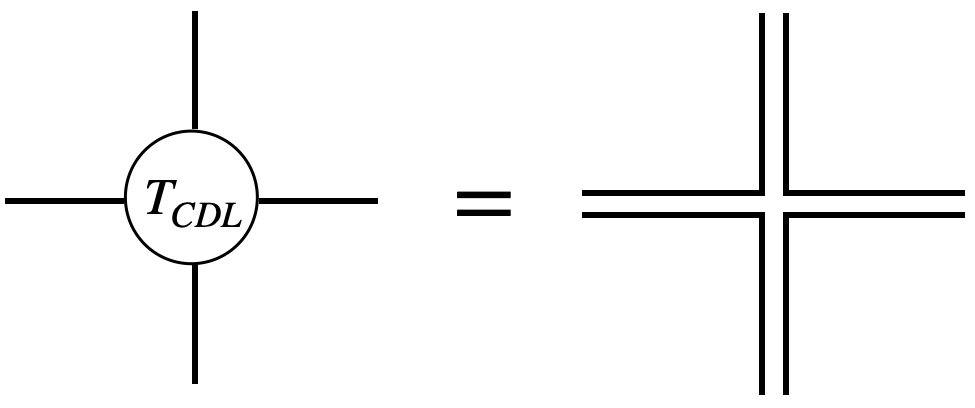}\label{CDL_figure}
\end{align}
\begin{figure}[tb]
\centering
\includegraphics[width=100mm]{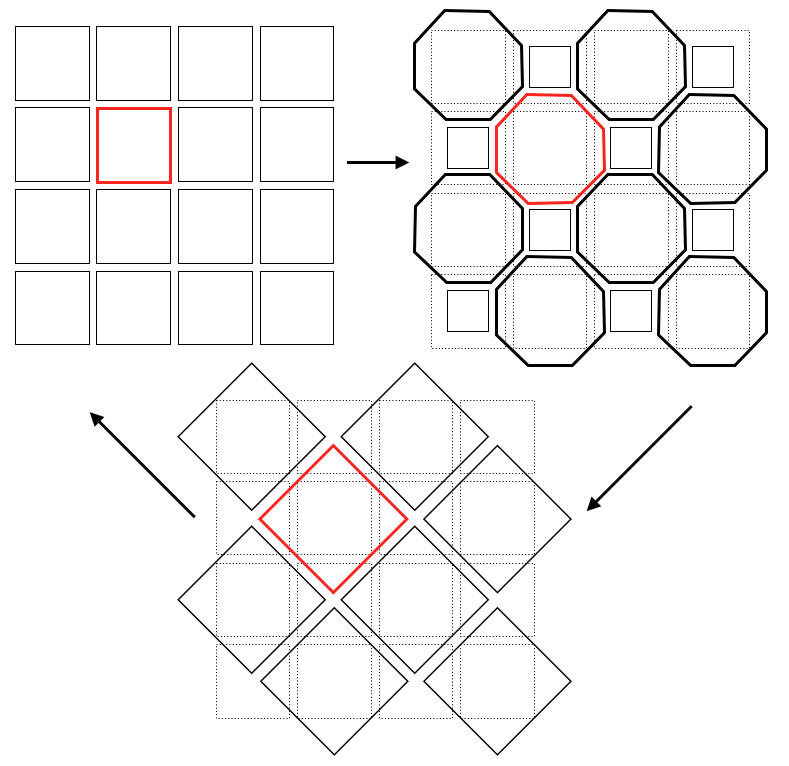}
\caption{The CDL tensors after TRG. The local loop, marked by a red square, persists after the TRG steps, indicating that ultraviolet information remains even after extensive coarse-graining.}
\label{CDL_tensors}
\end{figure}

In SVD-type TRG, there is a notable challenge in eliminating \textit{local} entanglement, as depicted by the red square in Fig.~\ref{CDL_tensors}. This figure demonstrates that ultraviolet information persists throughout successive coarse-graining steps. Consequently, these degrees of freedom consume valuable bond dimensions, leading to suboptimal approximations after multiple RG steps.

This inherent limitation of TRG underscores the need for more sophisticated coarse-graining algorithms that transcend the local tensor approximation presented in Eq.~\eqref{eq:svd_norm}. Such advanced algorithms are encompassed under the umbrella of TNR, which aims to address these specific challenges and improve the accuracy of coarse-graining in complex systems.

\subsection{Tensor network renormalization}
The field has witnessed the development of numerous algorithms aimed at overcoming the CDL problem, a known limitation in TRG applications~\cite{PhysRevLett.115.180405,PhysRevB.95.045117,PhysRevLett.118.110504,PhysRevLett.118.250602,PhysRevB.97.045111,homma2023nuclear}. While these algorithms vary in their technical details, they generally embrace two core principles: employing a larger unit-cell for optimization processes and effectively filtering out local entanglement.
Let us focus on Loop-TNR~\cite{PhysRevLett.118.110504}, known to be one of the best ones for two-dimensional TNR.

\underline{\textit{Optimization with a larger unit-cell}}

Loop-TNR aims to perform the approximation of the following:
\begin{align}
    \includegraphics[width=80mm]{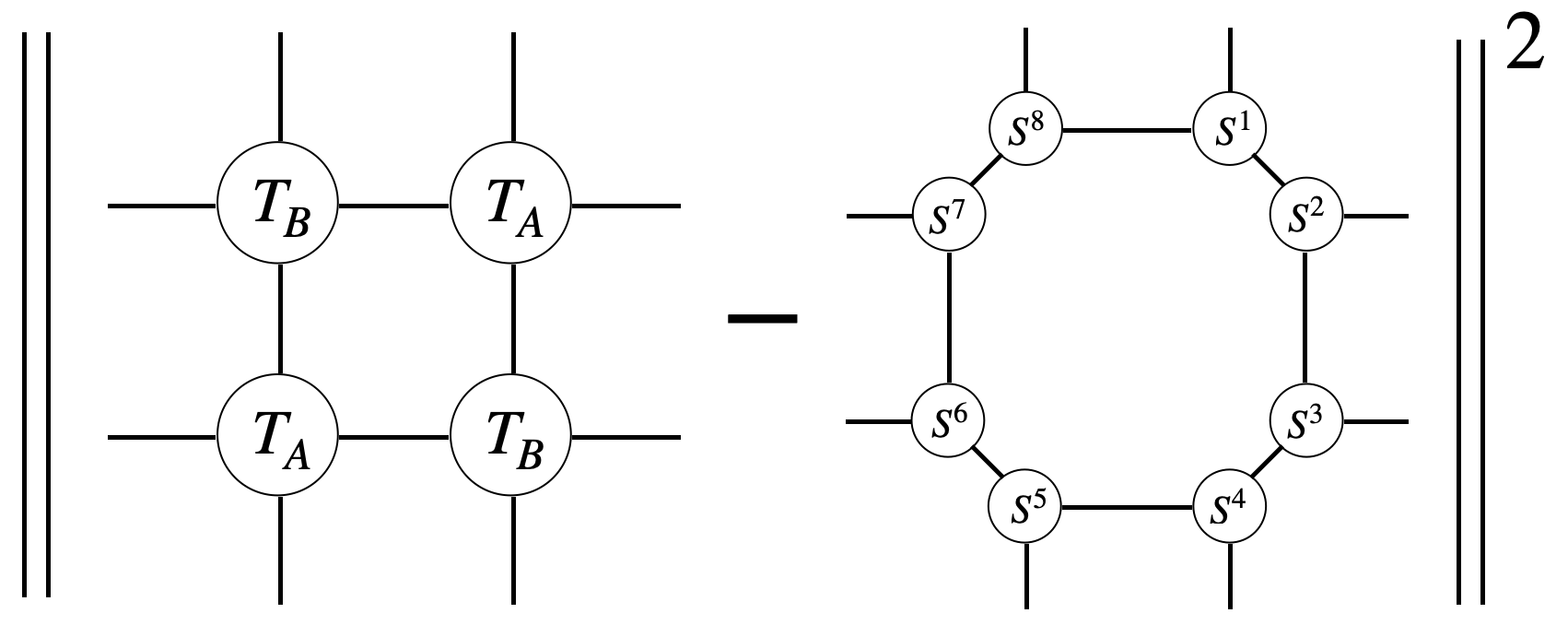}.\label{tnr_norm}
\end{align}
Contrasting with the approach described in Eq.~\eqref{eq:svd_norm}, TNR focuses on a two-by-two unit cell composed of two distinct tensors. In this process, an 8-leg tensor (as shown on the left side) is approximated through the contraction of eight 3-leg tensors, denoted as $S^1\sim S^8$(as depicted on the right side).

This methodology is central to the concept of TNR: rather than optimizing individual tensors, the focus is on optimizing contracted tensor networks. This approach allows for a more nuanced and effective handling of complex tensor structures, particularly in addressing the limitations of traditional SVD methods in TRG. By optimizing the entire network of tensors, TNR provides a more robust framework for accurately capturing the intricate interactions within these systems. 

In TRG, $S^1\sim S^8$ are constructed through SVD as explained in the algorithm. As those tensors from SVD are a good enough approximation of local tensors, we adopt them as initial tensors of optimizations. Let the left and right sides of Eq.~\eqref{tnr_norm} be $|\Psi_A\rangle$ and $|\Psi_B\rangle$. Then, the cost function is rewritten in the following forms.
\begin{align}
&||\Psi_A\rangle-|\Psi_B\rangle|^2
=\langle\Psi_A|\Psi_A\rangle + \langle\Psi_B|\Psi_B\rangle-\langle\Psi_A|\Psi_B\rangle-\langle\Psi_B|\Psi_A\rangle,\\
&\includegraphics[width=120mm]{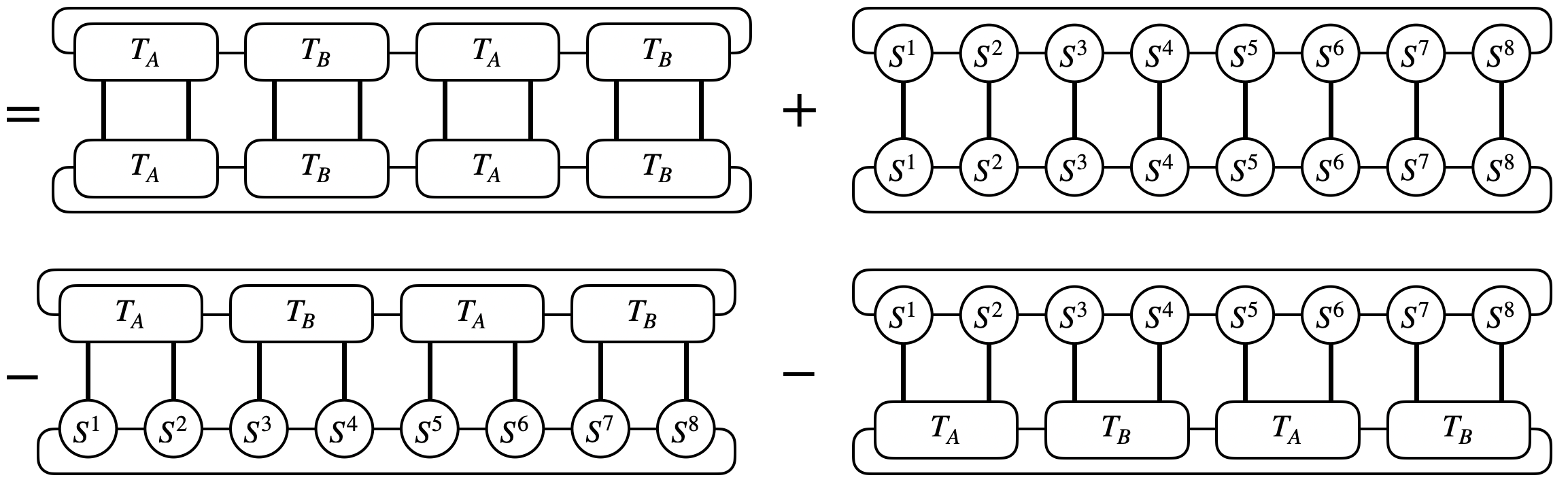}.\label{eq:looptnr_cost}
\end{align}
Now, we want to optimize $S^1\sim S^8$ to minimize the cost function. For convenience, we define $C$, $N_i$, $W_i$, and $W_i^\dag$ as followings:
\begin{align*}
    \includegraphics[width=135mm]{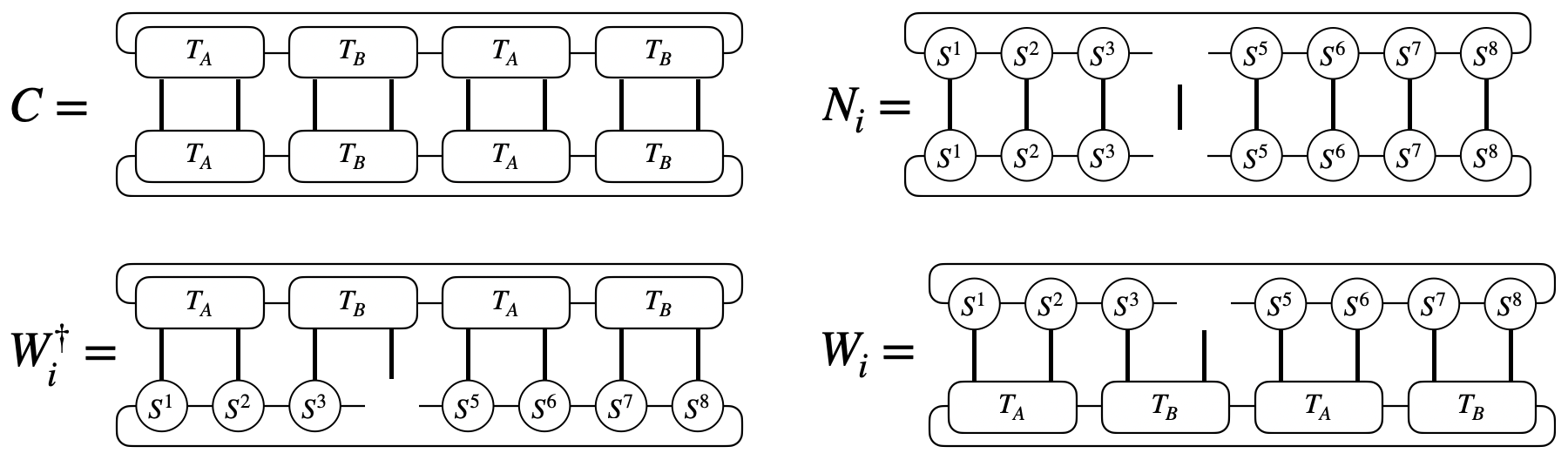}.
\end{align*}
This allows to rewrite Eq.~\eqref{eq:looptnr_cost} as 
\begin{align}
f(\{S^i\})&=||\Psi_A\rangle-|\Psi_B\rangle|^2\nonumber\\
&= C + (S^i)^\dag N_i S^i - W^\dag_iS^i - (S^i)^\dag W_i.
\end{align}
This function is quadratic if we fix 
 every tensor except for $S^i$. In this case, the minimum can be found by solving $\frac{\delta f(\{S^i\})}{\delta S^i}=0$. It is straightforward to check that it is equivalent to solving the following linear equation:
 \begin{align}
     N_i S_i = W_i.
 \end{align}

\underline{\textit{Entanglement filtering}}

Another important ingredient of TNR is entanglement filtering. This is a procedure to discard the local degrees of freedom under the existence of CDL tensors. When the tensor has the form of the CDL tensors shown in Eq.~\eqref{CDL_figure}, the contracted four tensors can be expressed as the following:
\begin{align*}
    \includegraphics[width=130mm]{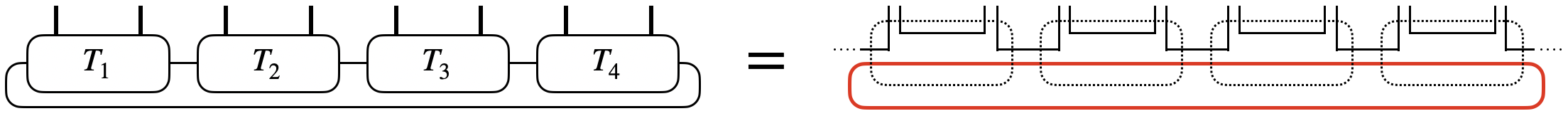},
\end{align*}
where $(T_1,T_2,T_3,T_4) = (T_A,T_B,T_A,T_B)$, and the red loop corresponds to the local loop in Fig.~\ref{CDL_tensors}. In scenarios where the red loop encompasses $n$ dimensions, each tensor within this red loop is associated with a rank-$n$ diagonal matrix, as illustrated in the following diagram:
\begin{align}
    \includegraphics[width=60mm]{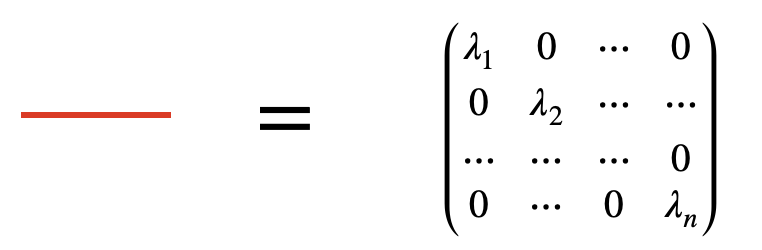}.\label{red_part}
\end{align}
Here, each $\lambda_i$ is arranged in descending order based on absolute values. The primary objective in entanglement filtering is to effectively compress this matrix to a rank-1 configuration. This compression is achieved by constructing a projector between $T_i$ and $T_{i+1}$ that targets the subspace corresponding to $\lambda_1$, the largest singular value. To do this, we use a QR decomposition. 
Consider the procedure of inserting a projector between tensors 
$T_4$ and $T_1$ in a TNR setup, where we denote $T_{i+4}=T_i$ for cyclic consistency. The first step involves placing a rank-$d$ identity matrix, denoted as $L_1^{[1]}$
 , to the left of $T_1$. Subsequently, we apply QR decomposition to the tensor product of $L_1^{[1]}$ and $T_1$, resulting in:
\begin{align}
L_1^{[1]} T_1 = \tilde{T}_1 L_1^{[2]},
\end{align}
where $\tilde{T}_1 $ is an orthogonal matrix and $L_1^{[2]}$ is an upper triangular matrix.
The next step involves normalizing $L_1^{[2]}$ appropriately and repeating a similar QR decomposition process with $L_1^{[2]}$ and $T_2$, and then proceeding with 
$L_1^{[3]}$ and $T_3$. This iterative process is continued until convergence is achieved, resulting in the final projector 
$L_1^{[\infty]}$ (The convergence is checked when $L_1$ comes back to between $T_4$ and $T_1$.
During this process, $L$ accumulates the matrix in Eq.~\eqref{red_part} to end up having 

\begin{align*}
\lim_{m\rightarrow\infty}
\begin{pmatrix}
\lambda_1^m & 0 &\cdots&0 \\
0 & \lambda_2^m&\cdots&\cdots \\
\cdots&\cdots &\cdots&0\\
0&\cdots &0&\lambda_n^m
\end{pmatrix}
\propto 
\begin{pmatrix}
1 & 0 &\cdots&0 \\
0 & 0&\cdots&\cdots \\
\cdots&\cdots &\cdots&0\\
0&\cdots &0&0
\end{pmatrix}.
\end{align*}
We repeat the same thing to the left starting from $R_4^{[1]}$ and $T_4$ to obtain $R_4^{[\infty]}$. Finally, we obtain the projectors using SVD as follows:
\begin{align}
    L_1^{[\infty]}R_4^{[\infty]} = U_{41}\Lambda_{41}V^\dag_{41}\nonumber,\\
    P_{4R} = R^{\infty}_4V_{41}\frac{1}{\sqrt{\Lambda_{41}}},\\
    P_{1L} = \frac{1}{\sqrt{\Lambda_{41}}} U^\dag_{41}L_1^{\infty},
\end{align}
Obtaining all projectors in hand, we redefine $T_{A/B}$ by contracting four projectors around as below:
\begin{align*}
    \includegraphics[width=60mm]{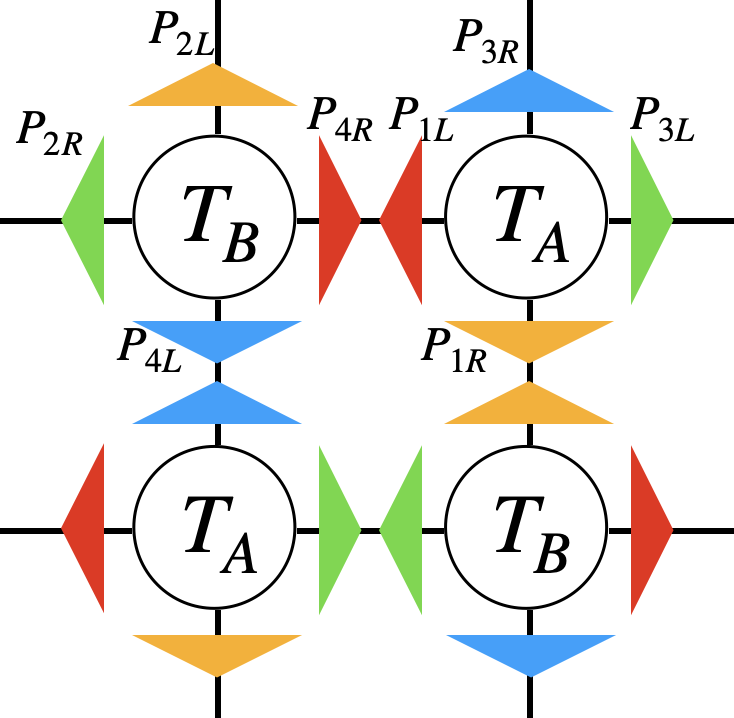}.
\end{align*}
For more details, readers shall consult the original paper~\cite{PhysRevLett.118.110504}. However, it is important to note that we can reduce the CDL loop structure through this procedure. Loop-TNR's algorithm is summarized in Algorithm.~\ref{tnr_algorithm}. The practical implementation in Python is in my GitHub repository(\url{https://github.com/dartsushi/Loop-TNR_RGflow}). 

\begin{algorithm}[H]
 \caption{Algorithm for Loop-TNR\label{tnr_algorithm}}
 \begin{algorithmic}
 \renewcommand{\algorithmicrequire}{\textbf{Input:}}
 \renewcommand{\algorithmicensure}{\textbf{Output:}}
 \REQUIRE Initial tensor $T_{A/B}^{(0)}$ representing local Boltzmann weights
 \ENSURE Renormalized tensor $T_{A/B}^{(n)}$
  \STATE Loop-TNR 1-step:
  \STATE 1. Entanglement filtering
  \STATE 2. Decomposition of $T_{A/B}$ into $S^1\sim S^8$ using SVD.
  \begin{center}
      \includegraphics[width=80mm]{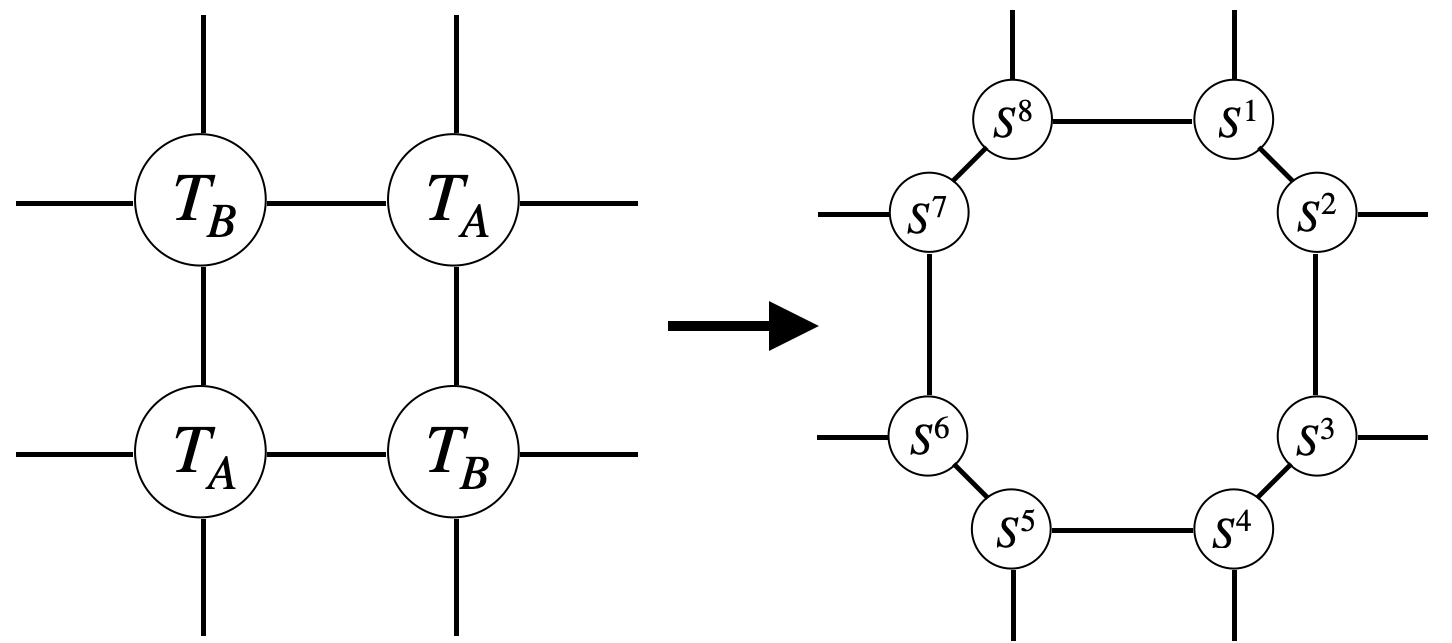}
  \end{center}
  \STATE 3. Optimize $S^i$ using from 1 to 8.
   \begin{align}
     N_i T_i = W_i.
 \end{align}
 \STATE 4. Repeat 3 until you reach the desired accuracy.
 \STATE 5. Combine $(S^1,S^4,S^5,S^8)$ and $(S^2,S^3,S^6,S^7)$ to make new $T_{A/B}$ respectively.\\
  \FOR {$i = 1$ to $n$}
  \STATE Repeat Loop-TNR 1-step
  \ENDFOR
 \RETURN $T^{(n)}_{A/B}$ 
 \end{algorithmic} 
 \end{algorithm}
\underline{Limitation of TNR}\\
While TNR effectively removes CDL tensors, another significant limitation arises from computational constraints, often termed finite bond dimension effects. For example, the computational cost of Loop-TNR scales as $O(D^6)$, limiting practical computations on a typical desktop computer to bond dimensions up to approximately $D\approx40$. Although TNR achieves exactness in the limit of $D=\infty$, the necessity of using a finite bond dimension inevitably introduces numerical errors. Prior to our research, the methodology for estimating these numerical errors was not well-established. Addressing this gap, we will delve into the strategies for estimating numerical errors in the context of tensor-network based methodologies in Chapter \ref{Sec:RGflow}.
 
\subsection{CFT data from TRG/TNR}
Up to this point, we have discussed the implementation of TRG and TNR methods in the context of real-space RG analysis. A key aspect of these methods is their ability to reveal the properties of fixed points in critical lattice models. In practice, when TRG and TNR are applied to simulate critical lattice models, the renormalized tensor $T^{(n)}$ is observed to converge rapidly to a specific tensor, denoted as $T^{*}$. This convergence behavior is indicative of the system approaching a fixed point in its parameter space.

The tensor $T^{*}$, aptly referred to as a fixed-point tensor, is believed to encapsulate the properties associated with the fixed-point of the system. Notably, Gu and Wen have proposed methodologies to calculate the scaling dimensions directly from this fixed-point tensor~\cite{PhysRevB.80.155131}. In our review, we will adopt a slightly different notation to facilitate a more seamless integration with the main content of our discussion~\footnote{While our explanation is fundamentally equivalent to that of Gu and Wen, we approach the topic using the energy basis, in contrast to their use of the character basis of CFTs.}.

The pivotal outcome of Gu and Wen's research concerns the analysis of the fixed-point tensor for a square lattice, which is a four-legged tensor mirroring the structure of the original lattice. A critical step in their methodology involves contracting the legs along the vertical axis of this tensor. This contraction process leads to a matrix, from which the scaling dimensions can be inferred based on the eigenvalues. The process can be represented as:
\begin{align}
\sum_n T_{\alpha n\beta n} =
\begin{matrix}\includegraphics[width=77mm]{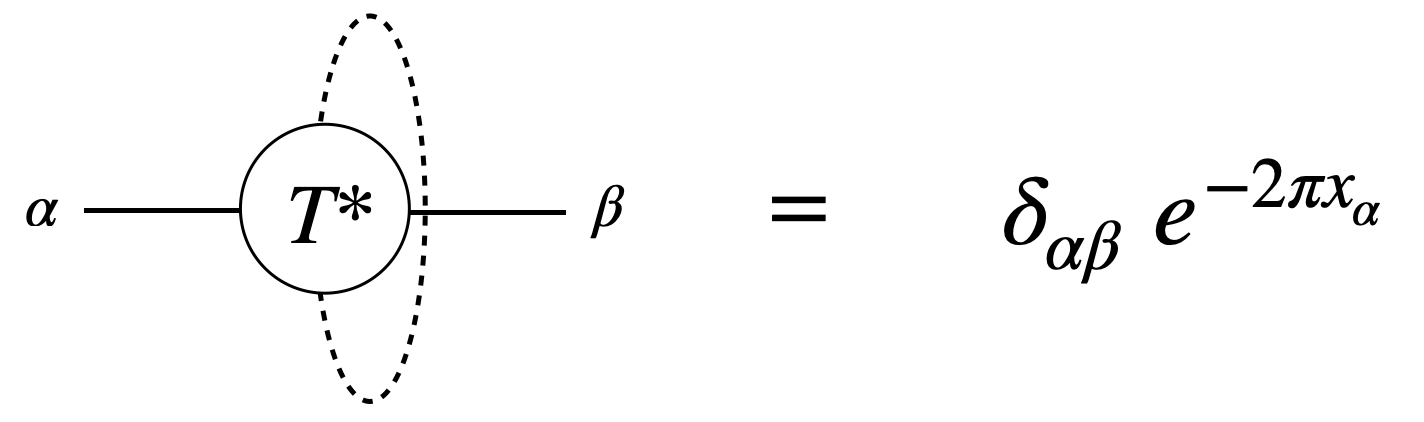},
\end{matrix}
\label{transfer_matrix_CFT}
\end{align}
In this formulation, the largest eigenvalue is normalized to one, aligning with the scaling dimension of the identity operator, $x_I = 0$. Subsequently, the scaling dimensions associated with other operators can be derived from the ratio of the eigenvalues, denoted as $\lambda_n$, using the equation:
\begin{align}
x_n=\frac{1}{2\pi}\ln\frac{\lambda_0}{\lambda_n}\label{eq:xn}.
\end{align}
Here, the eigenvalues $\lambda_n$ are arranged in descending order. This methodology offers a robust means to extract scaling dimensions from the fixed-point tensor, providing a valuable tool for analyzing the critical properties of the system. By utilizing the eigenvalue ratios, one can effectively determine the scaling dimensions corresponding to various operators, thereby gaining deeper insights into the nature of the critical points in the lattice model.

To understand why scaling dimensions are reflected in the eigenvalues of renormalized tensors, let us revisit the nature of renormalized tensors in TRG/TNR schemes, as discussed in the previous section. In these schemes, each RG step corresponds to a scale transformation with a factor of $b = \sqrt{2}$. Consequently, after $n$ steps of coarse-graining, the renormalized tensor effectively represents the tensor networks of a system whose size has been scaled to $L = \sqrt{2}^n$. This process can be visualized schematically as shown below:
\begin{align}
    \centering
    \includegraphics[width=80mm]{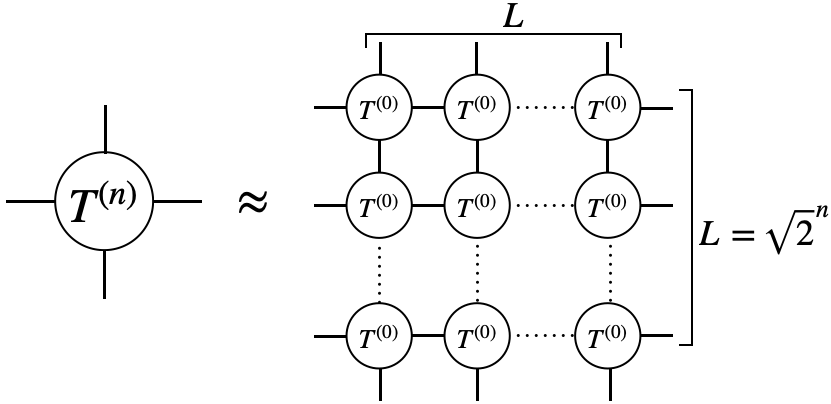}
\end{align}

In this $L \times L$ tensor network, contracting the vertical indices transforms the network into a cylindrical shape. The eigenvalues of the resultant matrix are essentially the $L$ repetitions of the column-to-column transfer matrix, which I will henceforth refer to simply as the 'transfer matrix'. In the context of classical-quantum correspondence, a single layer of the transfer matrix equates to a translation in imaginary time. In one-dimensional quantum systems, this translation corresponds to the Hamiltonian. Consequently, there exists a relationship between the eigenvalues of the transfer matrix and the spectrum of the one-dimensional Hamiltonian, $E_n(L)$, as follows:
\begin{align}
\lambda_n = e^{-LE_n(L)},\label{eq:lambda_E}
\end{align}
where the $E_n$ values are arranged in ascending order, such that $E_0$ represents the ground state. This perspective enables the subsequent chapters to analyze the transfer matrix spectrum within the framework of Hamiltonian formalism.

At criticality, the system's conformal invariance enables the computation of the energy spectrum, $E_n(L)$. Consider an infinitesimal transformation defined as $r'_\mu = r_\mu + \epsilon_\mu$. Within this framework, the variation of the action can be articulated as:
\begin{align}
\delta{S} = -\int\frac{d^2r}{2\pi}T_{\mu\nu}(r)\partial_\mu\epsilon_\nu(r),
\end{align}
where $T_{\mu\nu}$ denotes the stress tensor. In two-dimensional systems at criticality, conformal invariance is a key characteristic. This invariance includes transformations such as translations, rotations, dilatations, and special conformal transformations. As a result, conformal transformations in these systems can be represented through analytic functions on the complex plane, using the transformation $z \rightarrow w(z)$, where $z = x + iy$ and $\bar{z} = x - iy$.

The generators of these conformal transformations are expressed as $T(z) = \frac{1}{4}(T_{xx} - T_{yy} - 2iT_{xy})$ and $\bar{T}(\bar{z}) = \frac{1}{4}(T_{xx} - T_{yy} + 2iT_{xy})$. 
The variation of an operator $A$ under conformal transformations is effectively encapsulated by the Ward-Takahashi identity:
\begin{align}
\delta_\epsilon A(z,\bar{z}) = \oint_{C_z}\frac{d\zeta}{2\pi{i}}\epsilon(\zeta)T(\zeta)A(z,\bar{z}).
\end{align}

This equation implies that the variation of $A(z,\bar{z})$ can be computed by applying the product of $\epsilon$ and the stress tensor $T$ to $A$, followed by performing a contour integration around the point $z$. To facilitate this calculation, it is useful to express the stress tensor through Laurent expansions:
\begin{align*}
T(z) = \sum_{n\in\mathbb{Z}}z^{-n-2}L_n, \quad \bar{T}(\bar{z}) = \sum_{n\in\mathbb{Z}}\bar{z}^{-n-2}\bar{L}_n.
\end{align*}
Here, $L_n$ and $\bar{L}_n$ serve as the generators of conformal transformations and obey the Virasoro algebra, a cornerstone of CFT. The Virasoro algebra is given by:
\begin{align}
[L_m, L_n] = (m - n)L_{m + n} + \frac{c}{12}(m^3 - m)\delta_{m + n, 0},
\end{align}
where $c$ represents the central charge, a fundamental characteristic of the CFT. The central charge is a critical parameter that helps classify the universality class of the theory. It provides an intuitive measure of the number of bosonic excitations present: for instance, a theory with $n$ decoupled free bosons has a central charge of $c = n$, while a theory with $n$ decoupled fermions has $c = n/2$.\\

In this framework using the stress tensor, the dilatation $\hat{D}$ is defined as following:
\begin{align}
    \hat{D}&=\oint\frac{dz}{2\pi{i}}zT(z)+\oint\frac{d\bar{z}}{2\pi{i}}\bar{z}T(\bar{z}),\\
    &=L_0+\bar{L}_0.
\end{align}
The eigenvalues of $L_0$ and $\bar{L}_0$ are conformal weights denoted as $(h,\bar{h})$, and the scaling dimension $x =h+\bar{h}$ becomes indeed the eigenvalue of the dilatation.

\begin{figure}[tb]
\centering
\includegraphics[width=100mm]{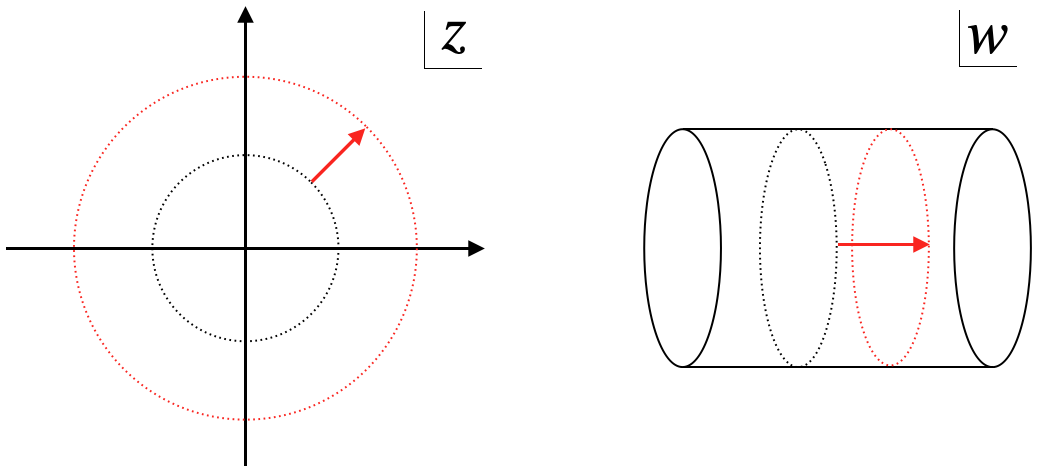}
\caption{The conformal mapping from a plane to a cylinder. The black and red dotted circles on the plane correspond to different time slices on the cylinder. As a result, the scale translation indicated by the red arrow on the $z$-plane transforms into a translation in imaginary time on the $w$-axis.}
\label{plane_to_cylinder}
\end{figure}

The intriguing aspect of dilatation in two-dimensional conformal field theory is that its generator, $\hat{D}$, can be reinterpreted as the generator of translation in imaginary time, denoted as $\hat{H}_P$, in a (1+1)-dimensional context. This relationship becomes evident when considering the conformal mapping $w = \frac{L}{2\pi}\ln{z}$, as depicted in Fig.~\ref{plane_to_cylinder}.

In this mapping, the original $z$-plane is transformed onto a cylinder with circumference $L$. The cylinder can be viewed as a quantum system with size $L$ and periodic boundary conditions in one dimension. The translation in scale, represented by the red arrow on the left panel of Fig.~\ref{plane_to_cylinder}, corresponds to a translation in time on the cylinder. Since the energy density in a quantum system is the diagonal $t$ component of the stress tensor, it follows that the energy of a finite quantum system can be calculated using $\hat{D}$. Conversely, if we know how the stress tensor $T$ transforms under conformal transformations, we can deduce the system's energy. This transformation is known and is given by:
\begin{align}
T(z) = \left(\frac{dw}{dz}\right)^2 \Tilde{T}(w) + \frac{c}{12}\{w,z\},
\end{align}
where $\{w,z\}$ is the Schwarzian derivative.

The Hamiltonian on the cylinder is then expressed as:
\begin{align}
H_P = \int_0^L \frac{dx}{2\pi} (T_{\text{cyl}}(w) + \overline{T}_{\text{cyl}}(w))\nonumber \\
= \frac{2\pi}{L}(L_0 + \overline{L}_0 - \frac{c}{12}),\label{eq:Hp}
\end{align}
yielding the energy $E_n = \frac{2\pi}{L}(x_n - \frac{c}{12})$~\cite{cardy1984conformal,cardy1986operator}. Therefore, the scaling dimension/the energy spectrum can be calculated from the energy spectrum of the transfer matrix in $y$ direction $\lambda_i=e^{-2\pi (x_i -\frac{c}{12})}$, being consistent with Eq.~\eqref{eq:lambda_E}.

In lattice models, it is crucial to consider the contribution of bulk energy when analyzing the energy spectrum. This consideration leads to the following expressions for the energy spectrum:
\begin{align}
E_n - E_0 &= \frac{2\pi}{L}x_n, \label{a}\\
E_0 &= \epsilon_0L - \frac{\pi c}{6L}. \label{b}
\end{align}

However, a challenge arises in determining the central charge $c$ due to the lack of a sufficient number of equations to separate the contribution of the bulk energy. To address this, one can utilize the partition function from the previous RG step, denoted as $Z(n-1)$:
\begin{align}
Z(n-1) = {\Tr}_{x_i} \exp\left(-2\pi\left(x_i - \frac{c}{12}\right) - \epsilon_0b^{2n-2}\right).
\end{align}
For the fixed-point tensor, it is reasonable to assume that both $c$ and $\epsilon_0$ remain constant. Under this assumption, the central charge can be determined as follows:
\begin{align}
c = \frac{6}{\pi}\frac{1}{b^2 - 1}\left(b^2\ln\lambda_0^{(n-1)} - \ln\lambda_0^{(n)}\right). \label{c}
\end{align}
A widely used formula for calculating the central charge is given by \cite{PhysRevB.80.155131}:
\begin{align}
c = \frac{6}{\pi}\left[\frac{b^2}{b^2 - 1}\left(\ln Z(n-1) - \frac{\ln Z(n)}{b^2}\right) + \ln\frac{\lambda_0^{(n)}}{Z(n)}\right]. \label{d}
\end{align}
In the critical case, Eqs.\eqref{c} and \eqref{d} are equivalent since $\frac{\lambda_0^{(n)}}{Z(n)} = \frac{\lambda_0^{(n-1)}}{Z(n-1)}$. However, Eq.\eqref{d} may become unstable when the system size surpasses the correlation length. In this thesis, we have calculated the effective central charge using Eq.~\eqref{d}.

Through this methodology, we are able to extract the central charge, a key parameter in conformal field theory that characterizes the universality class of the model. This approach enables a deeper understanding of the critical properties of lattice models, particularly in the context of tensor network renormalization and finite-size scaling theory.

\subsection{How to read CFT dictionaries: Character}
To determine the applicability of a specific Conformal Field Theory (CFT) to your models, one effective approach is to utilize the concept of the 'character', a tool that allows you to decipher the energy spectrum of the model. By comparing the energy spectrum of your model with known results in CFT literature, you can ascertain the universality class of the model in question. This method offers a practical solution to the often esoteric nature of CFT literature, facilitating its application to specific models.

In the previous section, we discussed how the partition function, excluding bulk energy contributions, is expressed as:
\begin{align}
Z(L,L) = \Tr \exp\left[-2\pi(L_0+\bar{L}_0-\frac{c}{12})\right],\label{Z_raw}
\end{align}
where the eigenvalue of $(L_0+\bar{L}_0)$ corresponds to the scaling dimension $x_i$. The key takeaway is that the trace in Eq.~\eqref{Z_raw} is calculated using the transfer matrix basis, meaning the eigenvalues of the transfer matrix in the $y$-direction are predicted to be $e^{-2\pi (x_i -\frac{c}{12})}$. For the partition function of a rectangular shape, such as $Z(L,2L)$, the eigenvalue becomes $e^{-4\pi (x_i -\frac{c}{12})}$, reflecting the squaring of eigenvalues due to the double transfer distance. Moreover, the spectrum of $Z(2L,2L)$ is identical to that of $Z(L,L)$, and $Z(2L,4L)$ mirrors $Z(L,2L)$. This uniformity results from the scale-invariance inherent in CFT, with the shape ratio $\frac{L_y}{L_x}$ being the critical factor. Therefore, the partition function is often represented in CFT literature as:
\begin{align}
Z(q) = \Tr q^{L_0-\frac{c}{24}}\bar{q}^{\bar{L}_0-\frac{c}{24}},\label{Z_q}
\end{align}
where $\tau = \frac{iL_y}{L_x}$, $q=e^{2\pi i\tau}$, and $\bar{q}=e^{-2\pi i\tau^{*}}$. Here, $\tau$ is known as the modular parameter. Defining this parameter allows for the generalization of the partition function concept to parallelograms, as illustrated below:
\begin{align*}
\includegraphics[width=60mm]{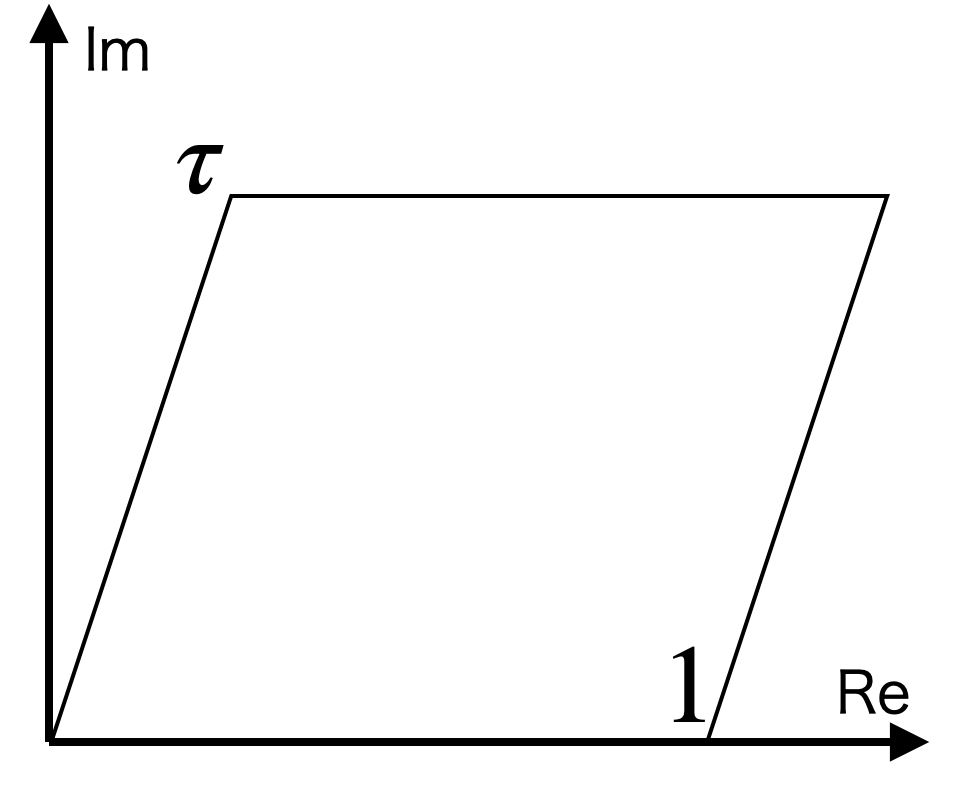}
\end{align*}
With periodic boundary conditions on both edges, this setup equates to the partition function on a torus, a phenomenon termed modular invariance. This principle imposes significant constraints on CFT. In the context of transfer matrix spectra, considering the spectrum on a parallelogram is insightful, as Eq.~\eqref{Z_q} encapsulates information about the conformal spin. This arises from the phase acquired by the operator due to momentum when the real part of $\tau$ is non-zero. As depicted in Fig.~\ref{plane_to_cylinder}, this shift on the plane corresponds to the additional phase acquired during operator rotation around the origin. Here, we specifically address the case where $\tau=i$. For a broader understanding encompassing more generic cases, readers are encouraged to consult comprehensive CFT literature~\cite{francesco2012conformal}.

\begin{figure}[tb]
    \centering
    \includegraphics[width=86mm]{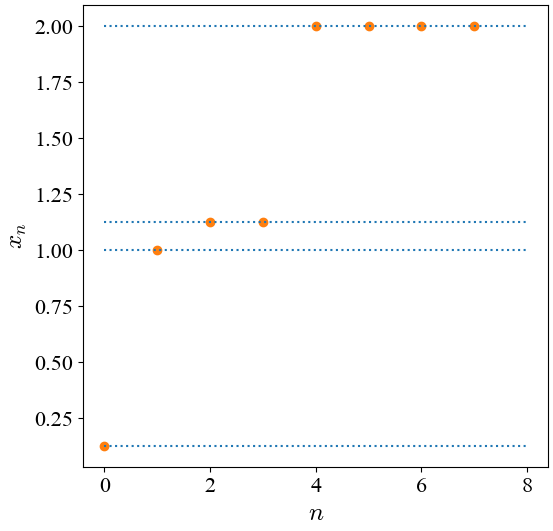}
    \caption{The scaling dimensions of the critical Ising model that are obtained from the transfer matrix spectrum of Loop-TNR. The blue dotted line are the theoretical values from the character.}
    \label{xn_Ising}
\end{figure}
"An essential aspect to understand is that Eq.\eqref{Z_q} possesses a universal form, and its specifics are precisely known for certain universality classes. For instance, the partition function of the Ising CFT is expressed as:
\begin{align}
Z(q) = \left|\chi_0(q)\right|^2 + \left|\chi_{\frac{1}{16}}(q)\right|^2 + \left|\chi_\frac12(q)\right|^2.
\end{align}
Here, $\chi_0(q)$, $\chi_{\frac{1}{16}}(q)$, and $\chi_{\frac{1}{2}}(q)$ correspond to the excitations of the $I$, $\sigma$, and $\epsilon$ families, respectively, and are known as characters. These characters often contain known quantities, allowing for the prediction of low-lying energies or scaling dimensions. For the Ising model, the characters are as follows:
\begin{align}
    q^{c/24}\chi_0(q) &= 1+q^2+q^3+2q^4+2q^5+\cdots,\label{chi_Ising1}\\
    q^{c/24}\chi_{\frac{1}{16}}(q) &= q^{1/16}(1+q+q^2+2q^3+2q^4+3q^5+\cdots),\label{chi_Ising2}\\
    q^{c/24}\chi_{\frac{1}{2}}(q) &= q^{1/2}(1+ q+q^2+q^3+2q^4+2q^5+\cdots).\label{chi_Ising3}
\end{align}
By substituting Eqs.(\ref{chi_Ising1}-\ref{chi_Ising3}) into Eq.\eqref{Z_q} and comparing it with Eq.\eqref{Z_raw}, we can identify low-lying scaling dimensions from the exponents of $q$. For the Ising model, we obtain $Z(q) = q^{-c/12}(1+q^{1/8}+q^{1}+2q^{9/8}+4q^2+\cdots)$, leading to the scaling dimensions $x_0,x_1,x_2,x_3,x_4,x_5,\cdots = 0,\frac{1}{8},1,\frac{9}{8},\frac{9}{8},2,\cdots$. This aligns with the numerical results of the critical Ising model, as demonstrated in Fig.~\ref{xn_Ising}.

In a similar vein, by comparing the character of a potential universality class from CFT literature with your numerical data, you can accurately determine the universality class of an unknown phase transition in lattice models.

\chapter{Finite-size and finite bond dimension effects of tensor network renormalization\label{Sec:RGflow}}
In this chapter, we introduce a comprehensive procedure aimed at extracting the running coupling constants denoted as $g_n(L)$ of the underlying field theory for classical statistical models on two-dimensional lattices. This approach synergizes TNR with the finite-size scaling principles of CFT. Our methodology extends Gu and Wen's analysis, originally focused on the transfer matrix spectrum of critical systems, to encompass off-critical systems.

In systems away from criticality, the spectral properties exhibit a departure from scale invariance during the RG steps, deviating from the universal values characteristic of CFT. We propose that these deviations are indicative of the RG flow. By meticulously analyzing these deviations, we can calculate the running coupling constants with extremely high precision at each scale. This process enables us to track the evolution of these coupling constants through successive scales, thereby offering a detailed visualization of the RG flow. To demonstrate the efficacy of our approach, we apply it to classical lattice models such as the Ising and three-state Potts models. This concept is also extended to determine the transition with extremely high precision.

Furthermore, we explore the potential of utilizing the eigenvectors of the transfer matrix to compute another critical component of CFT data: the OPE coefficients. This advancement in our methodology allows us to derive a complete set of CFT data from the TRG/TNR scheme. The ability to obtain both the running coupling constants and OPE coefficients marks a significant step forward in our understanding of these models, bridging the gap between tensor network approaches and the rich theoretical framework of CFT.

Finally, utilizing our new methodology, we reveal the limitations due to finite bond dimension $D$ on TNR applied to critical systems. We find that a finite correlation length is induced by the finite bond dimension in TNR, and it can be attributed to an emergent relevant perturbation that respects the symmetries of the system. The correlation length shows the same power-law dependence on $D$ as the "finite entanglement scaling" of the Matrix Product States. Using this, we can estimate the errors arising from TRG/TNR scheme, which was unclear before.

The following sections mainly discuss the Ising and three-state Potts models on the square lattice. The energy (classical Hamiltonian) of the Ising and three-state Potts models are 
\begin{align}
    \mathcal{E}_{Ising} &= -\sum_{\langle i,j\rangle}\sigma_i\sigma_j-h\sum_i\sigma_i,\\
    \mathcal{E}_{Potts} &= -\sum_{\langle i,j\rangle}\delta_{s_i,s_j},
\end{align}
where $\sigma_i=\pm 1$(Ising) and $s_i=0,1,2$(three-state Potts). The first terms and $h$ represent the nearest-neighbor interactions and the uniform magnetic field.
Employing the temperature $T$, the Boltzmann weight is defined as $e^{-\mathcal{E}/T}$, where we set the Boltzmann constant to unity.
Our primary focus in this chapter is the Ising model, while a detailed discussion of the three-state Potts model is provided in the appendix.
The Ising model reaches its critical point at $(T,h)=(T_c,0)$, where $T_c=2/\ln{(1+\sqrt{2})}$.
At this criticality, physical quantities like the spin-spin correlation function are governed by the Ising CFT, which comprises three primary operators: the identity operator $I$, magnetic operator $\sigma$, and energy operator $\epsilon$.

In the context of the lattice model, a shift from the critical temperature and the application of a magnetic field correspond to the perturbative insertion of $\epsilon$ and $\sigma$ into the effective Hamiltonian. As a result, $\sigma$ is odd in the $\mathbb{Z}_2$ spin-flip, while $I$ and $\epsilon$ are even. Given the operator structure of the CFT, certain quantities are consequently fixed.

\section{Operator product expansion coefficients}
Operator product expansion is another fundamental concept in field theory and statistical mechanics as explained in the previous chapter~\cite{kadanoff,wilsonOPE}. Since OPE coefficients determine the structure of the field theory, their computation is quite important. Numerical computation of OPE coefficients~\cite{PhysRevLett.116.040401,PhysRevResearch.4.023159} has not been so straightforward compared to that of
scaling dimensions. Here, we present a simpler way to compute them, which is applicable to TRG~\cite{PhysRevLett.99.120601}, HOTRG~\cite{PhysRevB.86.045139}, and Loop-TNR~\cite{PhysRevLett.118.110504}. 

The renormalized tensor $T^{(n)}$ contracted in $x$-direction is a transfer matrix in the $y$-direction. While the eigenvalues of the transfer matrix correspond to the energy or scaling dimension of the primary operators, the eigenvectors thereof are the wavefunctions of the corresponding
``primary states'' $|\psi_n(L)\rangle$. This is graphically represented below.
\begin{align*}
    \includegraphics[width=60mm]{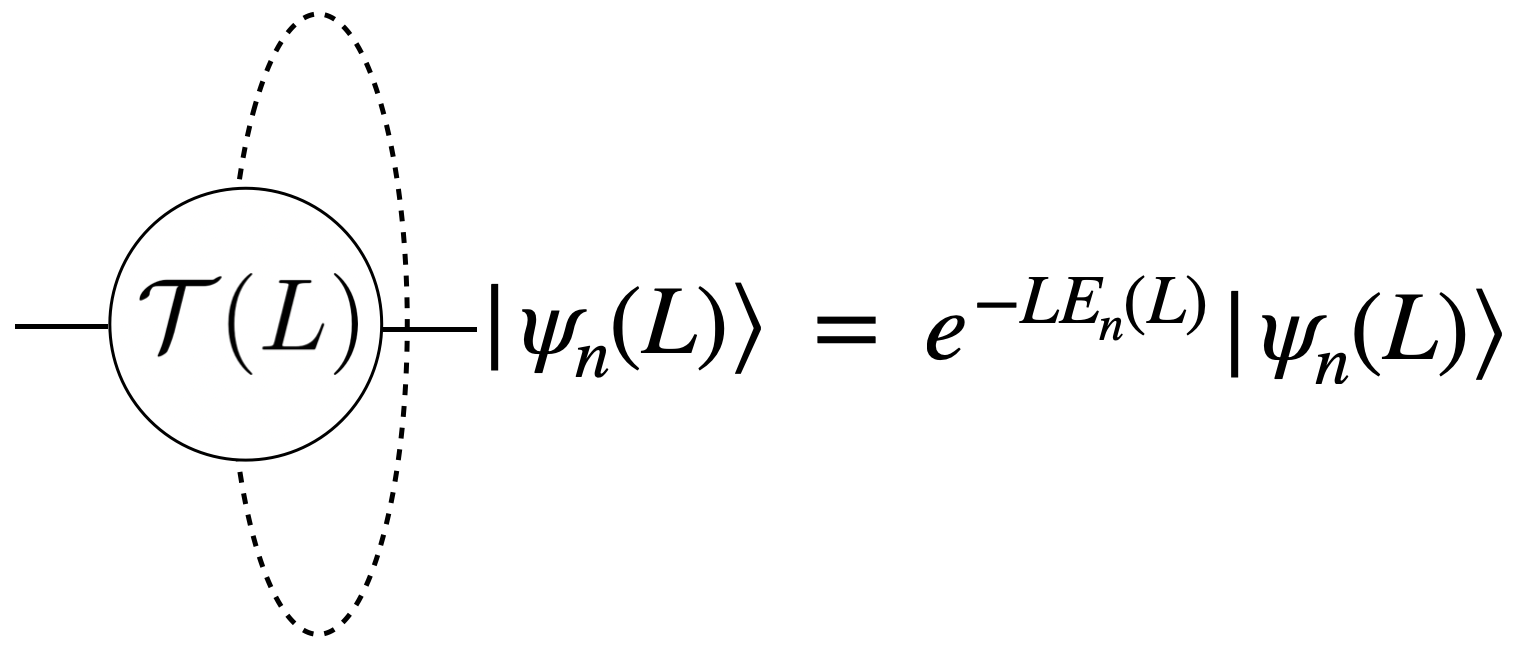}
\end{align*}
Note that the tensor has been rotated for ease of viewing. We do not change the contracted index. Likewise, we can compute the wavefunctions of the system size $2L$ as depicted below. 
\begin{align*}
    \includegraphics[width=60mm]{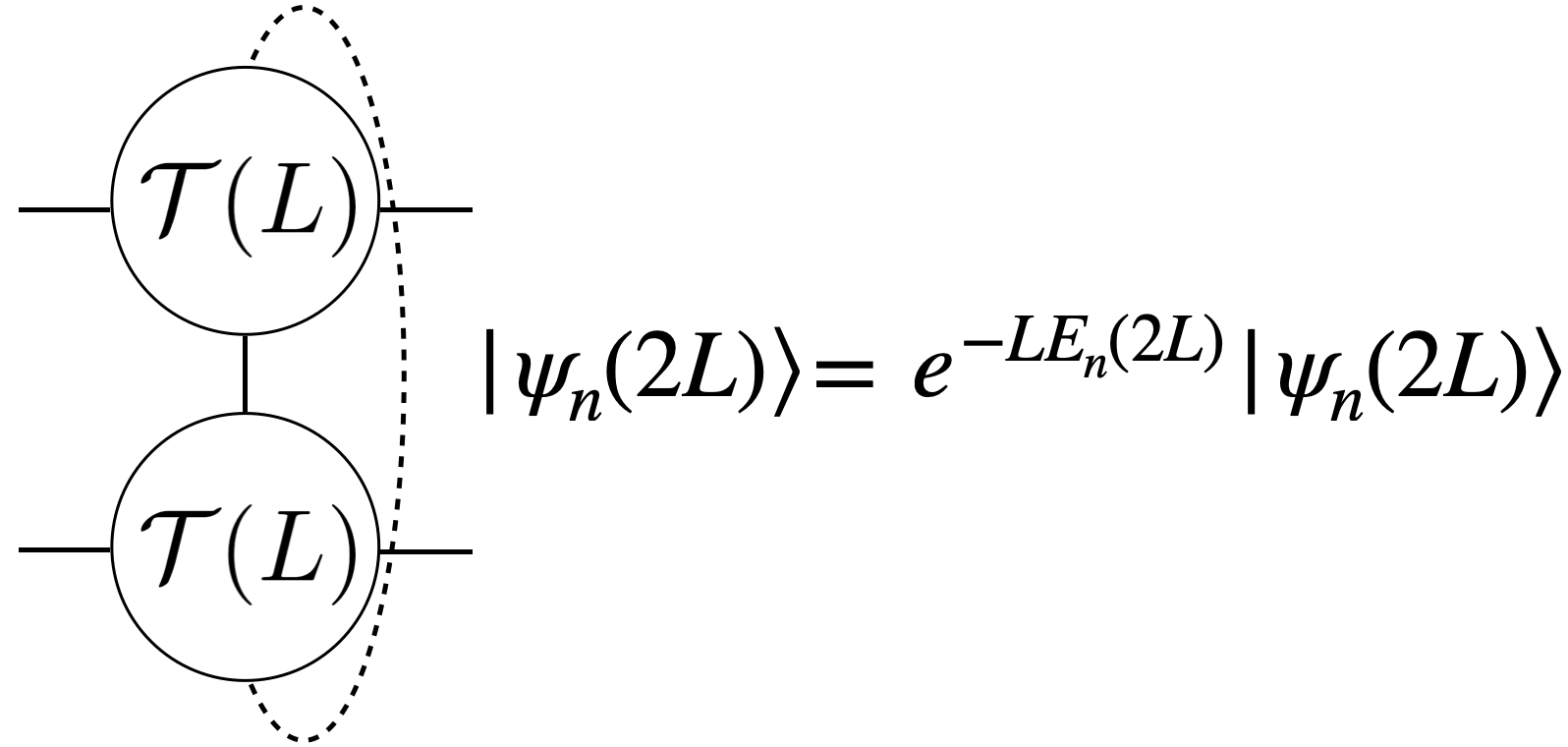}
\end{align*}
$|\psi_n(L)\rangle$ and $|\psi_n(2L)\rangle$ are one-leg and two-leg tensors, respectively. Thus, we propose a novel method for calculating OPE coefficients, utilizing the contraction of eigenstates derived from the transfer matrix of the renormalized tensor. Specifically, this method involves computing the overlaps between the states $\ket{\psi_\alpha(2L)}$ and the tensor product $\ket{\psi_\beta(L)} \otimes \ket{\psi_\gamma(L)}$ by contracting their respective indices. This computed quantity is directly proportional to the OPE coefficients. This approach aligns with the discussion on the overlap of quantum wave functions in Ref.~\cite{PhysRevB.105.125125,PhysRevB.105.165420,arxiv.2203.14992}.
\begin{table}[tb]
\begin{tabular}{lllll}
\multicolumn{1}{c}{$\psi_\alpha$} & \multicolumn{2}{c}{$\psi_\beta,\psi_\gamma$} & \multicolumn{1}{c}{$C_{\alpha\beta\gamma}$} & \multicolumn{1}{c}{$2^{2x_\beta+2x_\gamma-x_\alpha}{A_{\alpha\beta\gamma}}/{A_{III}}$} \\\hline
$I$                                 & \multicolumn{2}{l}{$\sigma,\sigma$}          & 1                                           & 0.8938                                                     \\
$\sigma$                          & \multicolumn{2}{l}{$\sigma,I$}               & 1                                           & 0.9473                                                     \\
$I$                                 & \multicolumn{2}{l}{$\epsilon,\epsilon$}      & 1                                           & 0.9966                                     \\
$\epsilon$                        & \multicolumn{2}{l}{$\epsilon,I$}             & 1                                           & 0.9968                                                     \\
$\epsilon$                        & \multicolumn{2}{l}{$\sigma,\sigma$}          & 0.5                                         & 0.5007                                                      \\
$\sigma$                          & \multicolumn{2}{l}{$\sigma,\epsilon$}        & 0.5                                         & 0.2705   \\
\hline
\end{tabular}
\caption{The numerically obtained OPE coefficients of the Ising CFT from TRG. The bond dimension and the system size are $D=56$ and $L=16\sqrt{2}$(9 RG steps), respectively.\label{OPE_table} }
\end{table}

In CFT, the overlap of wavefunctions, denoted as $\langle \psi_\alpha(2L) | \psi_\beta(L) \psi_\gamma(L) \rangle$, is proportional to the 'pants diagram' of path integrals. This relationship can be elucidated through a review of how eigenstates $|\psi_n(L)\rangle$ are expressed in CFT.

For the ground state, the process begins with a random initial state $|\psi_{ini}\rangle$, which undergoes imaginary time evolution:
\begin{align}
|\psi_0\rangle \propto \lim_{\tau\rightarrow\infty} e^{-\tau H}|\psi_{ini}\rangle,
\end{align}
where the dominance of the smallest eigenvalue of the Hamiltonian after sufficient imaginary time evolution ensures the ground state is attained. To obtain primary states corresponding to excited states, initial eigenstates are prepared, with state-operator correspondence in CFT allowing the creation of states by inserting the corresponding operator into the vacuum. On a cylinder, these operators acquire a prefactor $\left(\frac{2\pi}{L}\right)^{-x_n}$, derived from Eq.~\eqref{eq:conformal_mapping_primary_operator}. Additionally, an exponential factor $e^{\frac{2\pi}{L}x_n\tau}$ should be placed to ensure normalization relative to the ground state as in Eq.~\eqref{eq:Hp}\footnote{In essence, Eq.~\eqref{eq:Hp} states $E_n-E_0 = \frac{2\pi}{L}x_n$. We implement the factor in advance to compensate $e^{-\tau H} = e^{-\frac{2\pi}{L}x_n\tau}$.}. Consequently, the eigenstate at the $\tau=0$ slice is given by:
\begin{align}
|\psi_n(L)\rangle &= \lim_{\tau\rightarrow-\infty}e^{-\frac{2\pi}{L}x_n\tau}\left(\frac{2\pi}{L}\right)^{-x_n} \psi_n^{cyl}(\tau)|I^{cyl}\rangle,\label{eq:normalization_psin}\\
&= \psi_n(-\infty)|I^{cyl}\rangle\nonumber.
\end{align}
This formulation applies to $|\psi_\beta(L)\rangle$ and $|\psi_\gamma(L)\rangle$. Similarly, $|\psi_\alpha(2L)\rangle$ is constructed from the infinite future:
\begin{align}
|\psi_\alpha(2L)\rangle &= \lim_{\tau\rightarrow\infty}e^{-\frac{\pi}{L}x_\alpha\tau}\left(\frac{\pi}{L}\right)^{-x_\alpha} \psi_\alpha^{cyl}(\tau)|I^{cyl}\rangle\label{eq:normalization_psialpha},\\
&= \psi_\alpha(\infty)|I^{cyl}\rangle\nonumber.
\end{align}
In this setup, these three vectors meet at the $\tau=0$ slice, forming the basis for the 'pants diagram' path integral representation. (It does look like a pair of pants!)
\begin{align}
    \includegraphics[width=120mm]{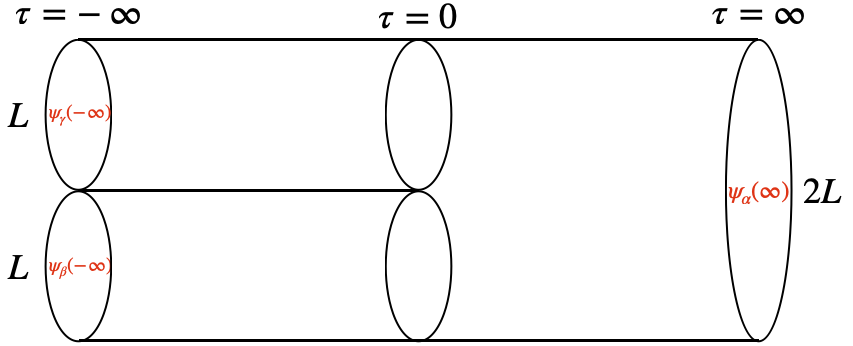}\nonumber
\end{align}

In this context, the overlap of the eigenstates can be conceptualized as a three-point function on the 'pants' manifold:
\begin{align}
\frac{A_{\alpha\beta\gamma}}{A_{III}} = \langle\psi_\alpha(\infty)\psi_\beta(-\infty)\psi_\gamma(-\infty)\rangle_{pants},\label{eq:corr_pants}
\end{align}
where $A_{\alpha\beta\gamma}$ represents the overlap $\langle{\psi_\alpha(2L)}|{\psi_\beta(L)}{\psi_\gamma(L)}\rangle$. To compute this quantity, a conformal mapping from the 'pants' manifold to a plane is required. This type of mapping, common in string theory calculations for string interactions, is well-understood. The conformal mapping, known as the Mandelstam mapping~\cite{Mandelstam:1973jk}, is defined as:
\begin{align}
z = \frac{L}{2\pi}[\ln(w-i)+\ln(w+i)-2\ln(w)],
\end{align}
where the points $z=-\infty$ and $z=\infty$ correspond to $w=\pm i$ and $w=0$, respectively~\footnote{The coefficients of the logarithmic of $w$ terms correspond to the length of the string, whereas its sign is negative for the states in the infinite future. The sum of the prefactors should be zero so that the total length of the strings from the infinite past is equal to that of the infinite future.}. This mapping, which effectively stitches three cylinders to a plane after opening them, calls for the additional factor $(\frac{dw}{dz})^{x_n}$ to the correlation function as in Eq.~\eqref{eq:conformal_mapping_primary_operator}. Thus, Eq.~\eqref{eq:corr_pants} is transformed to:
\begin{align}
\frac{A_{\alpha\beta\gamma}}{A_{III}} = |J_\alpha|^{x_\alpha}|J_\beta|^{x_\beta}|J_\gamma|^{x_\gamma}\langle\psi_\alpha(0)\psi_\beta(i)\psi_\gamma(-i)\rangle_{plane},
\end{align}
where the three-point function is now evaluated on the plane. The prefactor $J$ is derived from a combination of Eqs.~(\ref{eq:normalization_psin}-\ref{eq:normalization_psialpha}) and the $(\frac{dw}{dz})^{x_n}$ factor:
\begin{align}
|J_\alpha| &= \left|\lim_{z\rightarrow\infty}e^{-\frac{\pi}{L} z}\left(\frac{\pi}{L}\right)^{-1}\left(\frac{dw}{dz}\right)\right|_{w\rightarrow0},\\
|J_\beta| &= \left|\lim_{z\rightarrow-\infty}e^{-\frac{2\pi}{L} z}\left(\frac{2\pi}{L}\right)^{-1} \left(\frac{dw}{dz}\right)\right|_{w\rightarrow i},\\
|J_\gamma| &= \left|\lim_{z\rightarrow-\infty}e^{-\frac{2\pi}{L} z}\left(\frac{2\pi}{L}\right)^{-1} \left(\frac{dw}{dz}\right)\right|_{w\rightarrow-i}.
\end{align}
Upon evaluation, it is straightforward to verify that $|J_\alpha|=1$ and $|J_\beta|=|J_\gamma|=1/2$, and $\langle\psi_\alpha(0)\psi_\beta(i)\psi_\gamma(-i)\rangle_{plane}= 2^{x_\alpha-x_\beta-x_\gamma}C_{\alpha\beta\gamma}$. Consequently, the relationship between the OPE coefficient and the overlap becomes:
\begin{align}
\frac{A_{\alpha\beta\gamma}}{A_{III}} =
2^{x_\alpha-2x_\beta-2x_\gamma}C_{\alpha\beta\gamma},\label{OPE_eq}
\end{align}
illustrating how the OPE coefficients are intimately connected to the eigenstate overlaps within the 'pants' manifold framework. (For more generic cases, readers shall consult Ref.~\cite{PhysRevB.105.125125,PhysRevB.105.165420,arxiv.2203.14992}.)

 In most cases, the identity operator denoted as $I$, corresponds to the ground state or equivalently the leading eigenvector. Thus, the OPE coefficients $C_{\alpha\beta\gamma}$ can be computed from the ratio of the overlap $A_{\alpha\beta\gamma}$ and $A_{III}$, given the scaling dimensions from the transfer matrix.
We benchmark our method by the critical Ising model. Table.~\ref{OPE_table} shows the numerically obtained OPE coefficients by TRG~\cite{PhysRevLett.99.120601} at $L=16\sqrt{2}$ and $D=56$.
Naturally, there are finite-size corrections to Eq.~\eqref{OPE_eq}.
Since Eq.~\eqref{OPE_eq} is exact in the thermodynamic limit, using a very large system size $L$ might appear desirable.
However, as we will discuss later in Sec.~\ref{section_fixed_point}, corrections due to the finite bond-dimension effect appear for system sizes larger than a correlation length $\xi(D)$\footnote{This effect is even stronger and non-trivial for TRG due to the CDL tensors as discussed in the previous chapter.}. 
As reported in Ref.~\cite{arxiv.2203.14992}, the finite-size effects are significant for $C_{\sigma\sigma\epsilon}$ and $C_{\epsilon\epsilon I}$.
Nevertheless, even with the moderate size $L=16\sqrt{2}$, the obtained values $C_{I\epsilon\epsilon} = 0.9966$ and $C_{\epsilon\sigma\sigma}=0.5007$ are rather close to exact CFT results. While we tested our method by the simplest algorithm, Levin and Nave's TRG, the method for calculating OPE is straightforwardly applicable to other TRG and TNR algorithms, such as HOTRG~\cite{PhysRevB.86.045139}.

\section{Precise determination of the transition temperature}
\label{sec:LevelSpectroscopy}
\begin{figure}[tb]
    \centering
    \includegraphics[width=80mm]{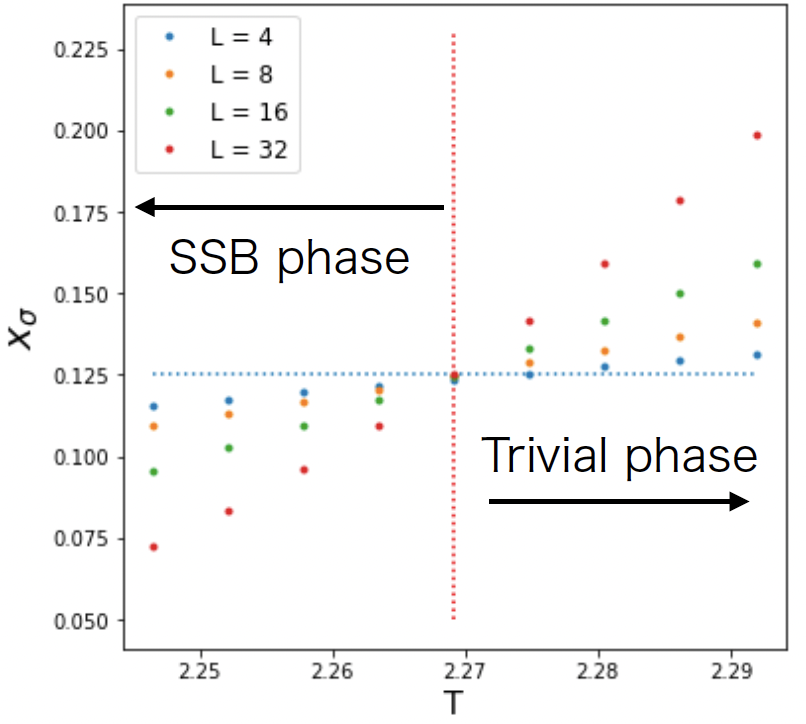}
    \caption{The scaling dimension obtained from the first and second leading eigenvalue of the transfer matrix of the Ising model as $x_\sigma(L) = \frac{1}{2\pi}\ln\frac{\lambda_0}{\lambda_1}$. At the critical temperature, denoted by a red dotted line, the value is consistent with the scaling dimension $\frac{1}{8}$ regardless of the system sizes. Away from criticality, however, the deviation from $\frac{1}{8}$ grows as $L$ increases.}
    \label{fig_xsigma_finite}
\end{figure}

As we have mentioned earlier, the ratios of the transfer matrix spectrum represent the scaling dimension as $\frac{\lambda_n}{\lambda_0}=\exp(-2\pi x_n)$ at criticality after sufficient coarse-graining. However, the rescaled energy levels of a lattice model generally depend on the system size $L$,
as the effective Hamiltonian of the system contains perturbations to the CFT. This lets us define a generalized concept of the scaling dimension, which depends on the system size. We denote it as “rescaled energy" defined as
\begin{align}
\frac{\lambda_n(L)}{\lambda_0(L)}=\exp(-2\pi x_n(L))\label{TNR_spectrum}.
\end{align}
Figure~\ref{fig_xsigma_finite} exhibits “the rescaled energy" of the first excited state of the Ising model, corresponding to $x_\sigma$.
At the critical temperature, which we denote with a red dotted line, this scaling dimension consistently aligns with the expected value of $\frac{1}{8}$, a characteristic feature of the Ising universality class in two dimensions. This consistency is observed regardless of the variations in system size $L$.

However, a notable shift in behavior occurs when the system deviates from the critical temperature. 
This observed shift in the behavior of the rescaled energy $x_\sigma(L)$ is a key aspect of critical phenomena. In off-critical scenarios, $x_\sigma(L)$ starts to diverge from its critical value of $\frac{1}{8}$. Notably, as the system size $L$ increases, this deviation becomes more significant. This trend is not just a simple anomaly; rather, it signifies the evolution of the running coupling constants, denoted as $g_n(L)$, in the system.

This relationship between the deviation in scaling dimensions and the running coupling constants is deeply rooted in the theoretical framework combining perturbation theory with CFT. The perturbation theory, when applied within the context of CFT, provides a robust explanation for this phenomenon. It elucidates how the changes in the system's parameters, as it moves away from criticality, influence the running coupling constants and consequently, the scaling dimensions.

For a detailed exploration of this relationship and the underlying theoretical principles, readers are directed to Sec.~\ref{sec:perturbation} in the appendix. Here, we only use Cardy's results~\cite{cardy1984conformal,cardy1986operator}.
The rescaled energy levels in a finite-size perturbed CFT are given as
\begin{align}
x_n(L)=x_n+2\pi\sum_jC_{nnj}g_j(L),\label{fss_scaling_dimension}
\end{align}
where $g_j(L)$ scales as $\propto L^{2-x_j}$~\footnote{The second term in Eq.~\eqref{fss_scaling_dimension} is the first-order perturbation term. In CFT, the unperturbed eigenstates and perturbations correspond to the primary states and $\Phi_j$. In this framework, the corrections to the energy are expressed as $\langle n|g_j\Phi_j|n\rangle$, which yields the OPE coefficients $C_{nnj}$}. 
Comparing Eq.~\eqref{TNR_spectrum} from TNR and Eq.~\eqref{fss_scaling_dimension} from the conformal perturbation theory,
we can obtain the running coupling constants $g_j(L)$ at each scale from the finite-size effect $\delta x_n(L)=x_n(L)-x_n$.

An immediate and practical application of our observations is the precise determination of critical points. This approach, often referred to as 'level spectroscopy,' was originally developed to address the complexities of the Berezinskii-Kosterlitz-Thouless (BKT) transition, particularly noted for its challenges in standard finite-size scaling analysis. Initially, this technique was applied to quantum spin systems in one dimension, as demonstrated by Nomura in 1994~\cite{nomura1994critical}. More recently we extended for classical statistical systems in two dimensions using TNR~\cite{PhysRevB.104.165132}.

The core principle of level spectroscopy is the careful analysis of the energy levels or eigenvalues, particularly how they shift and evolve as the system approaches and moves away from criticality. While this technique was conceived in the context of the BKT transition, its basic concept is broadly applicable to more conventional types of critical phenomena, such as those observed in the Ising model.

The RG fixed-point for the two-dimensional Ising model has two relevant operators, the energy density $\epsilon$
and the magnetization density $\sigma$.
The coupling constant $g_\epsilon$ for $\epsilon$ is proportional to the deviation of the temperature from the critical point,
and also scaled $\sim L$ in the small coupling limit $g_\epsilon \ll 1$ because $x_\epsilon=1$.
Thus
\begin{align}
g_{\epsilon}(L) \sim \alpha (T-T_{c}) L ,
\label{eq:linear_app}
\end{align}
when $g_\epsilon(L) \ll 1$.
Likewise, the coupling $g_\sigma$ is proportional to the magnetic field $h$ and scaled $\sim L^{15/8}$ because $x_\sigma=1/8$. When determining the critical point, we focus on the critical temperature with zero magnetic fields, where $g_\sigma = 0$.

Although the Ising critical phenomena are mostly described by the two relevant coupling constants $g_\epsilon$ and $g_\sigma$,
more accurate description can be obtained by including irrelevant perturbations.
Including the leading irrelevant operators, namely the irrelevant operators with the smallest scaling dimension permitted by the symmetries,
the effective Hamiltonian of the Ising model is described as following:
\begin{align}
    H=H^{*}_{Ising}+\int_0^Ldx&[g_\sigma\sigma(x)+g_\epsilon\epsilon(x)\nonumber\\
    &+g_{T^2} T_{\text{cyl}}^2(x)+g_{\bar{T}^2}\bar{T}_{\text{cyl}}^2(x)],
\end{align}
where $T_{\text{cyl}}$ and $\bar{T}_{\text{cyl}}$ are the holomorphic and anti-holomorphic parts of stress tensor on a cylinder~\cite{cardy1986operator}.
The holomorphic part $T_{\text{cyl}}$ of the stress tensor on a cylinder is related to that on the infinite plane
$T_{zz}(z)$ via the conformal mapping $z = e^{2\pi w/L}$, where $w= \tau + ix$ and $0 \leq x < L$. More explicitly, $T_{zz}(z)$ transforms as
\begin{align}
    T_{\text{cyl}}(w) &= \left( \frac{2\pi}{L}\right)^2 \left( z^2 T_{zz} (z)-\frac{c}{24} \right) .
\end{align}
This leads to
\begin{align}
    T_{\text{cyl}}(x) &= \frac{2\pi}{L} \left( \sum_{n=-\infty}^\infty L_n e^{2\pi i x/L} - \frac{c}{24} \right),
\end{align}
where $c$ is the central charge characterizing the CFT, and
$L_n$'s are generators of the Virasoro algebra defined by
\begin{align}
    T_{zz}(z) = \sum_{n=-\infty}^\infty \frac{L_n}{z^{n+2}},
\end{align}
in terms of the holomorphic part $T_{zz}$ of the energy-momentum tensor on the infinite plane.
Inserting the above $T_{\text{cyl}}$ and integrating over $0 \leq x < L$ with an appropriate regularization,
the $g_{T^2}$-term of the perturbation is given as~\cite{poghosyan2019shaping}
\begin{align}
   \int dx  T^2_{cyl}(x) =L_0^2-\frac{c+2}{12}L_0+2\sum_{n=1}^{\infty}L_{-n}L_{n}+\frac{c(22+5c)}{2880}
   \nonumber
\end{align}
Only the first and second terms affect the energy levels, and the contributions to $x_\sigma(L)$ and $x_\epsilon(L)$ are calculated to be
$-\frac{7}{768}g_{T^2}$ and $\frac{7}{48}g_{T^2}$ respectively.
The computation of the contributions from $\bar{T}^2$ is exactly the same, and we denote their sum as $g$.
These operators are the leading irrelevant operators for the Ising model on the square lattice. An important aspect to consider is the origin and implications of the squared terms of the stress tensor on the cylinder, $T_{\text{cyl}}^2$ and $\bar{T}_{\text{cyl}}^2$. These terms possess conformal spins of $+4$ and $-4$, respectively. The presence of these conformal spins is significant because they lead to the breaking of continuous rotational symmetry, which is the breaking of Lorentz invariance in Minkowski space-time. However, in the context of a square lattice, which only possesses discrete $C_4$ rotational symmetry, the inclusion of these terms is permissible. The presence of $T_{\text{cyl}}^2$ and $\bar{T}_{\text{cyl}}^2$ serves to adjust the symmetry of the continuum theory to match that of the discrete lattice model. Essentially, these 'irrelevant' operators play a crucial role in aligning the theoretical model's symmetry with the inherent symmetry of the square lattice. This aspect of conformal spin becomes particularly evident during odd-numbered RG steps. At these stages, the lattice undergoes a 45-degree rotation, leading to a sign change in these operators, as indicated by the factor $(e^{i\pi/4})^4=-1$. This rotation-induced sign change has observable consequences. For instance, when examining finite-size corrections at criticality, we notice an alternating sign in the corrections to the scaling dimension, $\delta x_\sigma$, at each RG step.

Including the contributions from relevant perturbations, the resulting finite-size corrections to $x_\sigma(L)$ and $x_\epsilon(L)$ are shown in Table.~\ref{RGscalingdimension}~\footnote{As $T_{\text{cyl}}^2$ and $\bar{T}_{\text{cyl}}^2$ are not primary operators, we need to pay special attention.}. 
While the exact critical point is known for the Ising model on the square lattice, let us demonstrate
the determination of the critical point from the TNR spectrum without using prior knowledge of the critical point
(but utilizing the CFT data, assuming that we identify the universality class).
Since we are interested in the critical point at zero magnetic fields, we can set $g_\sigma \propto h =0$.
The simplest way to determine the critical point is to look at the lowest rescaled energy level
$x_\sigma(L)$ in the lowest order of the relevant coupling constant $g_\epsilon$,
ignoring the irrelevant perturbation $g$.
Within this approximation, the shift $\delta x_\sigma(L) = x_\sigma(L) - x_\sigma$ vanishes at the critical point $T=T_c$ where $g_\epsilon = 0$.
Away from the critical point, $\delta x_\sigma(L)$ is non-zero and grows proportionally to $L$ because $g_\epsilon(L)$ scales as $L$.
Because of this, we can identify the critical point with the temperature where $\delta x_\sigma(L)=0$ is observed in the TNR spectrum.
However, this estimate suffers from the corrections due to the leading irrelevant perturbations $T_{\text{cyl}}^2$ and $\bar{T}_{\text{cyl}}^2$.
Since they have scaling dimension $4$, the corresponding coupling constant is renormalized as $g \propto L^{-2}$.
This leads to an error of $O(L^{-2})$ in the naive estimate of the critical point using $\delta x_\sigma(L)=0$.

\begin{figure}
    \centering
    \includegraphics[width=86mm]{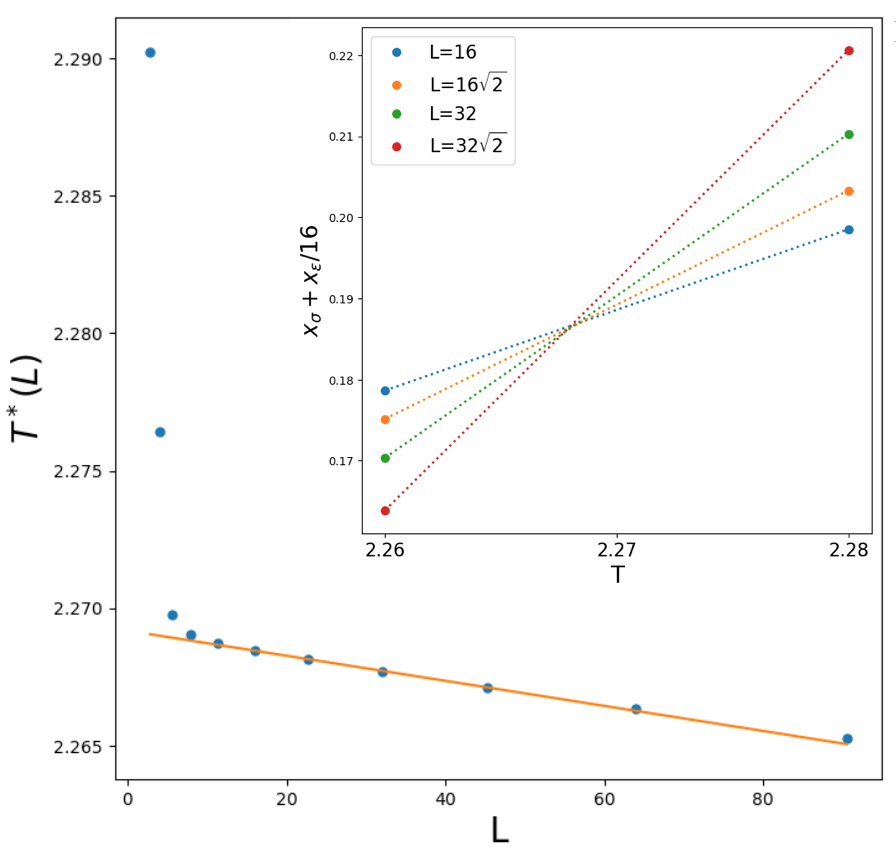}
    \caption{Example of estimating the transition temperature using Loop-TNR. We set $T^-=2.66$ and $T^+=2.68$ as an initial estimate. The level-crossing temperature $T^*(L)$ is linearly fitted to extrapolate the transition temperature. The insert shows how we compute $T^*(L)$ for various system sizes.}
    \label{Tc_extrapolation}
\end{figure}

We can improve the accuracy by removing the effects of the leading irrelevant perturbation $g$.
This can be done by combining the shifts of the rescaled energy levels $\delta x_\sigma(L)$ and $\delta x_\epsilon(L)$ following Table.~\ref{RGscalingdimension} as 
\begin{align}
\delta x_\text{cmb} \equiv & \delta x_\sigma(L)+\frac{1}{16}\delta x_\epsilon(L)\nonumber\\
    &=
    \pi g_\epsilon + 
    (\alpha^\sigma_\sigma+\frac{1}{16}\alpha_\epsilon^\sigma) g_\sigma^2 +(\alpha^\epsilon_\sigma + \frac{1}{16}\alpha^\epsilon_\epsilon ) g_\epsilon^2.
\label{eq:deltax_combined}
\end{align}
Note that the first-order correction in the irrelevant coupling $g$ is canceled out.
Now we can identify the critical point by finding the temperature for which $\delta x_\text{cmb} \propto g_\epsilon(L) = 0$.
Having eliminated the effects of the leading irrelevant perturbation $T_{\text{cyl}}^2, \bar{T}_{\text{cyl}}^2$, the dominant error is now caused by the next-leading irrelevant
operator with scaling dimension $6$ and thus should be scaled as $L^{-4}$.

In practice, the determination of the critical point can be efficiently implemented as follows.
First, we pick up one temperature from each phase: $T^{+} > T_{c}$ and $T^{-} < T_{c}$, and calculate the combined shift
$\delta x_\text{cmb}$ at these temperatures. 
The phase of the system can be confirmed by observing the growth of $\delta x_\text{cmb}$ as the system size increases because it increases/decreases if the system is in the high-temperature/low-temperature phase (if the initial choice of the temperature turns out to be wrong, change the temperature and restart the process).
Next, linear interpolations of the combined shift between the two temperatures $T^{\pm}$ are made,
and the crossing of the lines for system sizes $L$ and $\sqrt{2}L$ is found, as shown in the insert of Fig.~\ref{Tc_extrapolation}.
We denote the temperature where the two lines cross as $T^{*}(L)$.
Because of the second-order contribution $O({g_\epsilon}^2)$ in Eq.~\eqref{eq:deltax_combined},
the crossing temperature $T^*(L)$ obtained by the \textit{linear} interpolation deviates from the true critical point $T_c$
as $T^{*}(L) - T_{c} \propto g_\epsilon \propto L$, when $g_{\epsilon} \ll 1$\footnote{It is proportional to $L^{2-x_{\text{thermal}}}$, where $x_{\text{thermal}}$ is the scaling dimension of the thermal operator.}.
The critical point $T_c$ is estimated by fitting $T^{*}(L)$ by a linear function of $L$ as $T^*(L) \sim T_c + \text{const.} L$.
While the ``extrapolation'' to $L=0$ used here might look unusual, this procedure is done to remove the effect of the nonlinearity due to $O({g_\epsilon}^2)$
in Eq.~\eqref{eq:deltax_combined}, and the condition $\delta x_\text{cmb} = 0$ itself is accurate for $T_c$ up to the error of $O(L^{-4})$ due to the next-leading irrelevant
perturbations.
An example of the estimate of $T_c$ with the above procedure with the choice of the temperatures $T^{+}=2.68$ and $T^{-}=2.66$
and with system sizes $16 \leq L < 64$ is depicted in Fig.~\ref{Tc_extrapolation}.
The final estimate of the critical point is $T_c^\text{est} = 2.269177$.
Remarkably, even with the choice of two temperatures differ by $10^{-2}$ and the relatively low bond-dimension $D=20$, the estimated critical point is quite accurate:
$T_c^\text{est}- T_{c} = -8.11 \times 10^{-6}$.
This is thanks to the suppression of the error to $O(L^{-4})$ by eliminating the contributions from the leading irrelevant operators.
Once the critical point is estimated with good accuracy with this procedure, the accuracy can be further improved
by choosing $T^{\pm}$ closer to the estimated critical temperature and then applying the same procedure.

\begin{table}[tb]
\begin{tabular}{c|c|l|c}

           model                    & \multicolumn{2}{c|}{operator}             & Rescaled energy level                \\ \hline
\multirow{3}{*}{Ising model}
& \multicolumn{2}{c|}{ } & \\[0.5ex]
& \multicolumn{2}{c|}{
$x_\sigma(L)$ }&
$\frac{1}{8}+\alpha_\sigma^\sigma g_\sigma^2+\pi g_\epsilon+\alpha^\epsilon_\sigma g_\epsilon^2-\frac{7}{768}\pi g$\\[1.5ex]
&\multicolumn{2}{c|}{$x_\epsilon(L)$}&$1+\alpha^\sigma_\epsilon g_\sigma^2+\alpha^\epsilon_\epsilon g_\epsilon^2+\frac{7}{48}\pi g$\\[1.5ex] 
\hline
\end{tabular}
\caption{The finite-size scaling dimension of the Ising model. $\alpha$ is a constant determined from the second-order perturbation. Since $g_{T^2}$ and $g_{\bar{T}^2}$ decay in the same manner, we write them as $g$.\label{RGscalingdimension}}
\end{table}

\section{Renormalization group flow}
The comparison between the TNR spectrum in Eq.~\eqref{TNR_spectrum} and the conformal perturbation theory in Eq.~\eqref{fss_scaling_dimension}
can also be used to extract running coupling constants and their scale dependence, enabling visualization of the RG flow.
This analysis will be particularly useful in investigating the effects of finite bond dimensions in detail, a topic we plan to explore comprehensively in Sec.~\ref{section_fixed_point}.

\begin{figure*}[tb]
    \centering
    \includegraphics[width=140mm]{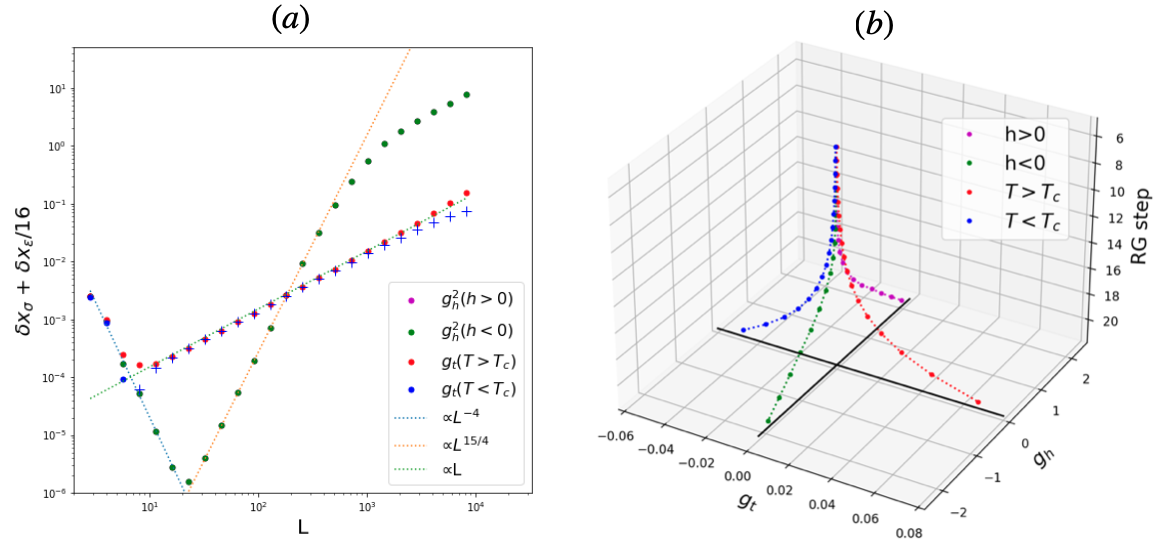}
    \caption{(Left panel) The system size dependence of $\delta x_\text{cmb} = \delta x_\sigma + \delta x_\epsilon/16$ for $h=\pm10^{-5}$(purple and green), $T= 1.0001T_c$(red) and $T= 0.9999T_c$(blue). The purple and green dots are on top of each other, and “$+$" denotes the data with a negative sign. After removing the $L^{-2}$ irrelevant perturbations, the next leading $L^{-4}$ perturbation shown with a blue dotted line appears. The data was obtained via Loop-TNR with a bond dimension of $D = 24$, which was deemed sufficient for the finitely-correlated systems being considered. (Right panel) The resulting renormalization group flow. Only data after six steps are exhibited, where the $L^{-4}$ perturbations disappear.}
    \label{Flow_Ising}
\end{figure*}

For the Ising model, the extraction of running coupling constants is based on observing shifts in the rescaled energy levels, as detailed in Table~\ref{RGscalingdimension}. It is also beneficial to consider the combined shift as described in Eq.~\eqref{eq:deltax_combined}. Given that $g_\sigma$ and $g_\epsilon$ are small near the criticality, we simplify our calculations by neglecting $g_\epsilon^2$ for $h=0$. Consequently, we redefine two relevant coupling constants for convenience: $g_t = \pi g_\epsilon$ and $g_h = \sqrt{(\alpha^\sigma_\sigma + \frac{1}{16}\alpha_\epsilon^\sigma)} g_\sigma$.
In this way, the combined shift Eq.~\eqref{eq:deltax_combined} simply gives $g_t$ when $h=0$ and ${g_h}^2$ when $T=T_c$, in the lowest order of $g_t, g_h$.
Using these relations, we can read off the relevant coupling constants $g_t$ or $g_h$ from the TNR data, as shown in Fig.~\ref{Flow_Ising}($b$).
As we have discussed in the previous subsection, the effects of the leading irrelevant perturbations $T_{\text{cyl}}^2, \bar{T}_{\text{cyl}}^2$ with scaling dimension $4$
are eliminated in the combined shift Eq.~\eqref{eq:deltax_combined}, and thus the finite-size correction is now of $O(L^{-4})$, due to the next-leading
irrelevant operators with scaling dimension $6$.
This $O(L^{-4})$ scaling is indeed observed in Fig.~\ref{Flow_Ising} near the critical point for small system size $L$ when relevant perturbations are still negligible.
Since it is safe to say that these contributions disappear after five RG steps, we can conclude that the origin of $g_t$ and $g_h$ are purely from $\epsilon$ and $\sigma$ after six steps. 
\par{} The right panel illustrates the scale-dependence of the coupling constants $g_t$ and $g_h$. It is nothing but the RG flow of the Ising critical point, and we conclude that we succeed in calculating the RG flow of the celebrated Ising fixed-point.
\par{}There is one thing to note on the left panel of Fig.~\ref{Flow_Ising}.
While the combined shift~\eqref{eq:deltax_combined}, which is an estimator for $|g_h|^2$,
scales as $L^{3.75}$ at $L<10^3$, it starts to flatten and scales as $L$ at $L>10^3$.
This behavior has a rather simple origin.
Since the magnetic perturbation is relevant, the system has a finite correlation length or equivalently, a non-zero gap $\Delta$.
This implies that the rescaled energy levels are proportional to $L$ for sufficiently large system size $L \gg \Delta^{-1}$.
As a consequence, the shift Eq.~\eqref{eq:deltax_combined} also grows proportionally to $L$.
In this regime, the conformal perturbation theory breaks down (higher-order contributions are important), and we no longer
identify the shift Eq.~\eqref{eq:deltax_combined} with $|g_h|^2$.
This should be distinguished from the $L$-linear behavior of the combined shift Eq.~\eqref{eq:deltax_combined}
observed for $L>10$ with $h=0$ and $T \neq T_c$, which corresponds to the renormalization of $g_t \propto L$ because of $x_\epsilon=1$.
The $L$-linear behavior due to the gap is observed in the non-perturbative regime $\delta x_{\epsilon, \sigma} \gg x_{\epsilon,\sigma}$, whereas
the $L$-linear behavior due to the scaling is observed in the perturbative regime $\delta x_{\epsilon, \sigma} \ll x_{\epsilon,\sigma}$.

\section{Finite bond-dimension effects\label{section_fixed_point}}

Let us examine the impacts of a finite bond-dimension $D$ on TNR from the perspective of our method.
In any computation that employs tensor networks, it is necessary to restrict the bond dimension to a finite value $D$ due to the increasing storage requirements and computational costs associated with larger bond dimensions.
The finiteness of the bond dimension inevitably leads to a loss of information in each step of renormalization after a certain number of iterations.
Although TNR can nominally handle arbitrary large systems, and the TNR-type calculations are often used to study extremely large systems,
we have to be careful about the limitations due to the finite bond dimension.

The limitation of the finite bond dimension $D$ on the matrix product state (MPS) is characterized by the finite (maximum) correlation length $\xi(D)$
of the MPS~\cite{luca_fes,pollmann2009theory,Pirvu_fes}.
The correlation length of MPS is known to obey the scaling law
\begin{align}
    \xi(D) \sim & D^\kappa ,
    \label{eq:xi_D_scaling}
    \\
    \kappa = & \frac{6}{c(1+\sqrt{\frac{12}{c}})} .
    \label{eq:kappa_c}
\end{align}
While the TNR-type calculation of two-dimensional statistical systems appears rather different from the MPS applied to one-dimensional quantum systems,
the emergence of the finite correlation length $\xi(D)$ obeying the similar scaling law~\eqref{eq:xi_D_scaling} was reported in
Ref.~\cite{PhysRevB.89.075116} for a HOTRG calculation of the critical Ising model in two dimensions.
The exponent $\kappa$ for the Ising model was estimated to be approximately $2$, which is close to the MPS exponent~\eqref{eq:kappa_c} $\kappa = 2.03425\ldots$
for the Ising CFT with central charge $c=1/2$.
A similar emergence of the finite correlation length $\xi(D)$ was also reported in our TNR finite-size scaling study of the two-dimensional XY model~\cite{PhysRevB.104.165132},
with the MPS exponent~\eqref{eq:kappa_c} for $c=1$.

In the following, using our TNR finite-size scaling methodology, we will demonstrate that the emergence of the finite correlation
length due to the finite bond dimension in TNR can be attributed to an emergent relevant perturbation.
Furthermore, we present evidences for the scaling~\eqref{eq:xi_D_scaling} with the MPS exponent~\eqref{eq:kappa_c}
in TNR of Ising and three-state Potts models.

\subsection{Emergent relevant perturbation}
\label{sec:emergent_perturbation}

\begin{figure}[tb]
    \centering
    \includegraphics[width=100mm]{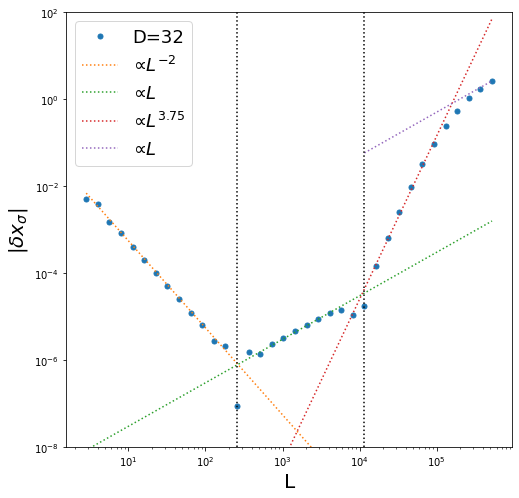}
    \caption{Shift $|\delta x_\sigma(L)|$ for the Ising model at $T=T_c$, $h =0$ computed by Loop-TNR with $D=32$.
    There is little finite-$D$ effect for small system sizes $L<256$. The emergent perturbations of $\epsilon$ and $\sigma$ appear at $L\sim 256$ and $L\sim10^4$, scaling as $L$ and $L^{15/4}$. The induced gap by finite-$D$ goes towards constant at $L>10^5$ as denoted with the purple dotted line.}
    \label{mixed_emergent_perturbation}
\end{figure}

If a finite correlation length emerges in the TNR, it would be
natural to identify the renormalized tensor with a Hamiltonian for the system away from the critical point,
that is, an RG fixed-point (CFT) Hamiltonian perturbed with relevant operators
\begin{align}
H_{\text{FB}}(D)=H^*_{\text{CFT}}+\sum_i\int_0^L dxg_i(D,L)\Phi_{i}(x,D),
\end{align}
where $H_{FB}$ is the effective Hamiltonian of the finite-$D$ system and $\Phi_{i}(x,D)$ are the scaling operators representing the perturbations.
In this view, we expect relevant perturbations to emerge in order to mimic the finite correlation length imposed by the finite bond dimension.

To demonstrate the emergence of the relevant perturbation, we investigate the system-size dependence of the
shift in the rescaled energy levels $\delta x_\sigma$.
In Fig.~\ref{mixed_emergent_perturbation}, we show the absolute value of the shift $|\delta x_\sigma|$ as a function of the system size $L$
used in calculating the transfer matrix spectrum in TNR exactly at the critical point $h=0, T=T_c$.
The conformal perturbation theory in Eq.~\eqref{fss_scaling_dimension} implies that
the shift $x_\sigma$ contains contributions from the irrelevant perturbations.
Since the leading irrelevant operators at the critical points are $T_{\text{cyl}}^2$ and $\bar{T}_{\text{cyl}}^2$ with scaling dimension $4$,
we expect $\delta x_\sigma(L)$ decays as $L^{-2}$. 
(This is to be contrasted with Eq.~\eqref{eq:deltax_combined} and Fig.~\ref{Flow_Ising}, in which the contributions from $T_{\text{cyl}}^2$ and $\bar{T}_{\text{cyl}}^2$
are eliminated.)
The expected $L^{-2}$ behavior in the shift $\delta x_\sigma(L)$ is indeed
observed for small system sizes $L<256$.
For larger system sizes, however, $|\delta x_\sigma(L)|$ starts to 
increase, deviating from the conformal perturbation theory scaling $L^{-2}$.
We identify the finite bond-dimension $D$ effects as the
origin of this deviation.
More remarkably, we can observe a clear scaling behavior of the deviation.
That is, the shift $|\delta x_\sigma(L)|$
 scales with the system sizes as $L$ and $L^{15/4}$ for $256<L<10^4$ and $10^4<L$, respectively.
Compared with the off-critical cases in Fig.~\ref{Flow_Ising}, we realize that these scalings are identical to
those induced by the thermal and magnetic perturbations.
In other words, the relevant perturbations emerge in the TNR calculation.

Let us first discuss the $L^{15/4}$ scaling of the shift, observed for $L > 10^4$.
This can be understood as the effect of an emerging magnetic perturbation $h$ because its second order perturbation scales as $g_\sigma^2\propto L^{15/4}$.
Although the magnetic perturbation $h$ is forbidden by the $\mathbb{Z}_2$ spin-flip symmetry,
the symmetry could be broken by the limitations in the machine precision.
Once the spin-flip symmetry is broken,
the magnetic field $h$, which is a relevant perturbation, is effectively generated.
Even if the effective magnetic field $h$ is extremely small, it will be enhanced at each RG step and
eventually dominate the system at sufficiently large length scales.
This is what we observe for $L > 10^4$.
This phenomenon should be related to machine precision and not intrinsic to the algorithm.
If we are interested in a $\mathbb{Z}_2$ symmetric system, we can impose the symmetry at each step of TNR
in order to avoid this effect.

In contrast, the $L$ scaling observed for $256 < L < 10^4$ is more intrinsic.
The most relevant perturbation allowed under the $\mathbb{Z}_2$ symmetry to the critical Ising fixed-point
is the thermal operator.
Thus, we expect that the finite bond dimension effect can be mimicked by the thermal perturbation $\epsilon$
to the fixed-point Hamiltonian $H^*_\text{CFT}$.
If this is the case, the effective coefficient $g_\epsilon$ grows proportionally to $L$ as the system size $L$
is increased, because the thermal operator $\epsilon$ has the scaling dimension $1$.
According to Eq.~\eqref{fss_scaling_dimension}, this will lead to a correction proportional to $L$
in the rescaled energy level $\delta x_\sigma(L)$.
This is indeed supported by the numerical result shown in Fig.~\ref{mixed_emergent_perturbation}.

In general, the finite-$D$ effect in TNR would be described in terms of the emergence of relevant perturbation(s) to the fixed-point Hamiltonian,
which induces the finite correlation length $\xi(D)$.
In addition to the emergence of the relevant operator $\epsilon$ in the critical Ising model discussed above, a similar emergence of the relevant operator is observed in the critical three-state Potts model, as demonstrated in Appendix~\ref{Potts_sec}.

\subsection{Scaling of the emergent correlation length}
\label{sec:scaling_xi}

\begin{figure*}[tb]
\begin{center}
\includegraphics[width=140mm]{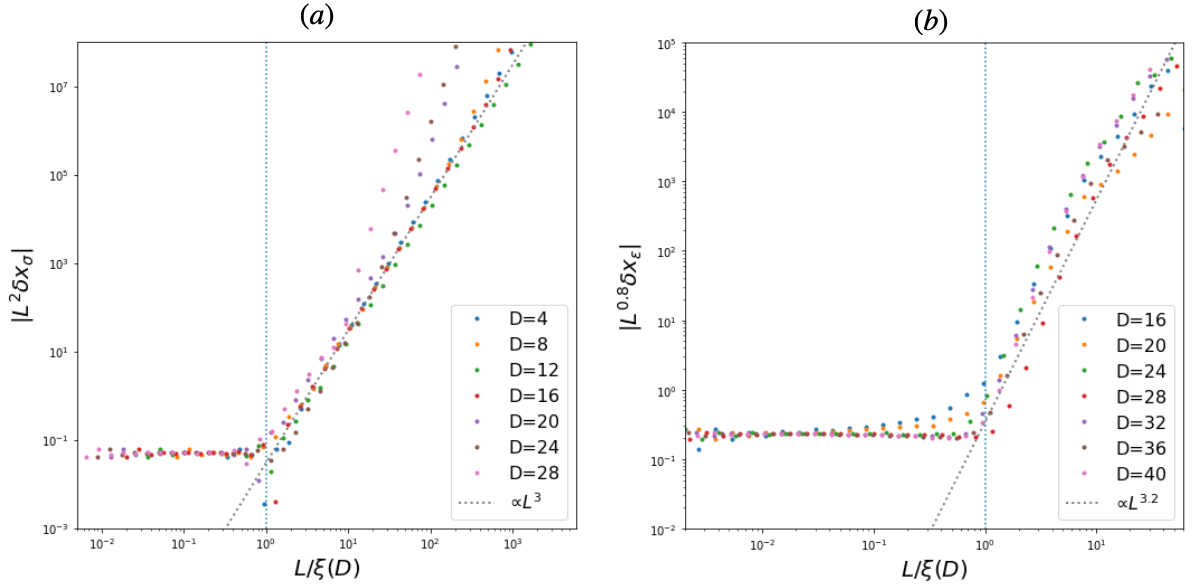}
 \caption{
 $(a)$ The scaling of the shift $\delta x_\sigma$ in TNR of the Ising model at the critical point, for various bond dimensions $D=4, \ldots, 28$.
 The vertical axis is scaled as $L^2 \delta x_\sigma$ so that it is constant when $L \ll \xi(D)$.
 When $L \ll \xi(D)$, the shift is dominated by the emergent relevant perturbation $\epsilon$; this is confirmed by the scaling $L^2 g_\epsilon \propto L^3$.
 The horizontal axis is scaled as $L/\xi(D)$, where the correlation length $\xi(D)$ is hypothesized as in Eqs.~\eqref{eq:xi_D_scaling} and~\eqref{eq:kappa_c}.
 The collapse of the data for different bond dimensions is strong evidence of the hypothesized scaling of the correlation length $\xi(D)$.
The blue dotted line indicates $L/\xi(D)=1$. We set $\xi(D) =2.0 D^\kappa$ so that $L/\xi(D)=1$ becomes the crossover scale between the finite-size scaling regime and the finite-$D$ scaling regime.
 $(b)$ Similar scaling analysis of the shift $\delta x_\epsilon$ in TNR of the three-state Potts model at the critical point, for various bond dimensions $D=16, \ldots, 40$ with $\xi(D) =0.067 D^\kappa$.
 The scaled shift $L^{0.8} \delta x_\epsilon$ behaves as a constant in the finite-size scaling regime $L/\xi(D) <1$, whereas it scales as $L^{3.2}$ in the finite-$D$ scaling regime
 $L/\xi(D) > 1$, as expected from the CFT analysis (see Appendix~\ref{Potts_sec} for details).
 The data for different bond dimensions collapse again, giving compelling evidence for the scaling of the correlation length~\eqref{eq:xi_D_scaling} and~\eqref{eq:kappa_c}
 }
 \label{FES_Ising}
 \end{center}
\end{figure*}

Now let us demonstrate that the finite correlation length $\xi(D)$ induced by the finite bond dimension $D$ in TNR
obeys the same scaling~\eqref{eq:xi_D_scaling} and~\eqref{eq:kappa_c} as in the MPS, as suggested in
Refs.~\cite{PhysRevB.89.075116,PhysRevB.104.165132}.

In Fig.~\ref{FES_Ising}, we demonstrate the scaling of the correlation length induced by the finite bond dimension in TNR of the critical Ising and the three-state Potts models.
In Fig.~\ref{FES_Ising}(a), we plot the shift $\delta x_\sigma$ in the Ising model obtained by the TNR of the Ising model at the critical point, which was also studied in Fig.~\ref{mixed_emergent_perturbation}, with the several different bond dimensions $D=4, \ldots, 28$.
Here, we rescaled the vertical axis as $L^2 \delta x_\sigma$ so that the constant behavior is observed for system size smaller than the correlation length, where the leading irrelevant perturbation (which causes $\delta x_\sigma \propto L^{-2}$) is dominant.
The deviation from the constant at larger system sizes $L$ can be attributed to the emergent relevant perturbation
$\epsilon$ induced by the finite bond dimension $D$, as discussed in the previous subsection.
This is confirmed by the $L^3$ scaling ($L^2$ times $\delta x_\sigma \propto g_\epsilon \propto L$).
Most importantly, the horizontal axis is the rescaled system size $L/\xi(D)$ using the hypothesized correlation length $\xi(D)=aD^\kappa$ given by Eqs.~\eqref{eq:xi_D_scaling} and~\eqref{eq:kappa_c}.
The collapse of the data for different bond dimensions strongly supports our hypothesis on the correlation length. Note that we roughly fit the prefactor $a$ so that the cross-over occurs at $L=\xi(D).$ 

In order to confirm the finite-$D$ scaling of the correlation length and its universality, we have also studied the three-state Potts model
at the critical point.
As an example, in Fig.~\ref{FES_Ising}(b), we plot the shift of the rescaled energy level corresponding to the energy operator $\epsilon$ in the three-state Potts model.
For this shift $\delta x_\epsilon$, the contribution from the leading irrelevant operator is $\sim L^{-4/5}$, and the dominant contribution from the emergent
relevant perturbation $\epsilon$ is expected to be proportional to ${g_\epsilon}^2 \propto L^{12/5}$. (See Appendix~\ref{Potts_sec} for details).
We rescaled the vertical axis as $L^{0.8} \delta_\epsilon$ so that it is constant in the finite-size scaling regime $L < \xi(D)$.
The horizontal axis is again the rescaled system size $L/\xi(D)$, with the correlation length $\xi(D)$ defined in Eqs.~\eqref{eq:xi_D_scaling} and~\eqref{eq:kappa_c} with
the central charge $c=4/5$ for the three-state Potts model.
The data for different bond dimensions again show a collapse, providing compelling evidence for our hypothesis on the correlation length scaling.
For $L/\xi(D) > 1$, the data fits well the expected behavior $L^{0.8} \times {g_\epsilon}^2 \propto L^{0.8} \times L^{2.4} = L^{3.2}$.

\chapter{Tensor network representation of fixed-point tensors\label{Sec:fptensor}}

TNR, while a powerful tool in the study of critical phenomena, encounters limitations when dealing with systems at criticality. A key challenge arises from the emergence of finite correlation lengths, which inherently restrict the effectiveness of TNR in capturing the true nature of the critical point. This limitation is compounded by the practical inability to simulate an infinite bond dimension ($D = \infty$), effectively creating a sort of 'no-go theorem' for TNR in accurately obtaining the true fixed-point tensor.

Fixed-point tensors, being invariant under RG transformations, are expected to exhibit certain distinct properties. Recognizing the constraints of TNR at criticality, we propose in this chapter an analytical approach aimed at uncovering the tensor elements of these fixed-point tensors. This method is designed to complement the RG techniques employed in tensor networks.

Our approach leverages analytical methods to probe deeper into the structure of fixed-point tensors, enabling us to bypass some of the limitations posed by finite correlation lengths and finite bond dimensions. By analytically determining the tensor elements of fixed-point tensors, we aim to provide a more comprehensive understanding of the behavior of systems at criticality. This method not only enhances our ability to study critical phenomena using tensor networks but also contributes to a more nuanced understanding of the universal properties associated with RG fixed points. 

\section{Fixed-point tensor}
To simulate two-dimensional statistical models, we use the tensor network methods, where the local Boltzmann weight is represented as a four-legged tensor $T^{(0)}$. We obtain the transfer matrix in the $y$-direction if we contract $L$ copies of the four-leg tensors along a circle in the $x$-direction; we obtain the partition function $Z(L, T^{(0)})$ if we contract $L\times L$ copies along the torus in the $x, y$-directions. 
In practical simulations, the exact contraction of two-dimensional tensor networks is notoriously challenging, often proving to be exponentially hard. To address this, we consider a tensor RG map that effectively coarse-grains the local tensors, as illustrated below:
\begin{equation}
    \includegraphics[width=80mm]{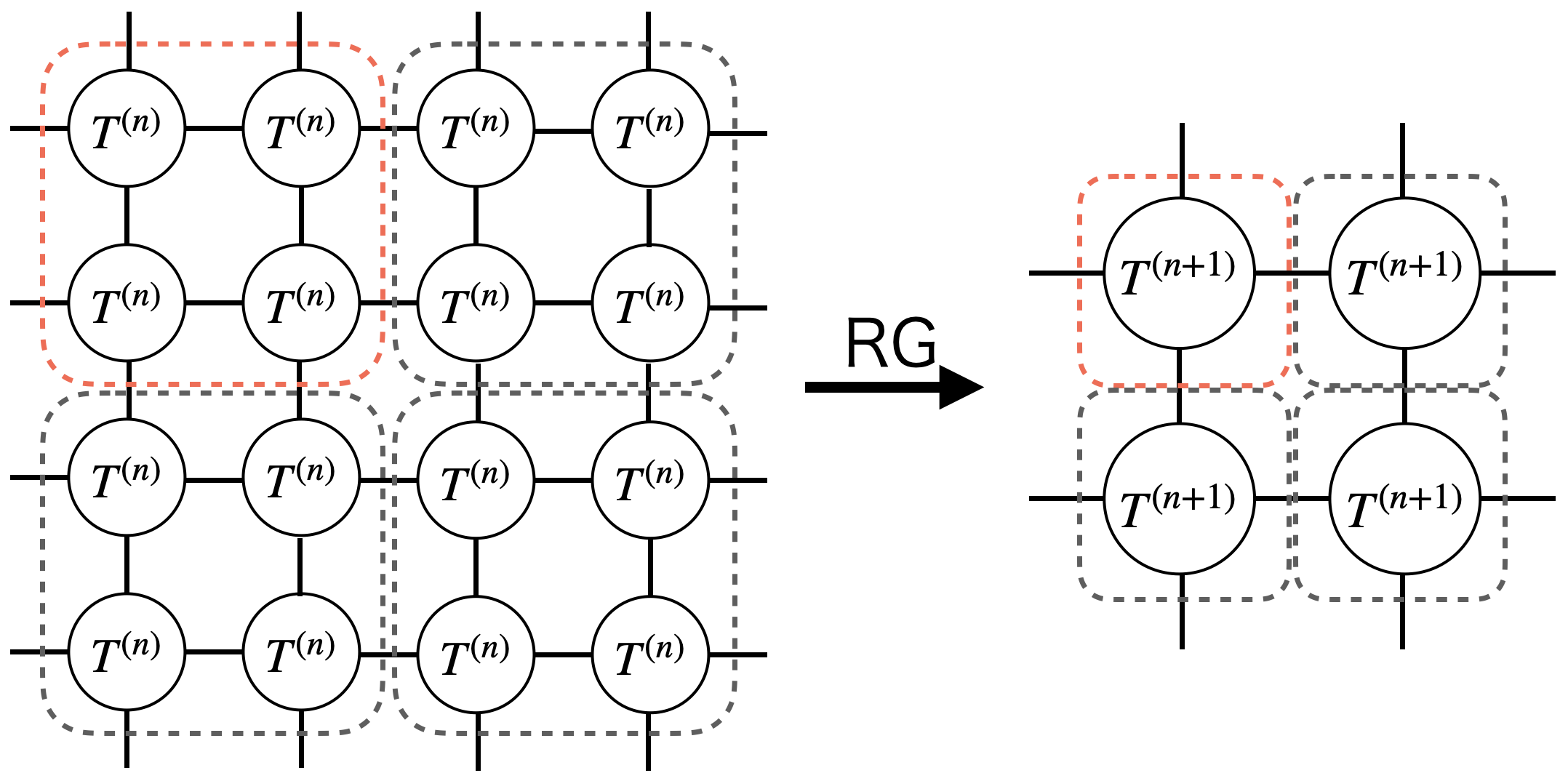}\nonumber
\end{equation}
In the initial RG step, we apply the RG map to the original Boltzmann tensor, denoted as $T^{(0)}$, to yield $T^{(1)}$. Subsequently, $T^{(1)}$ is used to generate the next tensor in the sequence. This RG map is designed to ensure that the renormalized tensor remains a close approximation of the original tensor group, while selectively discarding local entanglement. While the specifics of the technical implementation vary depending on the chosen algorithms, it is established that $T^{(n)}$ converges to a universal tensor, 
$T^*$, at critical points.
This universally convergent tensor, $T^*$, is referred to as the FP tensor. Its significance lies in its close association with the RG fixed-point, reflecting the underlying principles of scale invariance and universality in the renormalization group theory.

If the original tensor $T^{(0)}$ has $\mathrm{D}_4$ symmetry(reflection and $\pi/2$ rotation), $T^{*}$ also respects it. This allows the decomposition of the FP tensor into a pair of two identical three-leg tensors $S^*$:
\begin{align}
    \begin{matrix}
    \includegraphics[width=77mm]{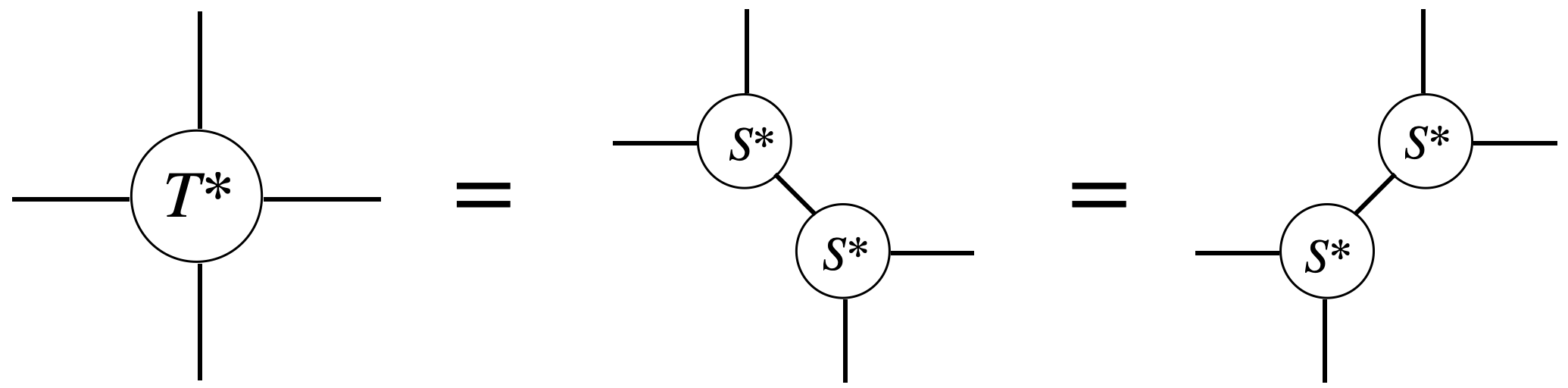}
    \end{matrix}
    \label{T_in_S}
\end{align}

The FP tensor $T^{*}$ has gauge degrees of freedom that change the basis of each leg. The insertion of the gauge transformation (unitary operators) does not change the spectral property of the FP tensor. In the following, we fix the gauge so that each index of the FP tensor is labeled by the eigenstates of the Hamiltonian $L_0+\bar{L}_0$ on a cylinder,
where $L_n$ ($\bar{L}_n$) are the standard generators of the left-moving (right-moving) Virasoro algebras.
By the state-operator correspondence, we can label these states by a set of operators $\phi_\alpha$,
among which we will find the identity operator $\phi_{1}$ with the lowest scaling dimension~\footnote{Note that the label $\alpha$ refers to both the primaries and the descendants of the Virasoro algebra.}. In tensor-network representations, the projector to this basis can be found by diagonalizing the transfer matrix as follows \cite{PhysRevB.80.155131}: 
\begin{align}
    \begin{matrix}\includegraphics[width=77mm]{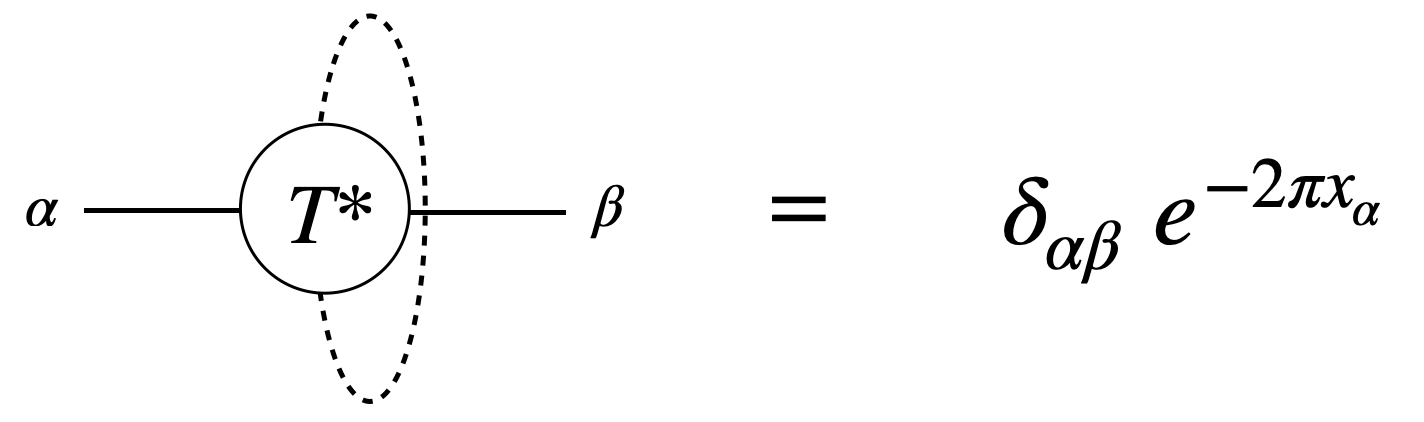} 
    \end{matrix}
    \label{x_two_point}
\end{align}
In the following, we choose the states $\alpha, \beta, \dots$ to be primary operators.
\section{Main results}
Let us now state the main results of this chapter.
First, the three-leg tensor $S^*$
 is proportional to the three-point functions of the FP CFT on the complex plane denoted as pl:
\begin{align}
    \frac{S^*_{\alpha\beta\gamma}}{S^*_{111}}&= \langle\phi_\alpha(-x_S)\phi_\beta(ix_S)\phi_\gamma(0)\rangle_{\textrm{pl}} .
    \label{x_three_point}
\end{align}
Second, the four-leg FP tensor 
 determines the four-point functions of the FP CFT as 
\begin{align}
    \frac{T^*_{\alpha\beta\gamma\delta}}{T^*_{1111}} &=\langle\phi_\alpha(-x_T)\phi_\beta(ix_T)\phi_\gamma(x_T)\phi_\delta(-ix_T)\rangle_{\textrm{pl}}.
    \label{x_four_point}
\end{align}
These equalities hold when we choose the values $x_S=e^{\pi/4}$ and $x_T=e^{\pi/2}/{2}$. $x_S$ and $x_T$ are just numbers. Do not confuse them with scaling dimensions. 

We can now reproduce the {\textit{full}} defining data for the FP CFT.
Recall that we can extract the scaling dimensions $x_{\alpha}$ operators from Eq.~\eqref{x_two_point}.
The remaining data is the OPE coefficients $C_{\alpha\beta\gamma}$ of the operators $\phi_\alpha$, which can be extracted by 
applying a conformal transformation to Eq.~\eqref{x_three_point}:
\begin{align}
\frac{S^*_{\alpha\beta\gamma}}{S^*_{111}}&=\frac{C_{\alpha\beta\gamma}}{x_S^{x_\beta+x_\gamma-x_\alpha}x_S^{x_\gamma+x_\alpha-x_\beta}(\sqrt{2}x_S)^{x_\alpha+x_\beta-x_\gamma}},\nonumber\\
&=\frac{2^{x_\gamma}C_{\alpha\beta\gamma}}{(\sqrt{2}x_S)^{x_\alpha+x_\beta+x_\gamma}} .
    \label{S_in_C}
\end{align}

Equation \eqref{T_in_S} represents
the equivalence of two different decompositions ($s$- and $t$-channels) of the four-point function into a pair of three-point functions, i.e.\ the celebrated crossing relation of the CFT.

To better understand Eqs.~(\ref{x_three_point}-\ref{x_four_point}),
we apply conformal transformations to the two equations to obtain 
\begin{align}
    \frac{S^*_{\alpha\beta\gamma}}{S^*_{111}}
    &= e^{-\frac{\pi}{4}(x_\alpha+x_\beta+x_\gamma)}\langle\phi_\alpha(-1)\phi_\beta(i)\phi_\gamma(0)\rangle_{\textrm{pl}}\label{S_three_point}, \\
    \frac{T^*_{\alpha\beta\gamma\delta}}{T^*_{1111}}
    &= \left(\frac{e^{\frac{\pi}{2}}}{2}\right)^{-x_{\textrm{tot}}}\langle\phi_\alpha(-1)\phi_\beta(i)\phi_\gamma(1)\phi_\delta(-i)\rangle_{\textrm{pl}}\label{four_point},
\end{align}
where $x_{\textrm{tot}} \equiv x_\alpha+x_\beta+x_\gamma+x_\delta$. 

Equations~(\ref{S_three_point}-\ref{four_point}) naturally arise from the following arguments. Once we fix the basis for the FP tensor, each index corresponds to the states of CFT.  Thus, the tensor elements of the FP tensor are the coefficients of each basis:
\begin{align}
    T^* = T^*_{\alpha\beta\gamma\delta}|\phi_\alpha\rangle|\phi_\beta\rangle|\phi_\gamma\rangle|\phi_\delta\rangle
\end{align}

On the other hand, the FP tensor itself is a lattice representation of the identity operator $1$. In Ref.~\cite{PhysRevLett.116.040401,PhysRevResearch.3.023048,guo2023tensor}, they confirmed that local scale-transformation could be realized using the FP tensors. 
\begin{center}
   \includegraphics[width=60mm]{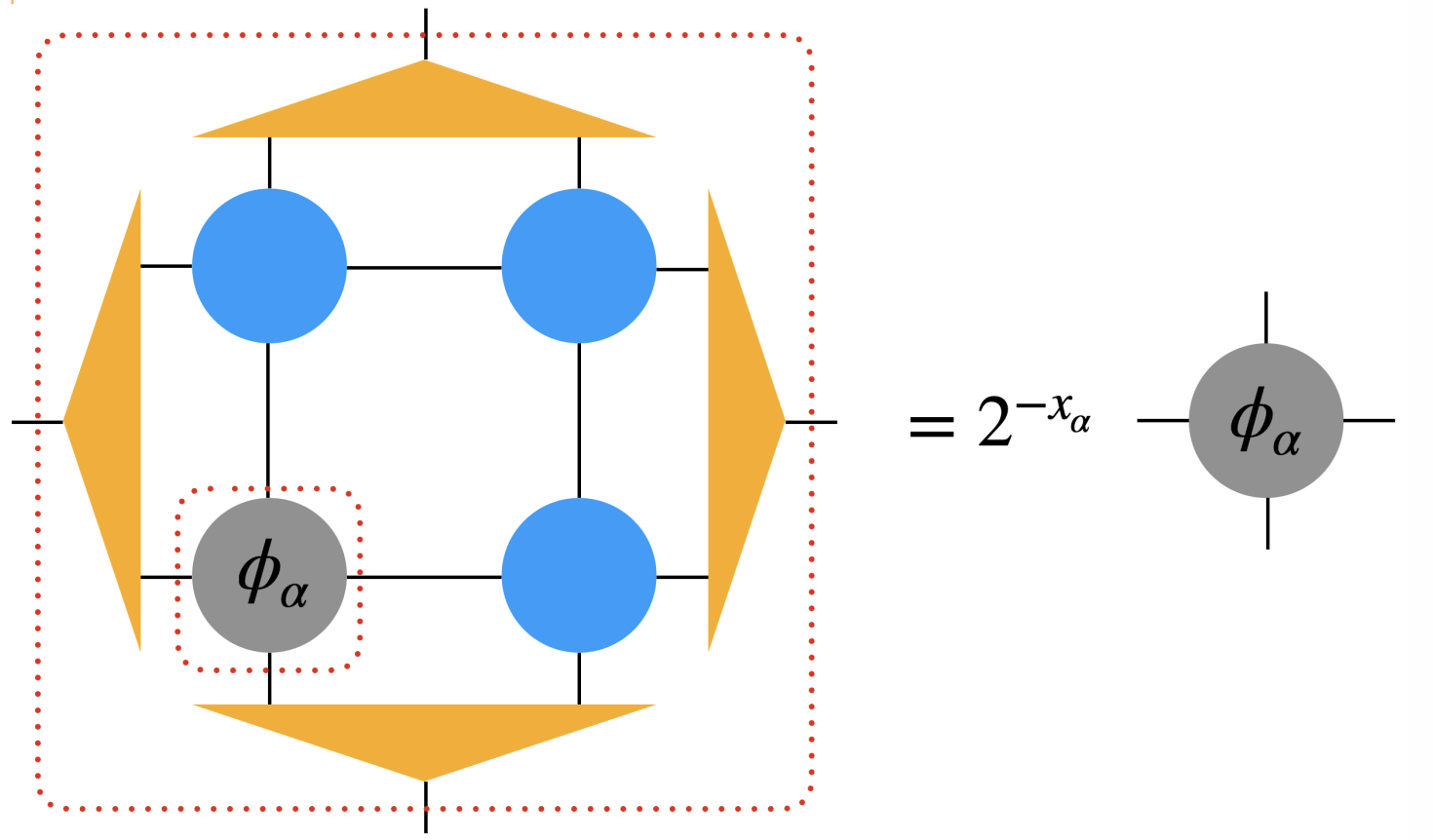} 
\end{center}
The scale transformation of a four-leg tensor, comprising the FP tensor (colored blue) and isometry (colored orange), results in primary operators emerging as eigenstates. Notably, the scale-invariant FP tensor corresponds to $x_\alpha=0$. This specific correspondence is significant, equating the FP tensor to the identity operator~\footnote{unitary CFTs have ground states corresponding to the identity operator}. This observation leads us to a conceptualization where the elements of the tensor can be expressed as an overlap between the four-leg identity operator and four one-leg primary operators as $T^*_{\alpha\beta\gamma\delta}=\langle\phi_\alpha\phi_\beta\phi_\gamma\phi_\delta|\phi_1^{4-leg}\rangle$. The same argument can be applied to the three-leg FP tensor $S^*$. To calculate these values, we employ a technique similar to the one described in the referenced literature, specifically in Ref.~\cite{PhysRevB.107.155124,PhysRevB.105.125125}.

\begin{figure}[tb]
    \centering
    \includegraphics[width=80mm]{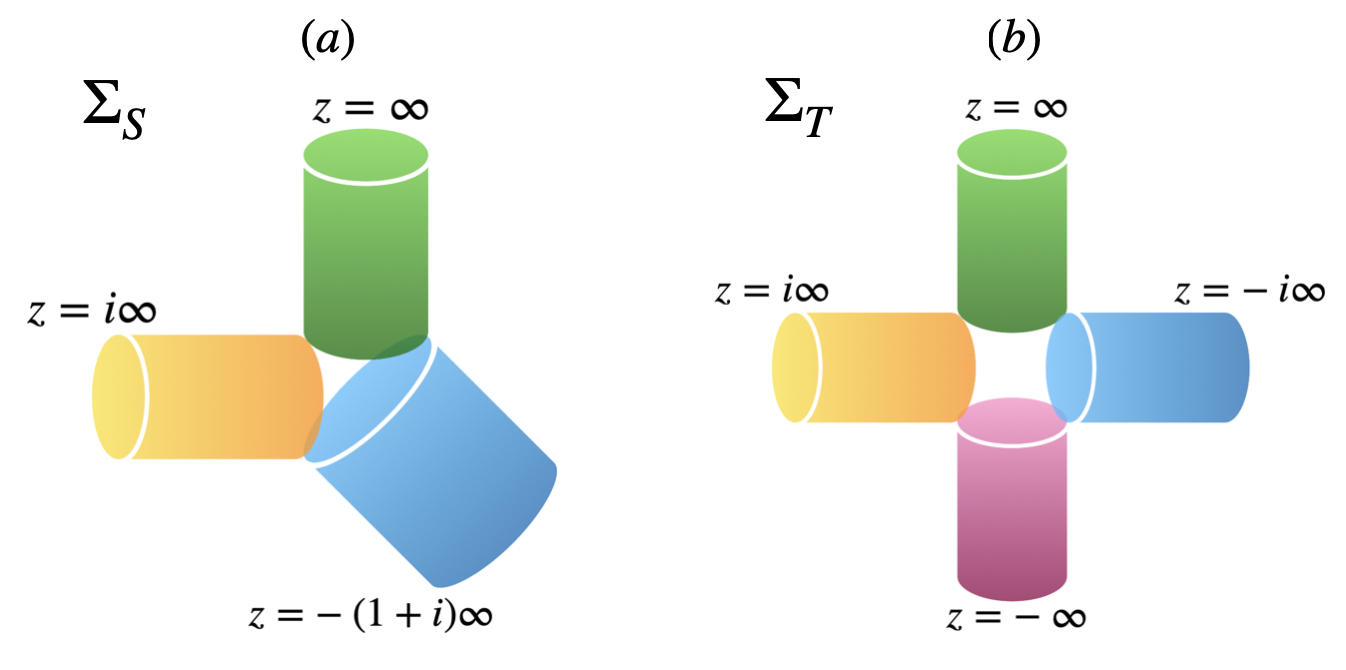}
    \caption{The path-integral representation of the tensor elements $(a)$  $S^*_{\alpha\beta\gamma}$ and  $(b)$ $T^*_{\alpha\beta\gamma\delta}$. The fixed-point tensor lives at the center of cylinders, and surrounding cylinders are bra vectors of primary fields. Since the FP tensor corresponds to the identity operator, “insertion of no operator" is illustrated as empty space. This identity operator at the origin in $z$ coordinate will be mapped to the infinity in $w$.}
    \label{Tensor_S}
\end{figure}

First, utilizing state-operator correspondence, the normalized wave function of the first index of $S^*$, for instance, is created by inserting $\phi_\alpha$ in the future infinity of the cylinder as follows:
$$|\phi^1\rangle=\left(\frac{2\pi}{L}\right)^{-x_\alpha}\lim_{z\rightarrow\infty}e^{2\pi zx_\alpha/L}\phi_\alpha(\infty)|I^{\textrm{cyl}}\rangle,$$
where $|I^{\textrm{cyl}}\rangle$ represents the ground state corresponding to the identity operator.
Subsequently, the FP tensors $S^*$ and $T^*$ can be expressed by the path integral on the manifolds $\Sigma_S$ and $\Sigma_T$, respectively, as illustrated in Fig.~\ref{Tensor_S}. Then, the FP-tensor elements are
\begin{align}
    \frac{S^*_{\alpha\beta\gamma}}{S^*_{111}}&= \langle\phi_\alpha(\infty)\phi_\beta(i\infty)\phi_\gamma(-(1+i)\infty)\rangle_{\Sigma_S},\\
    \frac{T^*_{\alpha\beta\gamma\delta}}{T^*_{1111}} &=\langle\phi_\alpha(-\infty)\phi_\beta(i\infty)\phi_\gamma(\infty)\phi_\delta(-i\infty)\rangle_{\Sigma_T}.
\end{align}
$\Sigma_S$ and $\Sigma_T$ can be mapped to the complex plane $w$ by using (generalized) Mandelstam mapping~\cite{Mandelstam:1973jk,Baba:2009ns},
\begin{align}
    z_S &= \frac{L}{2\pi}[-\ln(w-i)-i\ln(w+1)+(1+i)\ln w]\label{confmap_S},\\
    z_T &= \frac{L}{2\pi}\left[\ln\left(\frac{w+i}{w-i}\right)+i\ln\left(\frac{w-1}{w+1}\right)\right].
\end{align}
Each operator in the $z$-coordinate transforms accordingly as 
\begin{align}
\frac{S^*_{\alpha\beta\gamma}}{S^*_{111}}&=\langle\phi_\alpha(-1)\phi_\beta(i)\phi_\gamma(0)\rangle_{\textrm{pl}}\prod_{n\in(\alpha,\beta,\gamma)}|J_n|^{x_n},\nonumber\\
    \frac{T^*_{\alpha\beta\gamma\delta}}{T^*_{1111}}&=\langle\phi_\alpha(-1)\phi_\beta(i)\phi_\gamma(1)\phi_\delta(-i)\rangle_{\textrm{pl}}\prod_{n\in(\alpha,\beta,\gamma,\delta)}|J_n|^{x_n}\nonumber,
\end{align}
where $|J_n| = |\left(\frac{2\pi}{L}\right)^{-1}\lim_{z\rightarrow\zeta\infty}e^{2\pi z\zeta^*/(L|\zeta|)}|w'(z)|$, and $\zeta\infty$ is the coordinate of the index in the original manifold. The resulting $|J_n|$ are $e^{-\pi/4}$ and $2e^{-\pi/2}$, respectively, being consistent with Eqs.~(\ref{S_three_point}-\ref{four_point}). Detailed calculations are presented in the appendix.
\section{Numerical fixed point tensor}
Let us provide numerical confirmations of our main results using Levin's tensor renormalization group (TRG)~\cite{PhysRevLett.99.120601} and Evenbly's TNR~\cite{PhysRevLett.118.110504}.
TRG and TNR are numerical techniques that calculate effective $L \times L$ tensor networks. In our study, our interest lies in computing those of large system sizes to obtain a tensor that is as close as possible to the FP tensor. However, performing an exact contraction is exponentially difficult, prompting us to focus on extracting low-lying spectral properties. TRG/TNR seeks to circumvent this issue by employing the principles of the renormalization group theory. Each coarse-graining step entails decompositions and recombinations. Truncation, parameterized by the bond dimension $D$, is performed to maintain the tractability of numerical computation. However, it is important to note that this scheme is considered {\textit{exact}} when $D = \infty$, and thus, employing larger $D$ improves the numerical accuracy. Additionally, we impose special $D_4$ symmetry in TRG. The details can be found in the appendix. It is crucial to acknowledge that the TRG method is known to exhibit instabilities, primarily due to its inherent limitations in eliminating certain types of local entanglement. In contrast, TNR, which includes a local entanglement filtering process, typically demonstrates superior performance in extracting infra-red information. This enhanced capability of TNR is attributed to its more effective handling of local entanglement, making it a more robust approach for studying systems at criticality.
\begin{figure}[tb]
    \centering
    \includegraphics[width=86mm]{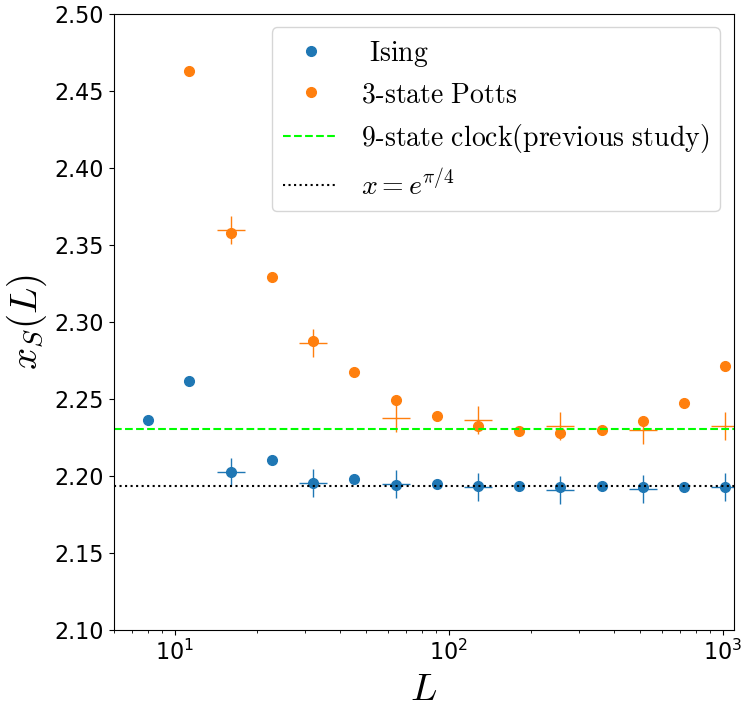}
    \caption{Estimation of $x_S(L)$ from Levin-TRG($D=96$) and Evenbly-TNR($D=40$). The values of $x(L)$ from the Ising and three-state Potts model converge to the theoretical value $x_S=e^{\pi/4}$ denoted by a black dotted line. We plot $x_S=2.23035$ obtained from Loop-TNR~\cite{PhysRevLett.118.110504} on the critical 9-state clock model~\cite{PhysRevResearch.4.023159} with a lime dashed line. The three-state Potts model exhibits a deviation for $L>100$ because simulating systems with higher central charges involves larger numerical errors.}
    \label{x_fs}
\end{figure}

\section{Tests on critical lattice models}
Let us first test the value $x_S=e^{\pi/4}$ in Eq.~\eqref{S_three_point}, by computing  $x_S$ from the critical Ising and three-state Potts models. 
Given Eq.~\eqref{S_three_point}, we can numerically compute the OPE coefficients $C_{\alpha\beta\gamma}$ from Eq.~\eqref{S_in_C}. 
We define $x_S(L)$ by solving Eq.~\eqref{S_in_C} to be 
\begin{align}
    x_S(L) \equiv \frac{1}{\sqrt{2}}\left(\frac{2^{x_\gamma}C_{\alpha\beta\gamma}}{S_{\alpha\beta\gamma}(L)}\right)^{1/(x_\alpha+x_\beta+x_\gamma)}.
\end{align}
Each model has a primary operator $\epsilon$, called the energy and the thermal operator, respectively. Since $C_{\epsilon\epsilon1}=1$, $x_S(L)$ can be computed from the finite-size three-leg tensor $S_{\epsilon\epsilon1}(L)$.

\begin{figure}[tb]
    \centering
    \includegraphics[width=86mm]{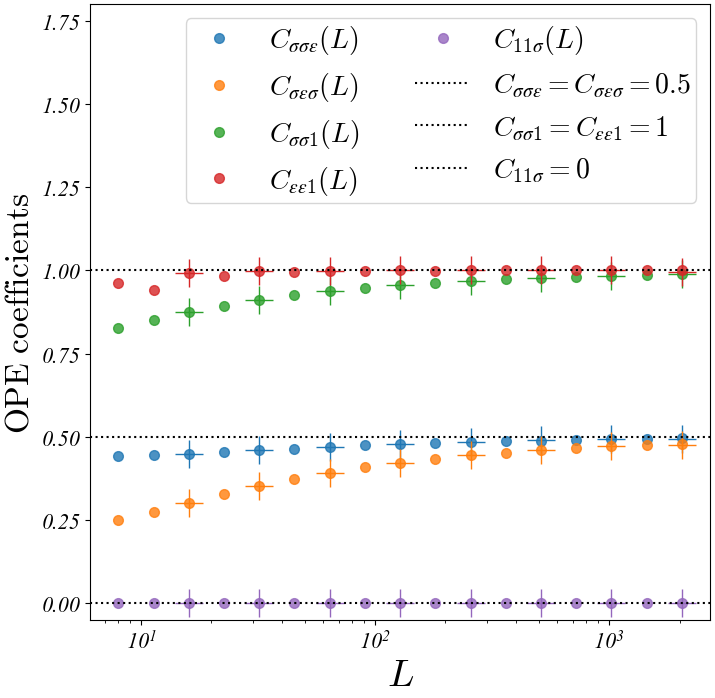}
    \caption{The OPE coefficients of the critical Ising model evaluated by setting $x_S = e^{{\pi/ 4}}$. The black dotted lines denote the theoretical values 0, 0.5, and 1. The data points, denoted by filled circles "$\circ$" and crosses "$+$," are obtained from Levin-TRG($D=96$) and Evenbly-TNR($D=40$), respectively. Relatively large finite-size effects have universal scaling as tested in Sec.~\ref{sec:finiteOPE}.}
    \label{ope_finitesize}
\end{figure}

Figure~\ref{x_fs} shows the value of $x_S(L)$ obtained from TRG and TNR at the bond dimension $D=96$ and $D=40$, respectively. The numerically derived $x_S(L)$'s for both models converge to the theoretical value of $e^{\pi/4}$. The noticeable increase in amplitude for the three-state Potts model by TRG at $L>10^2$ is attributed to the effect of the finite bond dimension and the remaining local entanglement. It is worth noting that our value for $x_S$ deviates slightly from the value $x_S=2.23035$~\footnote{Their paper showed that the three-leg FP tensor $S^{*}$ had the same structure as a three-point function by numerical experiments. In this process, they treat $x_S$ as a fitting parameter to reproduce known OPE coefficients. Our results show that they are indeed three-point functions and $x_S$ is a universal number and not a fitting parameter.} obtained in a previous study on the 9-state clock model~\cite{PhysRevResearch.4.023159}. We speculate that this minor deviation is due to the finite bond-dimension effect because higher central charges lead to more pronounced numerical errors~\cite{PhysRevB.108.024413}. For the system size $L=2048$ and bond dimension $D=96$, we ascertain $x_S = 2.193257$ for the Ising model, a value remarkably close to $e^{\pi/4} = 2.193280$.
Once we are certain of the value $x_S=e^{\pi/4}$, we can 
verify Eq.~\eqref{S_three_point} for all the OPE coefficients, which are computed from the three-leg tensor $S$ as 
\begin{align}
    C_{\alpha\beta\gamma}(L) = (\sqrt{2}\, e^{\pi/4})^{x_\alpha+x_\beta+x_\gamma}2^{-x_\gamma}S_{\alpha\beta\gamma}(L).\label{ope_fomula}
\end{align}
The results for the critical Ising model are exhibited in Fig.~\ref{ope_finitesize}. The obtained OPE coefficients are consistent with our theory with the finite-size effects of expected scaling. The finite-size effect originates from the twist operator at the branch points~\cite{PhysRevB.107.155124,PhysRevB.105.125125}, whose scaling is universal. The detailed analysis is discussed in the appendix. The same plot for the critical three-state Potts model is shown in the supplemental information in the appendix. While it has less accuracy due to the stronger finite bond dimension effect for higher central charges, the result is still consistent with the expected OPE coefficients.\\

\begin{figure}
    \centering
    \includegraphics[width=86mm]{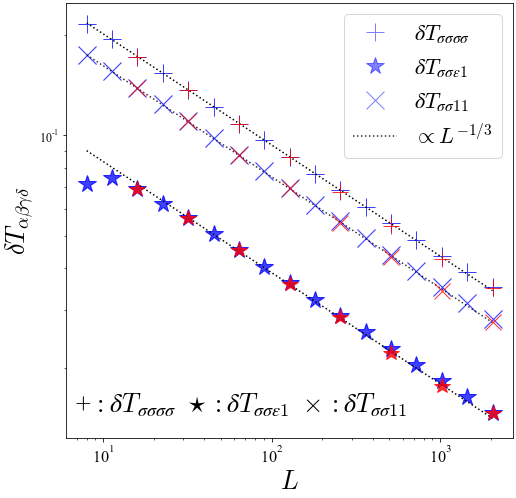}
    \caption{The finite-size effect of the fixed point tensor $\delta T_{\alpha\beta\gamma\delta} \equiv \langle\phi_{\alpha}\phi_j{\beta}\phi_{\gamma}\phi_{\delta}\rangle-T_{\alpha\beta\gamma\delta}(L)$ from Levin-TRG($D=96$, red) and Evenbly-TNR($D=40$, blue). We plot $\delta T_{\alpha\beta\gamma\delta}$ of $\sigma\sigma\sigma\sigma$(``$+$"), $\sigma\sigma\epsilon1$(``$\star$"), and $\sigma\sigma11$(``$\times$") with different colors depending on the algorithm. The difference converges to zero for $L\rightarrow\infty$ with the power-law $\sim L^{-1/3}$.}
    \label{4-point_fs}
\end{figure}

We next computed four-point tensors 
$T_{\alpha \beta \gamma \delta}$ and compared with the theoretical values from Eq.~\eqref{four_point},
where the explicit forms of the four-point functions of the critical Ising model are listed in the supplemental information in the appendix.
The result is consistent up to two digits for most tensor elements, as shown in Table~\ref{fptensor_table}. The exceptions are $T_{\sigma\sigma\sigma\sigma}$ and $T_{\sigma\sigma11}$, whose numerical values deviate approximately 5\% from the theoretical values. As for $T_{\sigma\sigma\epsilon1}$, the deviation is almost 24\%~\footnote{The TNR scheme has a similar performance at the same length-scale as seen in Fig.~\ref{4-point_fs}.}.
This discrepancy, however, can be attributed to finite-size effects and becomes negligible for infinite system sizes. To illustrate this, we define the finite-size deviation as (do not confuse with temperature)
\begin{align}
    \delta T_{\alpha\beta\gamma\delta}\equiv T^*_{\alpha\beta\gamma\delta}-T_{\alpha\beta\gamma\delta}(L).\nonumber
\end{align}
Figure~\ref{4-point_fs} presents the values of $\delta T_{\sigma\sigma\sigma\sigma}(L)$, $\delta T_{\sigma\sigma\epsilon 1}(L)$, and $\delta T_{\sigma\sigma11}(L)$ obtained from TRG calculations. A clear power-law decay with respect to the system size is observed, supporting the claim that the 
large deviations for those elements are finite-size effects. However, it is worth mentioning that the exponent closely approximates $\sim L^{-1/3}$, hinting at the existence of an underlying theory that might account for this.
\begin{table}[tb]
\caption{The comparison of the numerically-obtained fixed-point tensor $T_{\alpha \beta \gamma \delta}$ at $L=2048$ and the exact four-point function $\langle\phi_{\alpha}(-x_T)\phi_{\beta}(ix_T)\phi_{\gamma}(x_T)\phi_{\delta}(-ix_T)\rangle_{\textrm{pl}}$ of the Ising model with $x_T = e^{\pi/2}/2$. \label{fptensor_table} }
\begin{tabular}{|l|l|l|l|l}
\hline
                                  $\alpha \beta \gamma \delta$ & $T_{\alpha \beta \gamma \delta}\, (L=2048)$ & $\langle\phi_{\alpha}\phi_{\beta}\phi_{\gamma}\phi_{\delta}\rangle$ \\
                                   \hline
$1111$                             & 1          & 1                                        \\
$\sigma\sigma\sigma\sigma$         & 0.610       & 0.645                                    \\
$\sigma\sigma\epsilon\epsilon$     & 0.0714      & 0.0716                                    \\
$\sigma\epsilon\sigma\epsilon$     & 0.000      & 0                                    \\
$\epsilon\epsilon\epsilon\epsilon$ & 0.0168      & 0.0168 \\
$\sigma\sigma\epsilon1$         & 0.0618       & 0.0765                                    \\
$\sigma\epsilon\sigma1$         & 0.133       & 0.140                                    \\
$\sigma\sigma\sigma1$         & 0.000       & 0                                    \\
$\epsilon\epsilon\epsilon1$         & 0.001       & 0                                    \\
$\sigma\sigma11$         & 0.708       & 0.736                                    \\
$\sigma1\sigma1$         & 0.639       & 0.675                                    \\
$\epsilon\epsilon11$         & 0.0863       & 0.0864                                    \\
$\epsilon1\epsilon1$         & 0.0439       & 0.0432                                    \\
$\epsilon\sigma11$         & 0.000       & 0                                  \\
\hline
\end{tabular}
\end{table}

\chapter{Conclusion and discussion}
\begin{figure}[tb]
    \centering
    \includegraphics[width=140mm]{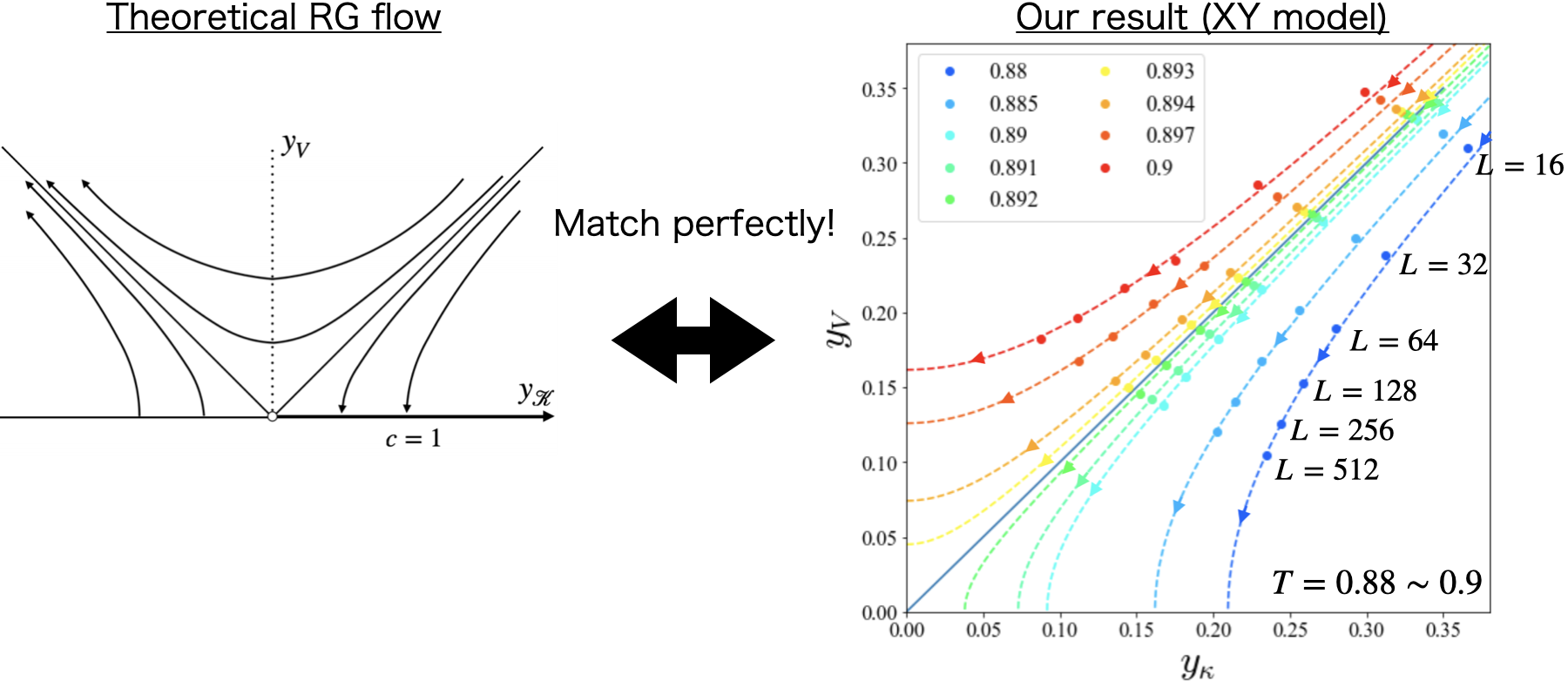}
    \caption{The RG flow of the classical XY model in two dimensions stands as a quintessential example of a topological phase transition. This particular type of RG flow is commonly referred to as the Kosterlitz RG flow. The right panel is numerically obtained RG flow in a similar manner. However, a key distinction lies in the consideration of up to third-order perturbations in our computational approach. The deviation in the smaller system size is due to the irrelevant perturbations. Further details can be found in Ref.~\cite{PhysRevB.104.165132}. }
    \label{XY_flow}
\end{figure}

In the first part of Chapter \ref{Sec:RGflow}, we discussed a method for computing the coupling constants using renormalized tensors based on the finite-size scaling theory of CFT. By plotting the resulting values at each scale, we were able to visualize the RG flow, and we confirmed that the theoretical RG flows, as shown in Fig.~\ref{Flow_Ising}, are consistent with the Ising models. Our methodology has undergone further validation through its application to the three-state Potts model and the XY model, as detailed in the appendix and in Ref.\cite{PhysRevB.104.165132}, respectively. In the case of the XY model, particular care is required in the perturbative calculations due to the marginal nature of the running coupling constants under consideration. Despite these complexities, we successfully demonstrate that the RG flow in the XY model aligns with the Kosterlitz RG flow\cite{Kosterlitz_1974}, as confirmed by our inclusion of third-order perturbative effects. This finding is depicted in Fig.~\ref{XY_flow}, offering a visual representation of the Kosterlitz RG flow in the context of the XY model.

Our method has the advantage of being able to extract both ultraviolet and infrared information, making it a valuable tool for investigating gapped and crossover systems. We applied this idea to reveal the asymptotic freedom of the Heisenberg and exotic cross-over behavior of $\mathrm{RP^2}$ models in Ref.~\cite{PhysRevE.106.014104}.

In the second part of the chapter, applying the methodology developed in the first part, we explored the impact of finite bond-dimension $D$ on the RG flow.
The finiteness of the bond-dimension results in a finite correlation length $\xi(D)$,
or equivalently in a non-zero gap in the energy spectrum of the corresponding one-dimensional quantum system.
We find that this gap formation can be attributed to the emergence of a relevant perturbation enforced by the finite bond dimension.
This is demonstrated by the RG flow of the emergent relevant coupling.

The finite-size scaling of TNR shows a crossover at $L \sim \xi(D)$, above which the system is governed by the finite correlation length.
The correlation length $\xi(D)$ induced by the finite bond dimension in TNR
shows the same scaling~\eqref{eq:xi_D_scaling},~\eqref{eq:kappa_c} as the correlation length of MPS.
While such scaling in TNR was suggested earlier in Refs.~\cite{PhysRevB.104.165132}, in this thesis, we presented more convincing evidence.

\begin{figure}[tb]
    \centering
    \includegraphics[width=86mm]{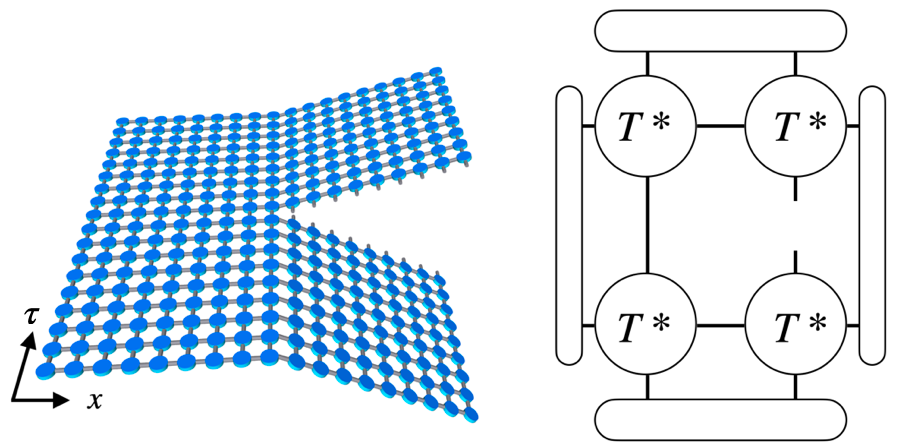}
    \caption{
    (Left panel) A schematic picture of the reduced density matrix $\rho_A$ for a bipartition of the system in the path integral picture.
    The uncontracted legs correspond to the indices of the reduced density matrix.
  (Right panel) Each of the four quadrants of the space-time in the left panel may be replaced by the renormalized tensor in TNR with appropriate boundary conditions.  }
    \label{reduced_rho}
\end{figure}

Although we do not have a mathematical proof for the scaling of $\xi(D)$ in TNR at this point, it may be natural from the following point of view.
Besides the construction of the transfer matrix by contracting horizontal legs, the renormalized tensor obtained in TNR can give the
corner transfer matrix by contracting the upper and left legs.
The same finite-$D$ scaling~\eqref{eq:xi_D_scaling},~\eqref{eq:kappa_c} as in MPS was observed in corner transfer matrix renormalization
group (CTMRG)~\cite{NISHINO199669,PhysRevE.96.062112,PhysRevE.101.062111}.
Moreover, the entanglement spectrum for the half-bipartition of the system of length $2L$ can be related to
a contraction of four renormalized tensors of linear size $L$~\cite{CalabreseLefevre2008}, as shown in Fig.~\ref{reduced_rho}.
These relations are suggestive of the identical scaling of $\xi(D)$ in MPS, CTMRG, and TNR as we have observed.

Our study highlights the importance of considering the impact of the finite bond dimension in the TNR-type approach.
In particular, a direct study of the thermodynamic limit with TNR would be prone to errors due to the finite correlation length $\xi(D)$ imposed by the finite bond dimension.
As a resolution of this problem, we have demonstrated that accurate data for the thermodynamic limit can be extracted by finite-size scaling of TNR spectra obtained for
system sizes smaller than $\xi(D)$, combined with conformal field theory. 
Even with this limitation, the tractable system size is greatly increased from $\sim \log{D}$ with exact diagonalization to $\xi(D) \sim D^\kappa$ in TNR. \\

\par{}Meanwhile, the unattainability of true critical fixed-point tensors in numerical simulations does not prohibit their calculation by analytical approach. In Chapter \ref{Sec:fptensor}, we have successfully computed explicit elements of the critical fixed-point tensor by field theory, which we have identified as the four-point function in CFT. This achievement enables us to directly extract the OPE coefficients from these tensor elements. When combined with the scaling dimensions derived from the transfer matrix, this approach allows us to determine a complete set of CFT data for any critical unitary lattice model. 

One of the key strengths of our method for extracting OPE coefficients is its simplicity and effective finite-size scaling properties. These characteristics make our approach particularly useful for extrapolating other, more complex OPE coefficients. As an example of this method's application, we have computed the OPE coefficients for the three-state Potts model, detailed in Sec.~\ref{sec:OPE_three_state_potts}.

Determining the infrared CFTs for lattice models stands as a cornerstone challenge in theoretical physics. This problem has garnered widespread attention across various domains of physics, including high energy physics, condensed matter physics, and statistical physics, due to its fundamental nature and broad implications.

In our research, we have introduced a novel solution to this long-standing issue, employing a synergistic blend of analytical and numerical techniques. This approach not only addresses the immediate problem at hand but also opens new avenues for exploration and inquiry in the realms of formal conformal field theories and numerical tensor networks. Our work, therefore, not only contributes to the resolution of a decade-old problem but also poses many intriguing questions for future research, potentially leading to significant advancements in both theoretical and applied physics

Looking to the future, exploring the tensor elements corresponding to descendants within the CFT framework presents an exciting avenue for further research. This exploration could deepen our understanding of the intricate structures within lattice models. In fact, following our work, Ref.~\cite{cheng2023exact} has examined descendants in a similar context, indicating the growing interest and potential in this area of study.

Ultimately, our approach, which combines the exact treatment of RG and fixed-point tensors in tensor networks, could significantly contribute to a more profound understanding of universality in lattice models. By leveraging these advanced methods, we open up new possibilities for exploring and elucidating the complex and fascinating behaviors inherent in critical systems.

\printbibliography[heading=bibintoc]
\printindex
\appendix
\chapter{Appendix}
\section{Perturbation theory on the finite-size spectrum}\label{sec:perturbation}
In the realm of effective field theory, particularly for lattice models or condensed matter systems, the theory often encompasses various perturbations to the Conformal Field Theory (CFT). Let us consider the effective Hamiltonian as follows:
\begin{equation}
 \hat{H}  = \hat{H}^* + \sum_j g_j \int dv \hat{\Phi}_j(v) .
\end{equation}
In this formulation, $H^{*}$ represents the Hamiltonian of the unperturbed CFT, and the additional terms correspond to various perturbations, each characterized by a coupling constant $g_j$ and a field operator $\hat{\Phi}_j(v)$.
When all these perturbations are deemed irrelevant in the RG context, the behavior of the system asymptotically approaches that described by the CFT in the large-distance or low-energy limit. However, it is crucial to note that these irrelevant perturbations can still significantly influence the finite-size spectrum of the system. This aspect was extensively studied by Cardy, who delved into the effects of such perturbations on the finite-size spectrum~\cite{cardy1984conformal}. His work provides a deeper understanding of how minor deviations from the idealized CFT can have measurable impacts in finite systems.

Now, we apply the standard Rayleigh-Schrödinger perturbation theory to our general Hamiltonian framework:
\begin{equation}
\calH = \calH_0 + V,
\end{equation}
where $V$ is treated as a perturbation. The unperturbed eigenstates are defined as
\begin{equation}
\calH_0 | n^{(0)} \rangle = E_n^{(0)} | n^{(0)} \rangle,
\end{equation}
and the perturbative expansion of the energy eigenvalue is given by
\begin{equation}
E_n = E_n^{(0)} + \epsilon E_n^{(1)} + \epsilon^2 E_n^{(2)},
\end{equation}
where the first-order correction is
\begin{align}
E_n^{(1)} = \langle n^{(0)} | V | n^{(0)} \rangle.
\end{align}

In this context, the unperturbed eigenstate corresponds to a primary state of CFT:
\begin{equation}
|n^{(0)} \rangle = | \Phi_n \rangle,
\end{equation}
and the perturbation $V$ is represented as
\begin{equation}
V = \sum_j g_j \int dv \hat{\Phi}_j(v),
\end{equation}

Using conformal mapping for the three-point correlation function, we obtain
\begin{equation}
\langle \Phi_i | \hat{\Phi}j(v) | \Phi_k \rangle = C_{ijk} \left( \frac{2\pi}{L} \right)^{x_j} e^{2\pi i (s_i-s_k)v/L}.
\end{equation}
Thus, for the matrix element of $V$, we have
\begin{align}
\langle l^{(0)} | V | n^{(0)} \rangle &= \sum_j g_j \int_0^L dv \langle \Phi_l | \hat{\Phi}j(v) | \Phi_n \rangle \\
&= \delta{s_l, s_n} \sum_j g_j C_{njl} L \left(\frac{2\pi}{L}\right)^{x_j} \\
&= \delta_{s_l, s_n} 2 \pi \sum_j g_j C_{njl} \left(\frac{2\pi}{L}\right)^{x_j-1}.
\end{align}
Hence, for $l=n$, the first-order energy correction is
\begin{align}
E_n^{(1)} = 2 \pi \sum_j g_j C_{nnj} \left(\frac{2\pi}{L}\right)^{x_j-1}.
\end{align}

Therefore, to the first order in perturbation theory, the energy difference relative to the ground state is
\begin{align}
E_n - E_0 = \frac{2\pi}{L} \left( x_n + 2\pi \sum_j g_j C_{nnj} \left(\frac{2\pi}{L}\right)^{x_j-2} \right).\label{eq.1storder}
\end{align}

When the perturbations are irrelevant ($x_j > 2$), they give
subleading corrections to the CFT scaling~\eqref{eq.1storder}.
In contrast, relevant perturbations with $x_j < 2$ eventually
dominates the CFT scaling for a sufficiently
large system size $L$, signalling
a breakdown of the perturbation theory.
Nevertheless, the perturbation theory can be useful for
small perturbations (at the microscopic scale) when the system size
is not very large.
When $x_j=2$, the perturbation is marginal.

The formula~\eqref{eq.1storder} may be interpreted in an
alternative way.
We can apply the scale transformation so that
the length of the system becomes $2\pi$.
The scaled gap is  $L (E_n - E_0)/(2\pi)$.
Applying the perturbation theory to the system on the ring
of radius $1$,
\begin{equation}
\frac{L}{2\pi} (E_n - E_0) = x_n + 2\pi \sum_j g_j(L) C_{nnj} .
\end{equation}
Using the scale-dependent coupling constant
\begin{equation}
 g_j(L) = \left(\frac{2\pi}{L}\right)^{x_j-2} g_j,
\label{eq.g_L}
\end{equation}
where $g_j$ is the ``bare'' value of the coupling constant that
we find in Eq.~\eqref{eq.1storder}.

\begin{table}[tb]
\begin{tabular}{ccc}
Symbol       & Dimension      & Meaning     \\ \hline
$I$ & 0              & identity    \\
$\epsilon$   & $\frac{2}{5}$  & thermal op. \\
$\sigma$     & $\frac{1}{15}$ & spin        \\
$X$          & $\frac{7}{5}$  &             \\
$Y$          & ${3}$  &             \\
$Z$          & $\frac{2}{3}$  &             \\ 
\hline
\end{tabular}
\caption{A set of primary operators of the three-state Potts model.\label{potts_operators}}
\end{table}
\section{Finite-entanglement scaling of the three-state potts model.\label{Potts_sec}}
\begin{figure*}[tb]
    \centering
    \includegraphics[width=130mm]{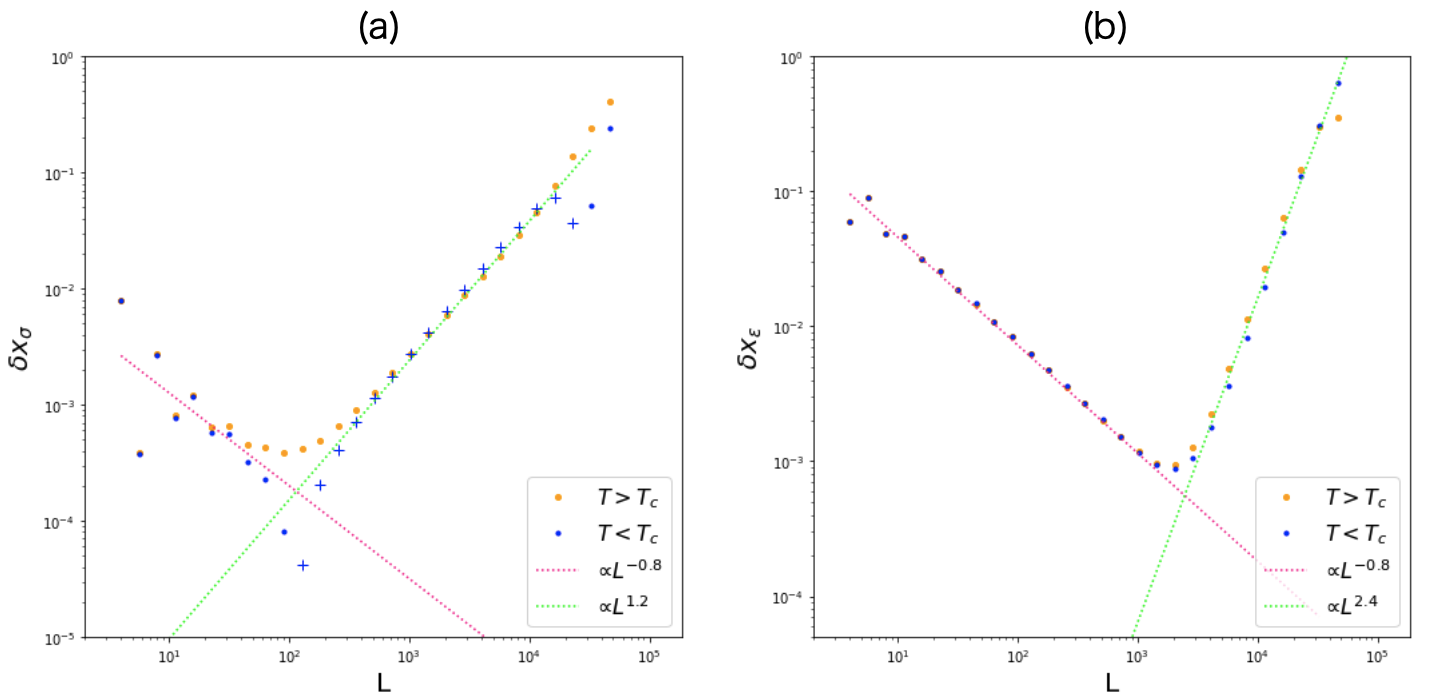}
    \caption{The size dependence of the (a)$\delta{x}_\sigma$ and (b)$\delta{x}_\epsilon$ at $T=0.999995T_c$ and $T=1.000005T_c$. The pink and green dotted lines denote $L^{-0.8}$, (a)$L^{1.2}$, and (b)$L^{2.4}$ fittings respectively. For the low-temperature phase, the sign of $\delta{x}_\sigma$ is negative at $L>100$. The dip on the left panel around $L\sim10^2$ corresponds to the zero point of Eq.~\eqref{Potts_perturbation}. (b) The finite-size effect to the $x_\epsilon$ suffers less from
    $T_{\text{cyl}}^2+\bar{T}_{\text{cyl}}^2$ in amplitude. The scaling of Eq.~\eqref{Potts_perturbation2} is clearly observed.}
    \label{Potts_sigma}
\end{figure*}
\subsection{Definition of the model and its CFT}
We can further verify the emergence of relevant perturbations by applying it to the three-state Potts model. It is a natural extension of the Ising model to the $\mathbb{Z}_3$ symmetry, and the Hamiltonian is 
\begin{align}
    H=-\sum_{\langle i,j\rangle}\delta_{s_i,s_j},
\end{align}
where $s_i$ takes 0, 1, and $-1$. It has a phase transition of $\mathbb{Z}_3$ symmetry breaking at $T_c=1/\log(1+\sqrt{3})$.
The critical theory of the three-state Potts model is another type of the minimal model $\mathcal{M}(6,5)$ with $c=\frac{4}{5}$\citep{francesco2012conformal,DOTSENKO198454}. A set of primary operators are shown in Table.~\ref{potts_operators}. 

As opposed to the Ising model, there are off-diagonal operators as $\Phi_{\frac{2}{5},\frac{7}{5}},\ \Phi_{\frac{7}{5},\frac{2}{5}}$ and $\Phi_{3,0},\ \Phi_{0,3}$(currents). 
\par{} Let us first examine the RG flow in a gapped system. Similar to the Ising model, the phase transition is identified by spontaneous symmetry breaking. The high-temperature phase is a trivial phase, whereas the low-temperature region is $\mathbb{Z}_3$ symmetry breaking phase. Thus, the fixed-point tensor is a stacking of three states with their $\mathbb{Z}_3$ charge $0,\ -1$, and 1.
\subsection{Construction of the effective Hamiltonian}
\par{}The RG flow can be seen by investigating the scaling dimensions. For instance, we can take the spin operator $\sigma=\Phi_{\frac{1}{15},\frac{1}{15}}$ and plot the value of $\delta{x}_\sigma=x_\sigma(L)-\frac{2}{15}$. Similarly, as in the Ising model, there is competition between irrelevant and relevant operators: $X=\Phi_{\frac{7}{5},\frac{7}{5}}$ and $\epsilon=\Phi_{\frac{2}{5},\frac{2}{5}}$. The thermal operator separates the $\mathbb{Z}_3$ symmetry-breaking phase from the trivial one. The finite-size corrections of $X$ and $\epsilon$ to $x_\sigma$ are $L^{-0.8}$ and $L^{1.2}$, respectively.
The fusion rules are $\sigma\times\sigma=1+\epsilon+\sigma+ X + Y + Z $, $\epsilon\times\epsilon=1+X$, and $\epsilon\times\sigma=\sigma+Z$.
Hence, $\delta{x}_\sigma$ has the following form:
\begin{align}
    \delta{x}_\sigma=2\pi C_{\sigma\sigma X} g_X\left(\frac{L}{2\pi}\right)^{-0.8}+2\pi C_{\sigma\sigma \epsilon} g_\epsilon\left(\frac{L}{2\pi}\right)^{1.2}. \label{Potts_perturbation}
\end{align}
On the other hand, the perturbation of $\epsilon$ appears as a second-order term for $\delta x_\epsilon$ because the fusion rule says $\epsilon\times\epsilon=1+X$. Consequently, $\delta x_\epsilon$ can be computed as
\begin{align}
\delta{x}_\epsilon=2\pi C_{\epsilon\epsilon X} g_X\left(\frac{L}{2\pi}\right)^{-0.8}+\alpha g^2_\epsilon \left(\frac{L}{2\pi}\right)^{2.4},\label{Potts_perturbation2}
\end{align}
where $\alpha$ is a constant determined from the second-order calculation.
\begin{figure}[tb]
    \centering
    \includegraphics[width=86mm]{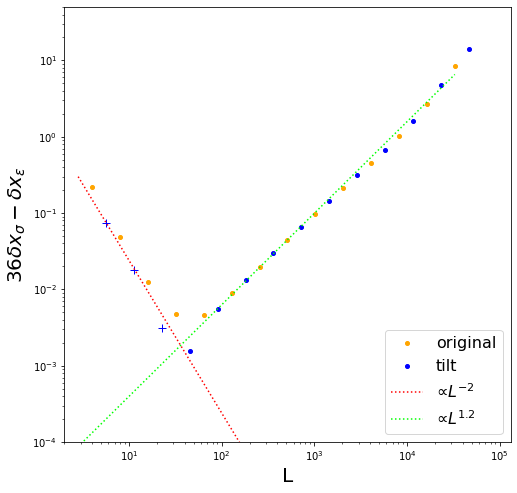}
    \caption{$36\delta x_\sigma-\delta x_\epsilon$ for the high temperature phase. “+” is used when the sign is negative. The red dotted line denotes the $L^{-2}$ fitting while the light green one is just a relevant $L^{1.2}$ contribution from $\epsilon$. Loop-TNR rotates the lattice by $\frac{\pi}{4}$ at each RG step, and the tilted system is plotted with the blue dots\label{Potts_irrelevant}.}
\end{figure}

Figure.~\ref{Potts_sigma} shows the computed $\delta x_\sigma$ by TNR. As expected, it exhibits the competition between irrelevant and relevant operators. The sign of $g_\epsilon$ is the opposite between two phases, which is a manifest indication of the RG flow in the opposite direction due to the thermal operator. $x_\sigma$ has doubly degenerate states with $\mathbb{Z}_3$ charge $\pm 1$. In the low-temperature phase, these two states flow to $x_\sigma(L)\rightarrow 0$, and the fixed-point tensor becomes three-fold degenerate.
As for the irrelevant perturbation, there seems to be a discrepancy between $\delta x_\sigma$ in Fig.~\ref{Potts_sigma} and Eq.~\eqref{Potts_perturbation}. The data points are scattered for small system sizes and not precisely on the fitting lines. This is due to the leading irrelevant operator we have not considered.
We can identify it as $T_{\text{cyl}}^2+\bar{T}_{\text{cyl}}^2$ as followings.  Just as we did in the left panel of Fig.~\ref{Flow_Ising}, the contributions from $g_X$ can be eliminated by combining $\delta x_\sigma$ and $\delta x_\epsilon$. The OPE coefficients for the three-state Potts model are known, and the ratio of the two OPE coefficients is ${C_{\epsilon\epsilon X}}/{C_{\sigma\sigma X}}=36$~\cite{DOTSENKO1984312,DOTSENKO1985691,DOTSENKO1985291,FUCHS1989303,esterlis2016closure}. Thus, the origin of the "scattering" shall be observed by plotting $36\delta x_\sigma-\delta x_\epsilon$.

\par{}Figure.~\ref{Potts_irrelevant} displays the result for the high-temperature phase. It is now obvious that the scattering of Fig.~\ref{Potts_sigma} comes from the $L^{-2}$ perturbation denoted with the red dotted line. Also, it has a conformal spin $s$ because it flips a sign at each step and $s\equiv4$ (mod 8)~\footnote{For each iteration, the lattice rotates by 45 degrees, and it corresponds to the conformal transformation $w=e^{\frac{i\pi}{4}}z$ on a complex plane.
As the irrelevant perturbations $T_{\text{cyl}}^2$ and $\bar{T}_{\text{cyl}}^2$ have a conformal spin $4$ and $-4$, they get an additional factor $(e^{\frac{i\pi}{4}})^4=(e^{\frac{i\pi}{4}})^{-4}=-1$ for an odd number of steps. We can see this by plotting the data from even steps (original) and odd steps (tilt) separately.}. As a result, we can conclude the irrelevant operator has the conformal weights as $(h,\bar{h})=(4,0)$ and $(0,4)$,
which are $T_{\text{cyl}}^2$ and $\bar{T}_{\text{cyl}}^2$. Finally, the effective Hamiltonian of the critical three-state Potts model on the square lattice can be constructed as 
\begin{align}
    H=H^*_{Potts}+\int_0^Ldx\left[g_X\Phi_{\frac{7}{5},\frac{7}{5}}(x)+g_T (T_{\text{cyl}}^2+\bar{T}_{\text{cyl}}^2)\right].
\end{align}

\subsection{Finite-Entanglement scaling}
\par{}At the critical temperature of the Ising model, the finite-$D$ effect proves to be a perturbation from the thermal operator. Let us verify it for the critical three-state Potts model. Due to the irrelevant perturbations from $T_{\text{cyl}}^2+\bar{T}_{\text{cyl}}^2$, the finite-$D$ effects are clearer for $x_\epsilon(L)$ as seen in Fig.~\ref{Potts_sigma}($b$). This is shown in Fig.~\ref{FES_Ising} of the main text. Here, we demonstrate that $\delta x_\sigma(L)$ also shows the universal behavior with $L/\xi(D)$.  
\begin{figure}[tb]
    \centering
    \includegraphics[width=86mm]{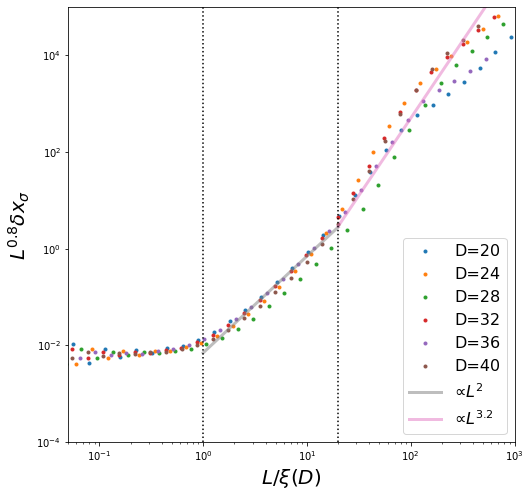}
\caption{Rescaled $\delta x_\sigma$ by $\xi(D)=D^\kappa$ at the critical temperature. The resulting data collapse onto a universal function that is independent of $L/\xi(D)$. If $L/\xi(D) < 1$, the system is in the FSS region, while if $L/\xi(D)\geq 1$, it is in the FES region. In the FES region, the scaling of the first-order and second-order perturbations are indicated by a gray and pink line, respectively. $x_\sigma$ is computed as an average value of the first and second excitation energy.}
    \label{FES_Potts}
\end{figure}
Figure.~\ref{FES_Potts} shows the rescaled correction to $\delta x_\sigma(L)$. For $L>\xi(D)$, the perturbation grows as $L^{2}$ denoted by a gray line, which means that the emergent perturbation scales as $L^{1.2}$.  Compared with Eq.~\eqref{Potts_perturbation}, it is clear that the emergent perturbation is from the thermal operator. However, as the system size increases, the second-order perturbation becomes predominant as shown with a pink line. As $\epsilon$ is the most relevant operator that is permitted by symmetry, it supports our conjecture stated in the main text.
\section{Supplemental information on the fixed-point tensor}
\subsection{Conformal mapping of S}
The three-leg tensor $S^*_{\alpha\beta\gamma}$ represents the three-sided thermofield double state~\cite{PhysRevB.105.125125} corresponding to the geometry in Fig. 1(a) in the main text. This manifold $\Sigma_S$ is mapped to the plane by a conformal mapping 
\begin{align}
    z = \frac{L}{2\pi}[-\ln(w-i)-i\ln(w+1)+(1+i)\ln w],
    \label{z_to_w}
\end{align}
which maps 
the three points in $\Sigma_S$, $(z_1,z_2,z_3) = (\infty,i\infty,-(1+i)\infty)$, to $(w_1,w_2,w_3)=(i,-1,0)$. 
Then, the tensor element is
\begin{align}
\frac{S^*_{\alpha\beta\gamma}}{S^*_{111}}=|J_1|^{x_\alpha}|J_2|^{x_\beta}|J_3|^{x_\gamma}\langle\phi_\alpha(-1)\phi_\beta(i)\phi_\gamma(0)\rangle_{\textrm pl},
\end{align}
where $J_i$ is the Jacobian of the conformal mapping \eqref{z_to_w}. The initial states are
\begin{align}
|\phi^1\rangle&=\left(\frac{2\pi}{L}\right)^{-x_\alpha}\lim_{z\rightarrow\infty}e^{2\pi zx_\alpha/L}\phi_\alpha(z)|I^{\textrm cyl}\rangle,\nonumber\\
|\phi^2\rangle&=\left(\frac{2\pi}{L}\right)^{-x_\beta}\lim_{z\rightarrow i\infty}e^{-i2\pi zx_\beta/L}\phi_\beta(z)|I^{\textrm cyl}\rangle,\nonumber\\
|\phi^3\rangle&=\left(\frac{\sqrt{2}\pi}{L}\right)^{-x_\gamma}\\
&\lim_{z\rightarrow{(-i-1)}\infty}e^{\frac{(i-1)}{\sqrt{2}}\frac{2\pi}{\sqrt{2}L}zx_\gamma}\phi_\gamma(z)|I^{\textrm cyl}\rangle\nonumber.
\end{align}
The Jacobian can be computed as 
\begin{align}
    |J_1| &= \left|\left(\frac{2\pi}{L}\right)^{-1}\lim_{z\rightarrow\infty}e^{2\pi z/L}w'(z)\right|\nonumber\\
    &=\left|\left(\frac{2\pi}{L}\right)^{-1}\lim_{w\rightarrow i}e^{2\pi z/L}\left(\frac{dz}{dw}\right)^{-1}\right|.
    \label{J1}
\end{align}
Using Eq. (10) in the main text, the first and second term is 
\begin{align}
    e^{2\pi z/L} &= \exp[\ln\frac{w}{w-i}+i\ln\frac{w}{w+1}],\\
    \frac{dz}{dw} &= \frac{L}{2\pi}\left[-\frac{1}{w-i}-\frac{i}{w+1}+\frac{(1+i)}{w}\right].
\end{align}
Substituting these into Eq.~\eqref{J1},
\begin{align}
    |J_1| &= \bigg|\lim_{w\rightarrow i}\frac{w}{w-i}\exp\left[i\ln\frac{w}{w+1}\right]
     \left(\left[-\frac{1}{w-i}-\frac{i}{w+1}+\frac{(1+i)}{w}\right]\right)^{-1}\bigg|\nonumber\\
    &=\left|\exp\left(i\ln\frac{i}{1+i}\right)\right|\nonumber\\
    &=e^{-\pi/4}.
\end{align}
In the same way, we can show $|J_2|=|J_3|=e^{-\pi/4}$. Thus, the 3-leg tensor is 
\begin{align}
    S^*_{\alpha\beta\gamma}
    &= e^{-\frac{\pi}{4}(x_\alpha+x_\beta+x_\gamma)}\langle\phi_\alpha(-1)\phi_\beta(i)\phi_\gamma(0)\rangle_{\textrm pl}.
\end{align}

\subsection{Conformal mapping of T}
The conformal mapping from the four-sided thermofield double state is 
\begin{align}
    z &= \frac{L}{2\pi}[-\ln(w-i)+\log(w+i)-i\ln(w+1)+i\ln(w-1)]\nonumber\\
    &=\frac{L}{2\pi}\left[\ln\left(\frac{w+i}{w-i}\right)+i\ln\left(\frac{w-1}{w+1}\right)\right].
\end{align}
To compute the Jacobian, we compute 
\begin{align}
    e^{2\pi z/L} &= \exp\left[\ln\frac{w+i}{w-i}+i\ln\frac{w-1}{w+1}\right],\\
    \frac{dz}{dw} &= \frac{L}{2\pi}\left[-\frac{1}{w-i}+\frac{1}{w+i}-\frac{i}{w+1}+\frac{i}{w-1}\right].
\end{align}
The Jacobian is then computed similarly as before:
\begin{align}
    |J_1|^{-1} &= \lim_{w\rightarrow i}\left|e^{-2\pi z/L}\left[-\frac{1}{w-i}+\frac{1}{w+i}-\frac{i}{w+1}+\frac{i}{w-1}\right]\right|\nonumber\\
    &=\frac{e^{\pi/2}}{2}.
\end{align}
The four-point function thus transforms as
\begin{align}
    \frac{T^*_{\alpha\beta\gamma\delta}}{T^*_{1111}}&=|J_1|^{x_\alpha}|J_2|^{x_\beta}|J_3|^{x_\gamma}|J_4|^{x_\delta}\langle\phi_\alpha(-1)\phi_\beta(i)\phi_\gamma(1)\phi_\delta(-i)\rangle_{\textrm pl}\nonumber, \\
    &= \left(\frac{e^{\frac{\pi}{2}}}{2}\right)^{-x_{\textrm tot}}\langle\phi_\alpha(-1)\phi_\beta(i)\phi_\gamma(1)\phi_\delta(-i)\rangle_{\textrm pl}.
\end{align}

\begin{figure}[tb]
    \centering
    \includegraphics[width=80mm]{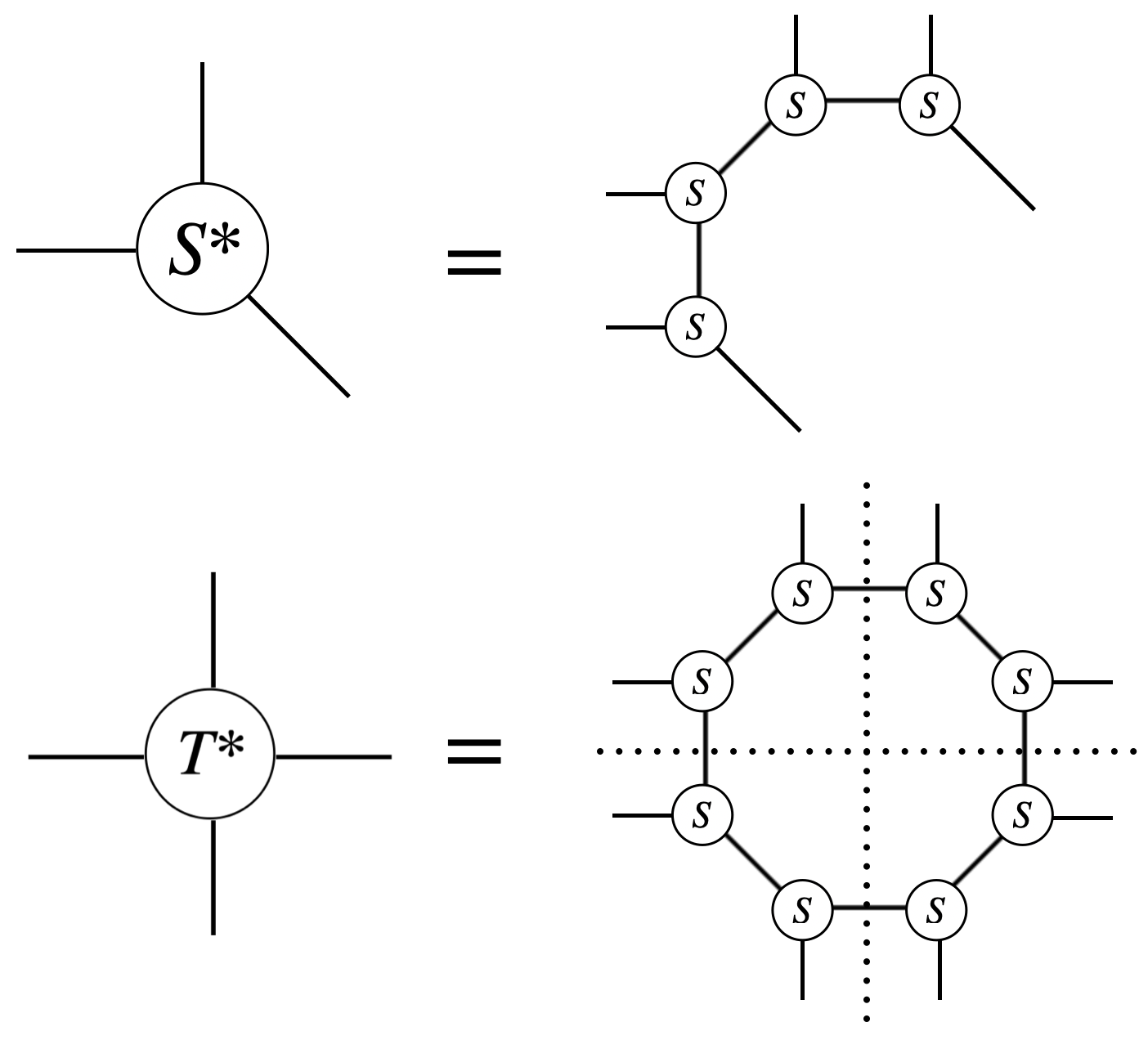}
    \caption{The contraction of the fixed-point tensors. We obtain $S$ from TRG and combine together to make $S^*$ and $T^*$. In this way, $T^*$ respects reflection symmetry along the dotted lines in addition to $C_4$ rotation symmetry. }
    \label{D4_TRG}
\end{figure}

\subsection{\texorpdfstring{${D_4}$-symmetric TRG}{D4-symmetric TRG}}

We use the TRG scheme which aligns closely with the original paper's methodology~\cite{PhysRevLett.99.120601}. In principle, singular-value decomposition (SVD) of the four-leg tensor should yield two identical symmetric tensors, given the $D_4$ symmetry of the original tensor. However, numerical errors sometimes make these two tensors non-identical. To mitigate this, we consistently select one of the three-leg tensors and supplement the other with its reflection. By adopting this approach, the fixed-point tensors, depicted in Fig.~\ref{D4_TRG}, maintain the $D_4$ symmetry at every RG step by construction.
\subsection{Four-point function of the critical Ising model}
Here, we list the four-point function of the Ising model. Given the four coordinates $z_i$ and its cross-ratio $x\equiv (z_{12}z_{34})/({z_{13}z_{24}})$, the four-point functions of the Ising CFT are
\begin{align}
\langle \epsilon^4\rangle &= \left|\left[\prod_{1\leq i<j\leq 4} z_{ij}^{-\frac13}\right] \frac{1-x+x^2}{x^\frac23(1-x)^\frac23}\right|^2\nonumber,\\
\langle \sigma^2\epsilon^2\rangle &=\left|\left[z_{12}^\frac14 z_{34}^{-\frac58}\left(z_{13}z_{24}z_{14}z_{23}\right)^{-\frac{3}{16}} \right]\frac{1-\frac{x}{2}}{x^\frac38(1-x)^\frac{5}{16}}\right|^2\nonumber,\\
\langle \sigma^4\rangle &= |z_{13}z_{24}|^{-1/4} 
\frac{|1+\sqrt{1-x}|+|1-\sqrt{1-x}|}{2|x|^\frac14 |1-x|^\frac14}\nonumber.
\end{align}
The functions above are used to evaluate the analytic FP tensor elements in the main text.
\begin{figure}[bt]
    \centering
    \includegraphics[width=86mm]{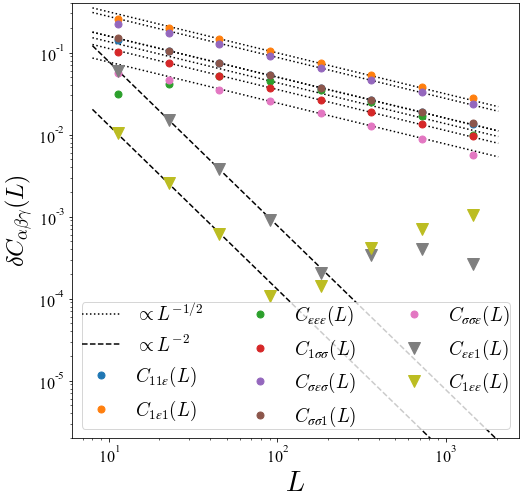}
    \caption{The finite-size corrections $\delta C_{\alpha\beta\gamma}(L)$ obtained from the numerical simulation of the critical Ising model. The numerical results for higher energy levels $\delta C_{\epsilon\epsilon 1}(L)$ and $\delta C_{1\epsilon\epsilon}(L)$ suffer from finite-$D$ effects for $L > 100$. The scalings of the finite-size corrections are nevertheless universal, which is consistent with Table III in Ref.~\cite{PhysRevB.107.155124}}
    \label{finite-size_ope}
\end{figure}

\subsection{Universal finite-size corrections}\label{sec:finiteOPE}
Here, we discuss the finite-size corrections to Eq. (13) in the main text. The finite-size corrections of the OPE coefficients are defined as 
\begin{align}
    \delta C_{\alpha\beta\gamma}(L) = |C_{\alpha\beta\gamma} - C_{\alpha\beta\gamma}(L)|,
\end{align}
where $C_{\alpha\beta\gamma}(L)$ is defined in Eq. (13) in the main text. We found that $\delta C_{\alpha\beta\gamma}(L)$ exhibits a universal power-law decay as
\begin{align}
    \delta C_{\alpha\beta\gamma}(L) \sim L^{-p_{\alpha\beta\gamma}}.\label{our_p}
\end{align}
Our numerical results suggest $p_{\alpha\beta\gamma} =1/2$ for $({\alpha,\beta,\gamma}) = (1,1,\epsilon),$ $(1,\epsilon,1),$ $(\epsilon,\epsilon,\epsilon),$ $(1,\sigma,\sigma),$ $(\sigma,\epsilon,\sigma),$ $(\sigma,\sigma,1),$ and $(\sigma,\sigma,\epsilon)$, and $p_{\alpha\beta\gamma} =2$ for  $({\alpha,\beta,\gamma}) = (\epsilon,\epsilon,1)$ and $(1,\epsilon,\epsilon)$ as shown in Fig.~\ref{finite-size_ope}. Similar universal scalings were discussed in Ref.~\cite{PhysRevB.107.155124}, where they considered the overlap of critical wavefunctions $A_{\alpha\beta\gamma}=\langle \phi^{3*}_\gamma|\phi^1_\alpha\phi^2_\beta\rangle$. The three wavefunctions are defined on a ring with a circumference of $L_1$, $L_2$, and $L_3=L_1+L_2$, respectively, and the lower indices are the label of the corresponding primary states. Ref.~\cite{PhysRevB.107.155124} found the overlap of wavefunctions to be
\begin{align}
\frac{A_{\alpha\beta\gamma}}{A_{111}} & \sim \left[\left(\frac{L_3}{L_1}\right)^{\frac{L_1}{L_3}}\left(\frac{L_3}{L_2}\right)^{\frac{L_2}{L_3}}\right]^{-\frac{L_3}{L_1}_\alpha-\frac{L_3}{L_2}_\beta+_\gamma}C_{\alpha\beta\gamma}
\quad +\quad \tilde{A}^{(p)}_{\alpha\beta\gamma}L_3^{-p_{\alpha\beta\gamma}},\label{previous_p}
\end{align}
where $p_{\alpha\beta\gamma}$ is the leading finite-size correction and $\tilde{A}^{(p)}_{\alpha\beta\gamma}$ is a prefactor that is independent of $L_3$. 

Our scaling exponents $p_{\alpha\beta\gamma}$ in Eq.~\eqref{our_p} coincide with those from the previous work in Eq.~\eqref{previous_p} for all fusion channels (see Table III of Ref.~\cite{PhysRevB.107.155124}). This universal scaling can be explained by considering rings 1 and 2 as an orbifold theory. The scaling $p_{\alpha\beta\gamma} =1/2$ is then attributed to the difference in the scaling dimensions of the orbifold theory, which is $x_\epsilon/2=1/2$. (See Ref.~\cite{PhysRevB.107.155124} for details.) 
Similarly, we conjecture that the universal scaling for $\delta T_{\alpha\beta\gamma\delta} \sim L^{-1/3}$ can be understood by considering the three of four legs to be an orbifold theory.

\subsection{The three-State Potts model\label{sec:OPE_three_state_potts}}
\begin{figure}
    \centering
    \includegraphics[width=86mm]{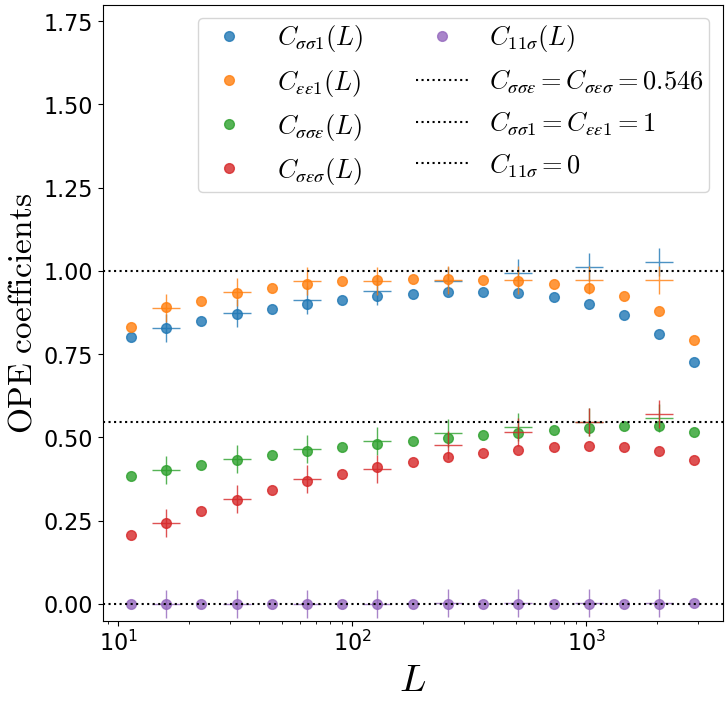}
    \caption{The OPE coefficients of the critical three-state Potts model evaluated by setting $x_S = e^{{\pi/ 4}}$. The black dotted lines denote the theoretical values 0, 0.546, and 1~\cite{mccabe1995critical}. The data points, denoted by filled circles "$\circ$" and crosses "$+$," are obtained from Levin-TRG($D=88$) and Evenbly-TNR($D=40$), respectively.}
    \label{OPE_3potts}
\end{figure}

Here, we present the OPE coefficients obtained from numerical simulations of the classical critical three-state Potts model. The low-lying primary states of this model are the identity operator "1", the two spin operators "$\sigma$," and the thermal operator "$\epsilon$," whose scaling dimensions are 0, 2/15, and 4/5. The non-trivial coefficients 
 is $C_{\sigma\sigma\epsilon} = 0.546$~\cite{mccabe1995critical}. Figure.~\ref{OPE_3potts} exhibits the numerical results from Levin-TRG and Evenbly-TNR as the Ising model in the main text. TRG/TNR schemes, generally speaking, have finite-$D$ effects for larger system sizes, and this effect is larger in higher central charges. Since the central charge $c=0.8$ of the three-state Potts model is larger than $c=0.5$ of the Ising model, these numerical errors manifest in the data plots. In particular, the TRG data is unstable due to CDL tensors and quickly diverts from the theoretical values. However, Evenbly-TNR's results still converge to the correct values.

\end{document}